%% file: jes_nim.tex
\documentclass[final,5p,twocolumn,10pt,preprint,longtitle,sort&compress]{elsarticle}

\usepackage{graphicx}

\usepackage[usenames]{color}

\usepackage{rotating}

\usepackage{changebar}

\usepackage{subfig}

\usepackage{ifthen}

\usepackage{amsmath, amssymb}
\usepackage{xspace}

\usepackage{epsfig}

\usepackage{placeins} 

\usepackage{slashbox}

\usepackage{multirow}

\usepackage{hyperref}
\usepackage{alphalph}
\hypersetup{
  colorlinks=false,
  pdfborder={0 0 0},
}

\let\LaTEXref\ref
\renewcommand{\ref}[1]
{{\LaTEXref{#1}}}

\usepackage[capitalise,nameinlink]{cleveref}
\crefname{section}{Sec.}{Secs.} 
\Crefname{section}{Section}{Sections} 
\creflabelformat{equation}{#2#1#3} 
\newcommand{\crefrangeconjunction}{--} 

\usepackage{comment}

\input{shortcuts}

\journal{Nuclear Instruments and Methods in Physics Research A }

\begin{document}


\begin{frontmatter}
\title{Jet energy scale determination in the \DZero experiment}

\input{author_list_nim}      

\date{December 24, 2013}


\begin{abstract}
The calibration of jet energy measured in the \DZero detector is presented,
based on \ppbar{} collisions
at a center-of-mass energy of \TeV{1.96} at the Fermilab Tevatron collider.
Jet energies are measured using a sampling calorimeter composed 
of uranium and liquid argon as the passive and active media, respectively. 
This paper describes the energy calibration of jets 
performed with \photonjet, \Zjet{} and \dijet{} events,
with jet transverse momentum $\pT > \GeVc{6}$ and pseudorapidity range $\meta < 3.6$.
The corrections are measured separately for data and simulation, achieving a
precision of $1.4\%-1.8\%$ for jets in the central part of the calorimeter and
up to 3.5\% for the jets with pseudorapidity $|\eta| = 3.0$.
Specific corrections are extracted to enhance the description of jet energy in
simulation and in particular of the effects due to the flavor of the parton originating the jet,
correcting biases up to $3\%-4\%$ in jets with low $\pT$ originating
from gluons and up to $6\%-8\%$ in jets from $b$ quarks.
\end{abstract}


\end{frontmatter}

\setcounter{tocdepth}{2}
\tableofcontents

\input{introduction}

\input{detector}
\input{objectid}

\input{overview}
\input{response_overview}

\input{datasets}
\input{offset}
\input{response}
\input{etadep}

\input{mpfbiases}
\input{showering}
\input{summary}
\input{closure}
\input{fdc_corrections}

\input{jssr}
\input{qcd_specific}

\input{correlation}
\input{conclusions}
\input{acknowledgement}
\input{biblio}

\end{document}

%% file: shortcuts.tex
\usepackage{ifthen}
\usepackage{slashed}
\usepackage{textcomp}

\newcommand{\DZero}     {D0\xspace} 
\newcommand{\TeVenergy} {\JESmathSym{\sqrt{s} = \TeV{1.96}}}

\newcommand{\I}       {Run~I\xspace}
\newcommand{\II}      {Run~II\xspace}
\newcommand{\IIa}     {Run~IIa\xspace}
\newcommand{\IIb}     {Run~IIb\xspace}
\newcommand{\IIbOne}  {Run~IIb1\xspace}
\newcommand{\IIbTwo}  {Run~IIb2\xspace}

\newcommand{\JESmathSym}[1]{{\ensuremath{#1}}\xspace} 

\newcommand{\abs}[1]{\ensuremath{\left|{#1}\right|}}
\newcommand{\norm}[1]{\ensuremath{\left\|{#1}\right\|}}
\newcommand{\average}[1]{\ensuremath{\left<{#1}\right>}}
\newcommand{\parentheses}[1]{\ensuremath{\left({#1}\right)}}
\newcommand{\sqbrackets}[1]{\ensuremath{\left[{#1}\right]}}
\newcommand{\braces}[1]{\ensuremath{\left\{{#1}\right\}}}
\DeclareRobustCommand{\brackets}[1]{\ensuremath{\left(#1\right)}}
\newcommand{\sqr}[1]{\ensuremath{\parentheses{#1}^2}}
\newcommand{\func}[2][f]{#1\brackets{#2}}
\newcommand{\hfrac}[2]{\JESmathSym{\left. #1 \middle/ #2 \right.}}
\DeclareMathOperator{\erf}{erf}

\newcommand{\Missing}[1]{\JESmathSym{\not\!\!#1}}

\newcommand{\MakeTransverseExt}[3]{\ensuremath{#1^{#2}_\textnormal{T#3}}}
\newcommand{\MakeTransverse}[2]{\MakeTransverseExt{#1}{#2}{}}
\newcommand{\pT}[1][]    {\JESmathSym{\MakeTransverse{p}{\textnormal{#1}}}} 

\newcommand{\pt}         {\pT}

\newcommand{\vecpT}[1][] {\JESmathSym{\MakeTransverse{\vec{p}}{\textnormal{#1}}}}
\newcommand{\ET}[1][]    {\JESmathSym{\MakeTransverse{E}{\textnormal{#1}}}}
\newcommand{\Eprime}     {\JESmathSym{E^{\prime}}} 
\newcommand{\pTprime}    {\JESmathSym{\MakeTransverse{p}{\prime}}} 
\newcommand{\MET}[1][]   {\JESmathSym{\Missing{\ET[#1]}}} 
\newcommand{\vecmet}[1][]{\JESmathSym{\MakeTransverse{\Missing{\vec{E}}}{\textnormal{#1}}}}
\newcommand{\METx}[1][]  {\Missing{E_{x}^{\textnormal{#1}}}}
\newcommand{\METy}[1][]  {\Missing{E_{y}^{\textnormal{#1}}}}
\newcommand{\METxy}[1][] {\Missing{E_{x/y}^{\textnormal{#1}}}}
\newcommand{\deltar}     {\JESmathSym{\Delta\mathcal{R}}}
\newcommand{\npv}        {\JESmathSym{n_{\textnormal{PV}}}}
\newcommand{\zpv}        {\JESmathSym{z_{\textnormal{PV}}}}
\newcommand{\mzpv}       {\abs{\zpv}}
\newcommand{\deta}[1][]  {\JESmathSym{\eta_\textnormal{det}^{\textnormal{#1}}}}
\newcommand{\dphi}[1][]  {\JESmathSym{\azim_\textnormal{det}^{\textnormal{#1}}}}

\newcommand{\meta}       {\JESmathSym{\abs\eta}}
\newcommand{\mdeta}      {\JESmathSym{\abs\deta}}

\newcommand{\radius}     {\JESmathSym{r}}
\newcommand{\azim}       {\JESmathSym{\phi}}
\newcommand{\ieta}       {\JESmathSym{i_{\eta}}}
\newcommand{\iphi}       {\JESmathSym{i_{\azim}}}

\newcommand{\rcone}      {\JESmathSym{\mathcal{R}_\textnormal{cone}}}

\newcommand{\InstLumin}  {\JESmathSym{\mathcal{L}}}

\newcommand{\Emeas}[1][] {\JESmathSym{E^\textnormal{meas}_{\textnormal{#1}}}}
\newcommand{\Ecorr}[1][] {\JESmathSym{E^\textnormal{corr}_{\textnormal{#1}}}}
\newcommand{\Eptcl}[1][] {\JESmathSym{E^\textnormal{ptcl}_{\textnormal{#1}}}}
\newcommand{\kO}[1][]    {\JESmathSym{k_{\textnormal{O}}^{\textnormal{#1}}}}
\newcommand{\kR}[1][]    {\JESmathSym{k_{\textnormal{R}}^{\textnormal{#1}}}}

\newcommand{\kOZS}       {\JESmathSym{\kO[ZS]}}
\newcommand{\kRZS}       {\JESmathSym{\kR[ZS]}}

\newcommand{\kRtopo}[1][]{\JESmathSym{k_{\textnormal{R}}^{\textnormal{topo#1}}}}
\newcommand{\MPF}[1]     {\JESmathSym{R_{\textnormal{MPF}}^{\textnormal{#1}}}}
\newcommand{\MPFCC}[1]   {\JESmathSym{R_{\textnormal{MPF,CC}}^{\textnormal{#1}}}}
\newcommand{\MPFeta}[1]  {\JESmathSym{R_{\textnormal{MPF,$\eta$}}^{\textnormal{#1}}}}
\newcommand{\pTgamma}[1][]{\JESmathSym{\MakeTransverseExt{p}{\textnormal{#1}}{\photon}}}
\newcommand{\vecpTgamma}[1][]{\JESmathSym{\MakeTransverseExt{\vec{p}}{\textnormal{ #1}}{\photon}}}
\newcommand{\pTtag}[1][] {\JESmathSym{\MakeTransverseExt{p}{#1}{\,tag}}}
\newcommand{\Feta}[1][]  {\JESmathSym{F_{\eta}^{\textnormal{#1}}}}

\newcommand{\RMSped}     {\JESmathSym{\sigma_{\textnormal{PED}}}}

\newcommand{\fpfm}[1][f]{\JESmathSym{#1^{+}#1^{-}}}

\newcommand{\photon}    {\JESmathSym{\gamma}}
\newcommand{\ffbar}[1][f]{\JESmathSym{#1\bar{#1}}}
\newcommand{\qqbar}     {\ffbar[q]}
\newcommand{\ppbar}     {\ffbar[p]}
\newcommand{\Xjet}[1][X]{\JESmathSym{#1+\textnormal{jet}}}

\newcommand{\photonjet} {\Xjet[\photon]}

\newcommand{\Zjet}      {\Xjet[Z]}

\newcommand{\dijet}     {\JESmathSym{\textnormal{dijet}}}
\newcommand{\Zee}       {\JESmathSym{Z \rightarrow \fpfm[e]}}
\newcommand{\MZ}        {\JESmathSym{M_{Z}}}

\newcommand{\Generator}[1]{\textsc{#1}}
\newcommand{\ALPGEN}{\Generator{alpgen}}
\newcommand{\PYTHIA}{\Generator{pythia}}
\newcommand{\ALPGENPYTHIA}{\ALPGEN+\PYTHIA}

\newcommand{\GEANT}{\Generator{geant}}

\DeclareRobustCommand{\Quantity}[2]{\JESmathSym{\ifthenelse{\equal{#2}{}}{}{#2\mskip 4mu plus 2mu minus 2.5mu}\textnormal{#1}}}
\DeclareRobustCommand{\TenTo}[2][]{\JESmathSym{\ifthenelse{\equal{#2}{}}{10^{#1}}{#1\,\cdot\,10^{#2}}}}

\newcommand{\rad}[1]{\Quantity{rad}{#1}}

\newcommand{\um}[1]{\Quantity{\textmu{}m}{#1}}
\newcommand{\mm}[1]{\Quantity{mm}{#1}}
\newcommand{\cm}[1]{\Quantity{cm}{#1}}
\newcommand{\sqcm}[1]{\Quantity{$\textnormal{cm}^{2}$}{#1}}

\newcommand{\kV}[1]{\Quantity{kV}{#1}}
\newcommand{\Hz}[1]{\Quantity{Hz}{#1}}
\newcommand{\kHz}[1]{\Quantity{kHz}{#1}}

\newcommand{\MeV}[1]{\Quantity{MeV}{#1}}
\newcommand{\GeV}[1]{\Quantity{GeV}{#1}}
\newcommand{\TeV}[1]{\Quantity{TeV}{#1}}

\newcommand{\MeVc}[1]{\Quantity{$\textnormal{MeV}$}{#1}}
\newcommand{\GeVc}[1]{\Quantity{$\textnormal{GeV}$}{#1}}
\newcommand{\GeVcc}[1]{\Quantity{$\textnormal{GeV}$}{#1}}

\newcommand{\MOhmSq}[1]{\Quantity{${\rm M}\Omega/\Box$}{#1}}

\newcommand{\ns}[1]{\Quantity{ns}{#1}}

\newcommand{\K}[1]{\Quantity{K}{#1}}

\newcommand{\T}[1]{\Quantity{T}{#1}}

\newcommand{\invfb}[1]{\Quantity{$\textnormal{fb}^{-1}$}{#1}}

\newcommand{\InstLumDot}[1]{\Quantity{$\cdot 10^{30}\,\textnormal{cm}^{-2}\textnormal{s}^{-1}$}{#1}}
\newcommand{\InstLumTimes}[1]{\Quantity{$\!\times 10^{30}\,\textnormal{cm}^{-2}\textnormal{s}^{-1}$}{#1}}

\def\addmore{0}
\newcommand{\extend}[1]{\ifnum \addmore=1 \emph{ (#1)}\fi}

\newcommand{\eg}{{e.g.}\@\xspace}
\newcommand{\ie}{{i.e.}\@\xspace}
\newcommand{\cf}{{cf.}\@\xspace}
\newcommand{\etc}{{etc.}\@\xspace}

\def\nthsuffix#1{{%
	\countdef\units=0 \countdef\tens=1 \countdef\q=2 
	\units=#1 \q=#1 \divide\q by10 \multiply\q by10 \advance\units by-\q 
	\tens=#1 \q=#1 \divide\q by100 \multiply\q by100 \advance\tens by-\q \divide\tens by10 
	\ifnum\tens=1 th \else \ifnum\units=1 st \else \ifnum\units=2 nd \else \ifnum\units=3 rd \else th \fi \fi \fi \fi 
}}

\newcommand{\RefCite}[1]{Ref.~\cite{#1}}


%% file: author_list_nim.tex
\def\collaboration {}
\def\noaffiliation {}
\address[md5cc7196d87cde26c813f890864b4a934b]{LAFEX, Centro Brasileiro de Pesquisas F\'{i}sicas, Rio de Janeiro, Brazil}
\address[md54859b9ee31a7df4ba30e7374859dd0f3]{Universidade do Estado do Rio de Janeiro, Rio de Janeiro, Brazil}
\address[md59b58cce363ddd7ae133b45d7ac8e34a1]{Universidade Federal do ABC, Santo Andr\'e, Brazil}
\address[md5816806e874e67b278e694889d375265a]{University of Science and Technology of China, Hefei, People's Republic of China}
\address[md52b964412fa23b11f599bddca8ef73c1f]{Universidad de los Andes, Bogot\'a, Colombia}
\address[md5277195eef8d143a7ed25967125282faf]{Charles University, Faculty of Mathematics and Physics, Center for Particle Physics, Prague, Czech Republic}
\address[md507fd366e73001d49b31eb82e3e4f300e]{Czech Technical University in Prague, Prague, Czech Republic}
\address[md5b78b2b946d0360da03a31502d37fc337]{Institute of Physics, Academy of Sciences of the Czech Republic, Prague, Czech Republic}
\address[md5aebc5d941d5ac941a53c7c4c551da08a]{Universidad San Francisco de Quito, Quito, Ecuador}
\address[md596c6923c189117d15c85349cc77077de]{LPC, Universit\'e Blaise Pascal, CNRS/IN2P3, Clermont, France}
\address[md50fdb8718c8a8ea4fdc96655cd287f801]{LPSC, Universit\'e Joseph Fourier Grenoble 1, CNRS/IN2P3, Institut National Polytechnique de Grenoble, Grenoble, France}
\address[md5725c78f97c9652163917c391ef9400b0]{CPPM, Aix-Marseille Universit\'e, CNRS/IN2P3, Marseille, France}
\address[md5fc48289704bd62a9a8337d91b2ed2121]{LAL, Universit\'e Paris-Sud, CNRS/IN2P3, Orsay, France}
\address[md544f75535f510ed4fbc6d466d69db0f3b]{LPNHE, Universit\'es Paris VI and VII, CNRS/IN2P3, Paris, France}
\address[md52ba807daf7d413d4fa8b75d33fec60a0]{CEA, Irfu, SPP, Saclay, France}
\address[md5d49e76bc74ab21cf96d05d86d8925206]{IPHC, Universit\'e de Strasbourg, CNRS/IN2P3, Strasbourg, France}
\address[md5b49b3dfd52480338587e551cf0a6e0af]{IPNL, Universit\'e Lyon 1, CNRS/IN2P3, Villeurbanne, France and Universit\'e de Lyon, Lyon, France}
\address[md592cdc2b721dc8d9dc10668ec30fcc09d]{III. Physikalisches Institut A, RWTH Aachen University, Aachen, Germany}
\address[md54a21222b0e95e6c06070cb332d329079]{Physikalisches Institut, Universit\"at Freiburg, Freiburg, Germany}
\address[md51d911a0d169192fb205779801b021f01]{II. Physikalisches Institut, Georg-August-Universit\"at G\"ottingen, G\"ottingen, Germany}
\address[md5c58e4f0e4eae729e47f40b9825362467]{Institut f\"ur Physik, Universit\"at Mainz, Mainz, Germany}
\address[md5545b26b934d52e7cc42a24e7af19e0b1]{Ludwig-Maximilians-Universit\"at M\"unchen, M\"unchen, Germany}
\address[md56aaf43a5aaac60795ff88cdecdee936c]{Panjab University, Chandigarh, India}
\address[md5326ca0a7d914cfc0a3680a973c58dfb4]{Delhi University, Delhi, India}
\address[md5fc386bbd67b34b140f8bc4643de8bc48]{Tata Institute of Fundamental Research, Mumbai, India}
\address[md553b1489c0c9b54a1147a5ccf3690a021]{University College Dublin, Dublin, Ireland}
\address[md5d52ef9d339581daeda4fa4710d718ac7]{Korea Detector Laboratory, Korea University, Seoul, Korea}
\address[md50095f63c5b7d7324a13b09c9326d0aa1]{CINVESTAV, Mexico City, Mexico}
\address[md524e259d70bb82a314af2869c8e6ade19]{Nikhef, Science Park, Amsterdam, the Netherlands}
\address[md518fd70a5e3413edaa86ded029df5c8bd]{Radboud University Nijmegen, Nijmegen, the Netherlands}
\address[md5306552cfe4e2d2ca3cc1d3ecc85cb870]{Joint Institute for Nuclear Research, Dubna, Russia}
\address[md59af13fd9e08e10f62d48d354d141c7dc]{Institute for Theoretical and Experimental Physics, Moscow, Russia}
\address[md5693e243e978ba579cfa107af43c5a30b]{Moscow State University, Moscow, Russia}
\address[md50fa920a2459f3dcc2c9e973db0840d5e]{Institute for High Energy Physics, Protvino, Russia}
\address[md538d5b2c440840d9dd033d11e592f5269]{Petersburg Nuclear Physics Institute, St. Petersburg, Russia}
\address[md5c887f4628beb512773b5c71cdb263630]{Instituci\'{o} Catalana de Recerca i Estudis Avan\c{c}ats (ICREA) and Institut de F\'{i}sica d'Altes Energies (IFAE), Barcelona, Spain}
\address[md53efd7e5e79140d92f93a9be02b5b4a75]{Uppsala University, Uppsala, Sweden}
\address[md5f50ce3b1367c4fea3e082769b1035988]{Lancaster University, Lancaster LA1 4YB, United Kingdom}
\address[md59f38e9ce1e213c2b848854ee630a5433]{Imperial College London, London SW7 2AZ, United Kingdom}
\address[md5a382df6e52dcbe8ebf66c19a8d8a7f3f]{The University of Manchester, Manchester M13 9PL, United Kingdom}
\address[md5c8896cd0e923b7c5e648052692828d3f]{University of Arizona, Tucson, Arizona 85721, USA}
\address[md51762b73071c846b960a5802965822e57]{University of California Riverside, Riverside, California 92521, USA}
\address[md5e2035f8d1fef0c6263b3c7d87fecc0f2]{Florida State University, Tallahassee, Florida 32306, USA}
\address[md5bb3834928f6eda6234448cc0b921307f]{Fermi National Accelerator Laboratory, Batavia, Illinois 60510, USA}
\address[md51a2d77b6457fed7f09090ccb5d034224]{University of Illinois at Chicago, Chicago, Illinois 60607, USA}
\address[md55dfbbd9ba3dd0c7d2c3cd1e20c3a010a]{Northern Illinois University, DeKalb, Illinois 60115, USA}
\address[md556791f9c01fa4d1fc52c2ff43c5ffe9f]{Northwestern University, Evanston, Illinois 60208, USA}
\address[md5454d61aaeaf94091167f8fce34e76fa9]{Indiana University, Bloomington, Indiana 47405, USA}
\address[md50d1129348fbd891bbca669e9a4107d24]{Purdue University Calumet, Hammond, Indiana 46323, USA}
\address[md5ef72b78eaf71ff17dc9bb50e3cbdb35f]{University of Notre Dame, Notre Dame, Indiana 46556, USA}
\address[md52e7abcfb66c79229acc43159a3a33e7e]{Iowa State University, Ames, Iowa 50011, USA}
\address[md5c08640bf74cabd71cf7bc9ed882b71ab]{University of Kansas, Lawrence, Kansas 66045, USA}
\address[md54ab02c3931e9d07b2084690f2c3ec017]{Louisiana Tech University, Ruston, Louisiana 71272, USA}
\address[md554fbc46752b66dfb1b41979d52117917]{Northeastern University, Boston, Massachusetts 02115, USA}
\address[md5e4064d55a722ab3bbfe0a99ffcc4f6d7]{University of Michigan, Ann Arbor, Michigan 48109, USA}
\address[md551cb7f6d81c5e31bd7a8348cf2ec9445]{Michigan State University, East Lansing, Michigan 48824, USA}
\address[md5e70e163a5e53569c43c7426bb0295ad9]{University of Mississippi, University, Mississippi 38677, USA}
\address[md5c2c7652879ae21aff975332e360757c7]{University of Nebraska, Lincoln, Nebraska 68588, USA}
\address[md51637761fc92038f1d2f785c729af7975]{Rutgers University, Piscataway, New Jersey 08855, USA}
\address[md5899b4beea5be1aa9e1b9740bc84c3340]{Princeton University, Princeton, New Jersey 08544, USA}
\address[md5d0ba23249990e81f799cde13a133c7de]{State University of New York, Buffalo, New York 14260, USA}
\address[md559b1addaa6f886b6cd013cb30d428de1]{University of Rochester, Rochester, New York 14627, USA}
\address[md548167b68bcb7c9b6961c523626919bc8]{State University of New York, Stony Brook, New York 11794, USA}
\address[md582494719b2bef33c80358f7d71924e25]{Brookhaven National Laboratory, Upton, New York 11973, USA}
\address[md510d9847b4d3a47ff5b4092afb0110474]{Langston University, Langston, Oklahoma 73050, USA}
\address[md54f830e8dc9ba7a58eb2af24f9caf2ed5]{University of Oklahoma, Norman, Oklahoma 73019, USA}
\address[md5634558bea5458fd72062d16ba0c32baa]{Oklahoma State University, Stillwater, Oklahoma 74078, USA}
\address[md5bad3354b8e2778ba42d27e9fade117d6]{Brown University, Providence, Rhode Island 02912, USA}
\address[md501ae293b042d193a66d06e015de37747]{University of Texas, Arlington, Texas 76019, USA}
\address[md5037b12823f24d7636ddba2fb14848c38]{Southern Methodist University, Dallas, Texas 75275, USA}
\address[md5e3540115023ff9a10f87ffa2010b16d5]{Rice University, Houston, Texas 77005, USA}
\address[md5589e3f7205f95307cf4059822dded820]{University of Virginia, Charlottesville, Virginia 22904, USA}
\address[md54e785e2c4e5e92b1b1c6c95ebb0e4317]{University of Washington, Seattle, Washington 98195, USA}
\author[md5306552cfe4e2d2ca3cc1d3ecc85cb870]{V.M.~Abazov}
\author[md54f830e8dc9ba7a58eb2af24f9caf2ed5]{B.~Abbott}
\author[md5fc386bbd67b34b140f8bc4643de8bc48]{B.S.~Acharya}
\author[md51a2d77b6457fed7f09090ccb5d034224]{M.~Adams}
\author[md5e2035f8d1fef0c6263b3c7d87fecc0f2]{T.~Adams}
\author[md5a382df6e52dcbe8ebf66c19a8d8a7f3f]{J.P.~Agnew}
\author[md5306552cfe4e2d2ca3cc1d3ecc85cb870]{G.D.~Alexeev}
\author[md538d5b2c440840d9dd033d11e592f5269]{G.~Alkhazov}
\author[md5e4064d55a722ab3bbfe0a99ffcc4f6d7]{A.~Alton$^{a}$}
\author[md5e2035f8d1fef0c6263b3c7d87fecc0f2]{A.~Askew}
% ...
\vskip 0.25cm

%% file: introduction.tex
\section{Introduction}
\label{sec:Introduction}

Jets are frequently produced in the parton interactions that occur at
hadron colliders.  They arise from the complex physical process of hadronization
that evolves a color-charged quark or gluon, collectively referred to as ``parton'', 
into a 
collimated set of final state colorless
hadrons, photons and leptons.  
Nearly all processes resulting from hard \ppbar{} collisions at the Fermilab
Tevatron Collider produce one or more jets.
The jet energy scale relates the measured energy of a jet to the energy of the
particles it contains.
Since many physics measurements involve events with jets, an accurate calibration
of the energy scale of jets is essential.
The jet energy scale is a major source of systematic uncertainty
in measurements involving jets, including inclusive jet and \dijet production cross
sections~\cite{Abazov:2008ae,IncJets,DiJetM}, top quark mass measurements~\cite{Top_mass,ttbar},
searches for the Higgs boson (see \eg, \cite{Higgs}), and for many
final states predicted by extensions to the Standard Model (SM).
Some of the particles predicted by such models would pass mostly undetected
through the detector (similar to neutrinos), 
manifesting themselves through an imbalance in the measured total momentum of the event.
Uncertainties on the jet energy scale can 
impede the ability to resolve these signatures
reducing the sensitivity of both SM measurements and new physics searches.

This paper describes new methods developed by the \DZero{} Collaboration
to determine an absolute energy calibration for jets reconstructed
with the \II{} Cone Algorithm~\cite{JetAlgo}.
This calibration corrects the reconstructed jet energy to the particle level.
Here the particle level includes all stable particles as defined in Ref.~\cite{particle}.
The development of the methods described here is based on previous studies performed
at the Tevatron during \I{}~\cite{JES_run1}, but those methods have been significantly extended
to meet new physics demands of \II.
The calibration is derived for jets reconstructed
with two different cone sizes, from data taken in \ppbar{} collisions
at \TeVenergy.

The characteristics and performance of the D0 
subdetectors are outlined
in \Cref{sec:detector}. The definition of a jet
and the description of object reconstruction and selection
are provided in \Cref{sec:objectid}.  An overview of the correction procedure 
is given in \Cref{sec:overview,sec:response_overview}.
\Cref{sec:datasets} describes the data and Monte Carlo (MC) samples used 
for the calibration.
\Cref{sec:offset,sec:response,sec:etadep,sec:mpfbiases,sec:showering}
discuss 
different corrections involved in the jet energy calibration, explain the 
methods used in their derivation, and present the numerical results. 
The jet energy corrections and their uncertainties are summarized in 
\Cref{sec:summary}, and the MC-based consistency checks 
and data-based verifications of the method are presented in 
\Cref{sec:closure}.
\Cref{sec:fdc,sec:jssr} describe additional tuning applied 
to jet energy in MC events to more precisely model data. \Cref{sec:qcd_specific} 
discusses specific corrections applied to jets in \dijet{} events. Correlations 
between systematic uncertainties are described in \Cref{sec:correlation}. 
Finally, \Cref{sec:conclusion} presents concluding remarks.

%% file: detector.tex
\section{The \DZero{} detector}
\label{sec:detector}

\subsection{Overview}
\label{ssec:detector:overview}

The \DZero{} detector, as upgraded for the 2001--2011 \II{} of the Tevatron,
is presented in detail elsewhere~\cite{bib:RunI_nim,bib:RunII_nim,l1cal,l0}. Here, we focus
on aspects of the detector and its calibration of particular importance for
understanding the jet energy scale.

In \II, the Tevatron operated with proton and antiproton beams
of 36 bunches each, grouped in three bunch trains.
The bunches within each train are spaced at \ns{396} intervals.  In normal operation,
the state of colliding beams suitable for data taking is 
maintained for 10 to 15 hours at a time (called a ``store'').  
The distribution of the location of \ppbar{} collisions is
approximately Gaussian along the beam line ($z$ axis) with 
a standard deviation around 
the nominal collision point in the center of the detector of about \cm{25}.
The typical number of \ppbar{} interactions per bunch crossing ranged from 2 to 12, 
depending on the instantaneous luminosity. 

The trajectories of charged particles from the \ppbar{} interaction region
are measured with a silicon microstrip tracker (SMT)~\cite{SMT} that is surrounded by a
scintillation fiber tracker. Both are located within a \T{2} solenoidal field.  The
tracking detectors are used to reconstruct the position of the 
vertex of the \ppbar{} interaction, and, in case of multiple \ppbar{} interactions, 
to associate calorimeter activity with the observed interaction vertices.
The tracking system provides \um{35} resolution on the position of the
\ppbar scattering along the beam line and \um{15} 
impact parameter resolution in the $r$-$\azim$ plane~\cite{d0_coordinate}
near the beam line for tracks with 
momentum transverse to the beam line $\pT > \GeVc{10}$.  While the amount
of material traversed by a charged particle depends on its trajectory, the
typical material traversed is about 0.1 radiation lengths ($X_{0}$)
in the tracking system.
In 2006 the \DZero{} detector was upgraded with the addition of an extra layer
of the silicon detector close to the beam pipe. This ``Layer~0''
is described in \RefCite{l0}.
The period before this upgrade is referred to as \IIa and the period after as \IIb.

Outside the tracking system, preshower detectors and the solenoidal magnet 
present $\approx\!2~X_0$ of material.  
The central preshower detector (CPS),
used also for photon identification, consists of $\approx\!1~X_0$  of
lead absorber surrounded by three layers of scintillating strips.
The preshower detectors are in turn surrounded by sampling
calorimeters constructed of depleted uranium absorbers and liquid argon as 
active medium enclosed in separate cryostats.
The central calorimeter (CC) covers pseudorapidities~\cite{d0_coordinate}
up to $\mdeta \approx 1.1$; and
two end calorimeters (EC) extend coverage up to $\mdeta \approx 4.2$.  While
the basic structure of the calorimeter was retained from \I,
new electronics was developed to accommodate the greatly reduced time
between beam crossings in \II.

Outside of the calorimetry, three layers of tracking and scintillation
detectors~\cite{bib:RunII_muon}, in conjunction with an \T{1.8} toroidal field, 
are used to identify muons.
The measurement of the momentum of muons, that may deposit up to 2.5 \GeV{} of their
energy in the calorimeter by ionization \cite{bib:RunII_muon}, combines the information from the muon system with the
independent and more accurate measurement from the central tracking system.
Luminosity is measured using plastic scintillator arrays placed in front of the EC cryostats,
covering $2.7 <\mdeta < 4.4$~\cite{LumiNIM}.
The average instantaneous luminosity (\InstLumin) was \InstLumDot{45}
for \IIa data taking period, and \InstLumDot{140} for \IIb.

The \DZero detector has a three level trigger system, with each successive
level examining and passing on fewer events selected according to tighter
criteria.  The first level (Level~1) is constructed from trigger elements
implemented in hardware and has an accept rate of about \kHz{2}.  In Level~2,
hardware engines and embedded microprocessors provide subdetector level
information to a global processor.  The Level~2 system accepts events at
about \kHz{1}. Those events are then sent to a farm of Linux-based 
processors.  This Level~3 farm identifies high-level objects
(jets, electrons, etc.) with greater precision and selects events for later detailed 
reconstruction at a rate of about \Hz{100}.

\subsection{Calorimeters}
\label{calorimeters}

The innermost part of the CC and EC is optimized for the measurement of
electromagnetic energy. These four (three) layers of the CC (EC) calorimeter are termed the
EM section.  They are approximately 20.0 (21.4)~$X_{0}$ thick in the
CC (EC) and use depleted uranium plates of \mm{3-4} thickness.
The following three layers (four in the EC) of the calorimeter are
constructed of 6-mm thick uranium-niobium alloy, and are
optimized for the measurement of hadronic energy. These layers comprise the fine
hadronic (FH) section.  In the CC, the FH section is about 3.1 nuclear interaction lengths ($\lambda_{\rm int}$)
thick, and in the EC it ranges from about 3.6 to 4.4~$\lambda_{\rm int}$.
In the outer layers of the calorimeter, thick plates of copper
or stainless steel are used in place of uranium.  This coarse hadronic
(CH) calorimeter ranges from 3.2 to 6.0~$\lambda_{\rm int}$ in depth.

The geometry of the region between the CC and EC is complicated.  In some
places there are substantial amounts of material other than the 
calorimeter layers.
In these areas, separate single cell structures called ``massless gaps''
are installed to sample the development of showers from interactions with
uninstrumented materials. 
To provide additional coverage there is also a
plastic scintillator inter-cryostat detector (ICD) between the CC and EC, covering
the pseudorapidity range $1.1 < \mdeta < 1.4$.  Because of the complicated geometry and
relatively rapidly changing response of the plastic scintillator
detector system during data collection, data from this inter-cryostat region (ICR) need to
be analyzed separately.

\begin{figure*}[t!]
	\centerline{\includegraphics[width=0.8\textwidth]{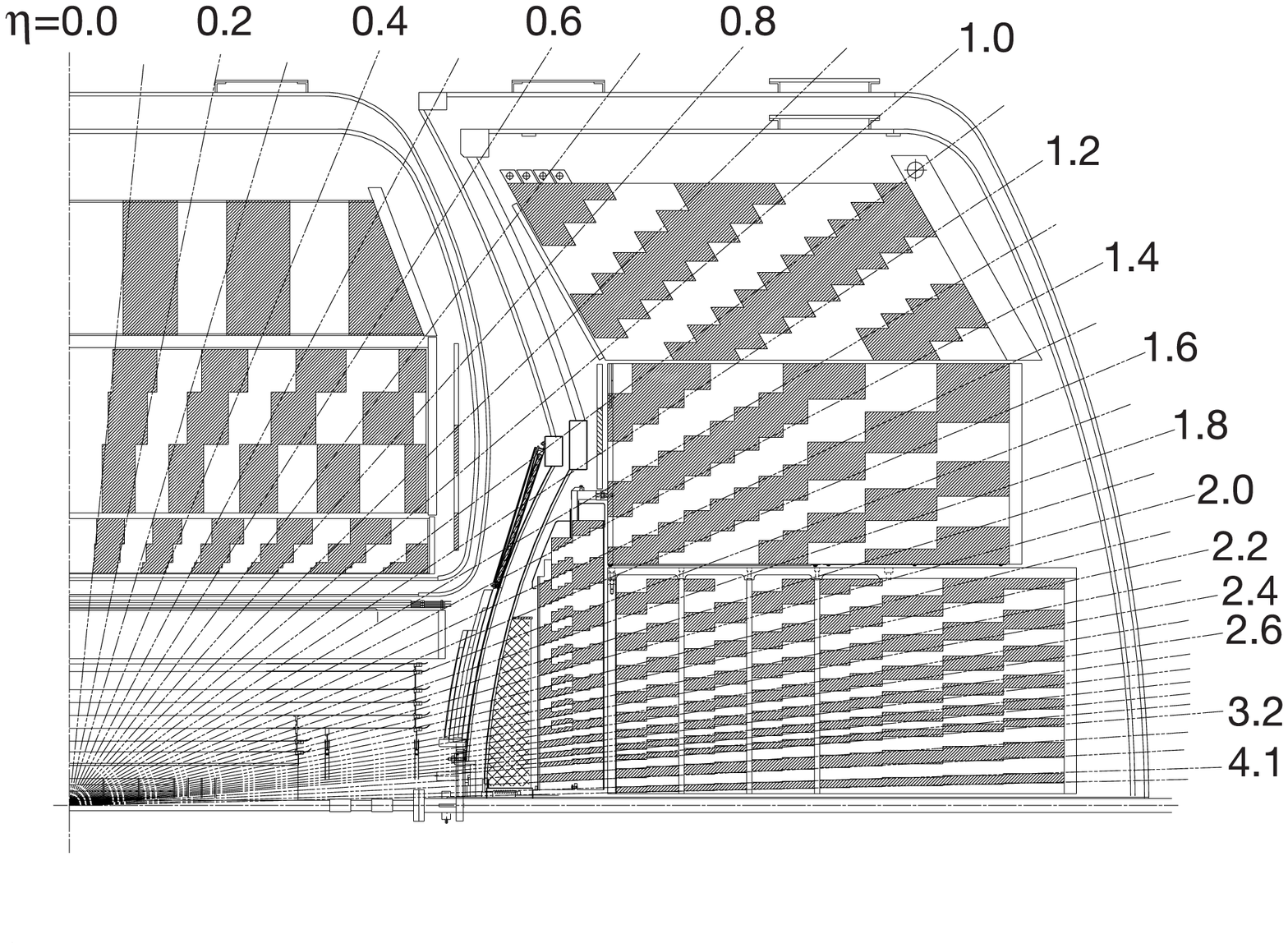}}
	\caption{
		Side view of a quadrant of the \DZero{} calorimeters showing the 
		transverse and longitudinal segmentation. The alternating shading pattern 
		indicates the cells for signal readout. The lines indicate the pseudorapidity 
		intervals defined from the center of the detector.  
		The inter-cryostat detector (ICD) is visible as a thin dark shaded tile 
		between the cryostats, within $1.1< |\deta| <1.4$.
	}
	\label{fig:projective}
\end{figure*}

Each layer of the calorimeter is segmented into 64 sectors in azimuthal angle
$\azim$ and in segments of $\Delta\eta = 0.1$. 
The segmentation is
thus about $\Delta\eta \times \Delta\azim = 0.1 \times 0.1$.
To allow for more precise location of electromagnetic showers,
the segmentation is doubled to $\Delta\eta \times \Delta\azim\approx 0.05 \times 0.05$ in 
the third layer of the EM calorimeter.  For $\mdeta>3.2$, the segmentation becomes
0.2 or more in both $\eta$ and $\azim$.  The segmentation in each layer is 
arranged so as to construct towers
that project back to the center of the interaction region,
as shown in \cref{fig:projective}.

Each tower can be identified with two indices (\ieta, \iphi) that reflect the
projective nature of the segmentation. 
For example, the region $\mdeta < 3.2$ is segmented, as shown in \cref{fig:projective},
in 64 towers labeled with \ieta running from $-32$ to $32$, with a tower with most forward 
boundary at $\eta$ numbered as $\ieta = 10 \eta$.
The 64 segments in $\azim$ are numbered with \iphi running from $1$ to $64$.

As in any sampling calorimeter, incoming particles shower in absorber layers
and ionize the active material (liquid argon). The electrons from the ionization are 
collected on anodes formed of carbon-coated epoxy films on G-10 substrates%
\footnote{The resistivity of this film is required to be not less than 
\MOhmSq{50} at room temperature. Some earlier documents
incorrectly reported the requirement as \MOhmSq{40}.}.
Typical surface resistance when
cold\footnote{Typical temperature is \K{90} at a pressure of \Quantity{p.s.i.a.}{20}} is \MOhmSq{150-180}.  
The voltage across the typical \mm{2.3} liquid argon gap is \kV{\sim 2.0}.
At this field, the electron drift velocity in
argon varies by about $0.3\%$ for a $1\%$ variation in electric field
strength.  Since the charge collection time of \ns{\approx 450} is larger than
the Tevatron bunch crossing time, only about two thirds of the electrons produced
in the gap are used for charge measurement.  As a result, changes
in electric field in the gaps create a change in the detector response.

During the decade of \II{} data collection, ``dark'' currents in the CC both with and
without beam increased. 
The cause is attributed to a layer of uranium oxide on the surface of the
CC absorber plates that is not present on the EC plates. This current
increase is only seen in the CC.  Migrating ions adhere to the 
surface of the oxide, creating a large potential across that material.  A
current through the layer could be caused by these large fields, and its flow
could change the electric properties of the oxide, 
increasing its conductivity \cite{GS,Malter}.
This additional current draw through the resistance of the carbon-coated epoxy film results in a lower
voltage across the argon gaps in the center of the CC than at the edges of the
CC where high voltage connections are made to the resistive film.  The
spatial variation in collected charge (on the scale of 1\%) is corrected by the
offline calibration process. Calibrations had to be performed more
frequently towards the end of \II.

\subsection{Calorimeter calibration}
\label{calibration}

There are a number of steps for the conversion of a collected charge at a
preamplifier into an amount of energy deposited in the
calorimeter.  We describe below the three steps applied concurrently with data taking:
baseline subtraction, zero suppression, and electronics calibration. 

To remove the baseline, the signal corresponding to a sampling occurring 
one bunch crossing earlier (by \ns{396}) is subtracted in
analog circuitry before analog-to-digital conversion of the signal.

Due to residual uranium activity, the pedestal distribution around the
baseline is asymmetric, with a larger tail towards more positive values,
contributing to positive energies.
Between stores, pedestal runs are taken to measure noise levels and to set
zero suppression thresholds for each readout cell at $1.5$ times the RMS of the
cell noise with no beam (\RMSped).  This zero
suppression results in a net positive average cell energy, even in the absence
of a particle flux, which is included in the jet energy scale corrections.

The stability and non-linear behavior of the electronics is measured and
corrected by calibrating pulses at the inputs
to the preamplifiers.  This ``NLC'' calibration was done every two to three
weeks during data taking.  To extend the range of the analog to
digital conversion, there are two gain paths ($\times 1$ and $\times 8$) in
the readout electronics. The NLC runs calibrate both paths.  There is a 
nonlinearity that remains due to saturation for extremely large signals, 
which becomes a significant effect when \GeV{\gtrsim 400} of electromagnetic
energy appears in a single calorimeter tower.  No correction is applied
for this saturation,
but the results of the \GEANT-based \cite{geant} calorimeter 
simulation are modified to describe this effect.

We use an algorithm called ``T42''~\cite{Vlimant:2005ur} to identify 
possible clusters of signal cells while suppressing isolated cells that are
likely to arise from fluctuations in noise.
Cells with an energy less than $2.5\,\RMSped$
are considered to contain only noise and are rejected.  Cells with an energy
between $2.5$ and $4\,\RMSped$ are considered 
only if adjacent to a cell with an energy at least $4\,\RMSped$,
since cells with little energy that are near cells with large
signals are likely to measure the edges of a shower, while such low energy cells,
when isolated, are likely due to noise.  The T42 algorithm
leads to a better rejection of noise cells, and hence improves jet energy 
resolution.


At periodic intervals, typically once per year but more frequently towards the
end of \II, specific data samples are taken and analyzed to provide
a relative response calibration on a cell-by-cell basis uniform throughout the calorimeter.
In each cell, the distribution of deposited energy, taken over all the events,
is exponentially falling.
The response of all the cells with the same \ieta
value is adjusted so that the occupancy above a selected energy threshold is uniform in $\azim$.
Events containing \Zee{} decays were used to remove \ieta response variations
within each module in the electromagnetic layers of the calorimeter
and \dijet{} events were used
for the same purpose for the hadronic layers of the calorimeter. For the
ICD detector, only $\azim$ uniformity
is enforced by this procedure. The absolute response variation of ICD channels
relative to the CC and EC is simulated in the physics analyses.
This procedure corrects not only for the difference in response
from the electronics, but also for the different amount of inactive material in
front of the calorimeter cells, which varies with \ieta and \iphi~\cite{D0WMass2012}.
Finally, the overall scale of the calorimeter calibration
is fixed using \Zee{} decay events and the known $Z$-boson mass \MZ{}~\cite{PDG}.

\begin{figure}[t]
	\begin{center}
		\includegraphics[width=\columnwidth]{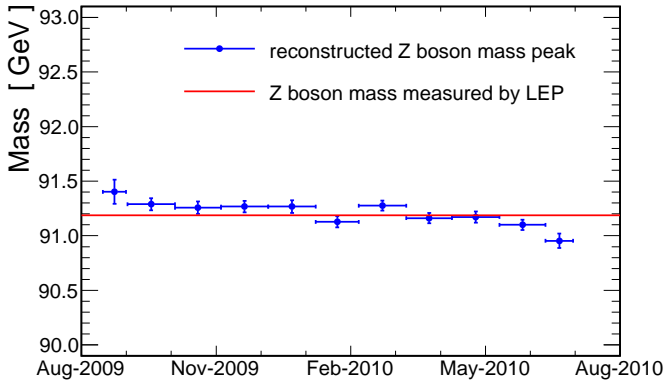}
	\end{center}
	\caption{
		Typical variation with time of the mass of $Z$ boson reconstructed in the
		\fpfm[e] final state for a fraction of the \II{} data
		(reference value from \RefCite{PDG}).
	}
	\label{fig:Zpeak}
\end{figure}

In most cells, the variations between these calibrations are of the order of 1\%.
The stability of the measured value of \MZ{} after 
calibration was typically on the scale of a few hundred \MeV{} or less in the CC, and
somewhat larger in the EC.  \Cref{fig:Zpeak} shows the measured mass
of the $Z$ boson in \fpfm[e] decays for a fraction of \II{} data.  This is the last
fraction of the data for which a year passed between calibrations. After the
summer of 2010, we began to calibrate more frequently because the rate of high voltage 
current increase was increasing with time, leading to a corresponding increase in
the shifting of the observed $Z$ mass peak toward lower masses. 
In addition to energy, $\eta$ and $\phi$ dependences, electron energy is also
corrected as a function of instantaneous luminosity.

%% file: objectid.tex
\section[Reconstruction and identification of objects]{Reconstruction and identification of objects}
\label{sec:objectid}

This section describes the procedures used to reconstruct and identify
the basic objects used 
to calibrate the jet energies.


\subsection{Primary vertex}
\label{pvsel}

The first step is the reconstruction of vertices using prompt tracks.
These vertices, referred to as ``primary
vertices'' (PV), correspond to the locations of inelastic \ppbar{} collisions.
A significant fraction of the PVs are at $z$ positions considerably
displaced from the center of the detector, and it is important to reconstruct
the vertices with high efficiency and accuracy.

The reconstruction of vertices involves three steps: track
selection, vertex fitting, and vertex selection.
Tracks are selected with $\pT \geq \GeVc{0.5}$, with at least two SMT hits,
and transverse impact parameter (with respect to the beam axis) smaller than
three times its uncertainty \cite{bID_nim}.
Starting from the track with highest \pT, the tracks are clustered 
based on the $z$ position of  their closest approack to the beam axis. 
Tracks are added to the cluster if they are 
within \cm{2} in $z$ from the $z$ position of the seed track. 
By constraining all tracks in a cluster to a common
vertex, the track parameters and vertex position are recalculated using
a Kalman Filter technique~\cite{bID_nim,Kalman}. The algorithm is repeatedly 
applied to the remaining tracks to build a list of vertex candidates.

The presence of multiple \ppbar{} interactions during the bunch crossing
typically leads to the reconstruction of several vertices in the event.
For each reconstructed vertex, the probability that it originates from a
soft \ppbar{} inelastic interaction (``minimum bias probability'') is
computed from the tracks associated with the vertex,
making use of a template of the distribution of $\log_{10}(\pT)$.
The vertex with the lowest minimum bias probability is chosen as the hard-scatter PV. 
To ensure that a hard-scatter vertex of high quality is selected, it is required
to be reconstructed from at least three tracks, and
to be located at $\mzpv \le \cm{50}$.

\subsection{Calorimeter objects}

The jet energy calibration procedure relies on calorimeter
objects (photons, jets, and missing transverse energy), which are
reconstructed starting from the individual calorimeter cells. 
This section presents a discussion of the reconstruction and 
identification algorithms used for the relevant calorimeter objects.

\subsubsection{Electromagnetic clusters}
\label{sssec:ElectromagneticClusters}

EM clusters are formed from the towers in
electromagnetic calorimeter which have $\pT > \MeVc{500}$ (``seed towers'')
starting from the highest \pT tower.
Neighboring towers are added if they have $\pT > \MeVc{50}$ and if they are within
$\deltar = \sqrt{(\Delta\eta)^2+(\Delta\phi)^2} < 0.3$ of the seed tower in the CC,
or within a cone of radius \cm{10} in the third layer of the
EM calorimeter in the EC. Such preclusters are used as starting
points for the final clusters if their energy exceeds \GeV{1}. Any EM
tower within $\deltar<0.4$ is added, and the center of the final
cluster is defined by the energy-weighted mean of its cells in the
third layer of the EM calorimeter.

\subsubsection{Jets}
\label{objectid_jets}
Jets resulting from the hard \ppbar scatter usually involve a large number of
particles that deposit energy in numerous calorimeter cells. 
The reconstruction of jets, either from stable particles or calorimeter
towers, involves a clustering algorithm to assign particles or calorimeter 
towers to jets. We define jets using the
Run~II Midpoint cone algorithm~\cite{Blazey:2000qt}, which is a fixed-cone algorithm. 
The jet centroid is defined as $(y_{\textnormal{jet}},\azim_{\textnormal{jet}})$ \cite{d0_coordinate}, and objects are clustered if
their distance relative to the jet axis, 
$\deltar = \sqrt{(y_{\textnormal{obj}}-y_{\textnormal{jet}})^2+(\azim_{\textnormal{obj}}-\azim_{\textnormal{jet}})^2}<\rcone$,
where $\rcone$ is the cone radius. Jet energy scale corrections and uncertainties have been determined 
for $\rcone=0.5$ and $0.7$.

The reconstruction of jets in the detector involves a number of steps.
First, pseudo-projective calorimeter towers 
are reconstructed by adding the four-momenta of their associated calorimeter cells that are 
above threshold, treating each cell four-momentum as massless. The momentum of each cell is defined with 
respect to the PV, as reconstructed by the tracking system. As a result,
calorimeter towers are treated as massive objects. In a second step, the
calorimeter towers with $\pT \geq \GeVc{1}$ are used as seeds to find pre-clusters, which are 
formed by adding neighboring towers within $\deltar<0.3$ with respect to the seed tower.
The pre-clustering step reduces the number of seeds passed to the main algorithm,
keeping the analysis computationally feasible.
A cone of radius $\rcone$ is formed around each pre-cluster, centered at its centroid,
and a new proto-jet center is computed using the $E$-scheme:
\begin{equation}
\label{Escheme}
\begin{split}
   p^{\mu} &= \left(E, \vec{p} \right) = \sum_{i} \left(E_{i}, \vec{p}_{i} \right), \\
         y &= \frac{1}{2} \ln \left( \frac{E+p_{z}}{E-p_{z}} \right), \\
     \azim &= \tan^{-1} \left( \frac{p_{y}}{p_{x}} \right), \\
       \pT &= \sqrt{p_{x}^{2} + p_{y}^{2}}, 
\end{split}
\end{equation}
\noindent where the sums are over all towers (or, in MC, particles or partons) contained in the cone. 
The proto-jet center is modified to the location ($y, \phi$).
The direction of the resulting four-vector is used as the center point for a new cone. 
When the proto-jet four-momentum does not
coincide with the cone axis, the procedure is repeated
using the new axis as the center point until a stable
solution is found. The maximum number of iterations
is 50 and the solution is considered to be stable if the
difference in $\deltar$ between two iterations is smaller than
0.001. In the rare cases of bistable solutions, the last iteration
is retained. Any proto-jets falling below the threshold of
$\pT < \GeVc{3}$ are discarded.

The presence of a threshold requirement on the cluster seeds introduces a dependency on infrared and collinear
radiation. The sensitivity to soft radiation is reduced by the addition of \pT-weighted midpoints between pairs of 
proto-jets and repeating the iterative procedure for these midpoint seeds.
The last step of the algorithm involves splitting and merging to treat overlapping proto-jets, \ie{} proto-jets
separated by a distance of $\deltar < 2\rcone$. Overlapping proto-jets are merged into a single jet
if more than $50\%$ of the \pT{} of the lower-energy jet is contained in the overlap region. Otherwise, the
energy of each cell in the overlap region is assigned to the nearest jet. Finally, the four-momentum of the jet is
recomputed using the $E$-scheme and jets with $\pT < \GeVc{6}$ are discarded.

The jet algorithm described above can also be applied to stable particles
in MC events.
Stable particles are defined as those reaching the \DZero{} detector volume.
All stable particles produced in the interaction
are considered, including not only the ones from the hard scattering process, but also from
the underlying event. The exceptions are muons and neutrinos that are not included. 
Jets clustered from the list of considered stable particles (particle jets) 
are used to define the particle level jet energy. The goal of the jet energy scale calibration 
procedure is to correct the measured energy of
calorimeter jets to the particle level.

Small modifications in the jet-finding algorithm (pre-cluster selection and merging/splitting treatment)
are applied to
\IIb data to meet conditions with higher instantaneous luminosity. 

\subsubsection{Missing transverse energy}
\label{met_description}

The missing energy in the transverse plane $\MET[cal]$ is defined by its
components in $x$ and $y$ projections:
\begin{displaymath}
	\METx[cal] = -p_x^\textnormal{vis} \quad \text{and} \quad \METy[cal] = -p_y^\textnormal{vis},
\end{displaymath}
where $p_{x/y}^\textnormal{vis}$ are the components of the visible transverse momentum, computed
from all the calorimeter cells that pass the T42 selection:
\begin{displaymath}
	p_{x/y}^\textnormal{vis} = \sum_{i \in \textnormal{cells}} p_{x/y,i}.
\end{displaymath}
For the measurements presented in this article, CH cells are excluded from \MET[cal]
due to their limited energy resolution.

The \MET{} is adjusted for energy scale corrections
that are applied to reconstructed electromagnetic objects. 
The corrections of electromagnetic objects that pass the photon identification
criteria described in \cref{photonid} are subtracted:
\begin{equation}
	\Missing{E}_{x/y} = \METxy[cal] - \sum_{i\in\textnormal{photons}} 
\parentheses{E_{x/y,i}^{\gamma,\textnormal{corr}} - E_{x/y,i}^{\gamma,\textnormal{uncorr}} }.
	\label{eq:METcorrEM}
\end{equation}

\subsection{Photon identification criteria}
\label{photonid}
A cluster in the electromagnetic calorimeter is identified as a photon
if it satisfies the following criteria:

\begin{itemize}
\item The object is an isolated electromagnetic cluster.
\item The object is reconstructed in the central region
  ($\mdeta < 1.0$) 
  and in the fiducial regions of the calorimeter (objects near module
  boundaries are excluded).
\item The fraction of energy deposited in the electromagnetic part of
  the calorimeter ($f_{\rm EM}$) must be greater than $0.96$.
\item The probability to have a spatially matched track must be less than $0.1\%$.
\item The cluster is isolated in the calorimeter in a cone of radius $\deltar = 0.4$ by 
$\sqbrackets{E_{\textnormal{tot}}(0.4)-E_{\textnormal{EM}}(0.2)}/E_{\textnormal{EM}}(0.2)<0.07$,
where $E_{\textnormal{tot}}(0.4)$ $[E_{\textnormal{EM}}(0.2)]$ is the total [EM only] energy in the cone of $\deltar = 0.4$ [$0.2$].
\item The scalar sum of the \pT of all tracks originating from the \ppbar hard-scatter vertex 
in an annulus of $0.05 < \deltar < 0.7$ around the EM cluster must be less than \GeVc{1.0}.
Tracks are considered if their transverse momentum
  exceeds \GeVc{0.4} and if their distance of closest approach in $z$
  to the vertex is less than \cm{1}.
\item The square of the energy-weighted cluster width in
  $\radius \times \azim$ in the third layer
  of the EM calorimeter must be less than \sqcm{14}.
%
 \item The weighted sum of energy depositions in the CPS strips around
  the line connecting the PV and EM cluster must correspond to a single EM object.


\end{itemize}

This set of criteria is further referred to as a ``tight photon selection'',
and an object satisfying these criteria as a ``tight photon''.
For the purpose of background studies,
namely the measurement of contamination from dijet events where one of the jets
is misidentified as a photon, two additional slections with less stringent criteria
are considered.
A loose photon selection follows the same criteria, but the requirement on
the scalar sum of transverse momenta of associated tracks is removed,
and no information from the preshower detector is used.
A medium selection is also based on the tight one, but the cut on the scalar 
sum of transverse momenta of associated tracks is relaxed to \GeVc{2}
and the outer radius of the annulus is reduced to 0.4.

\subsection{Jet identification criteria}
\label{jetid}
Jets reconstructed in the calorimeter must satisfy 
the following selection criteria:

\begin{itemize}
\item The fraction $f_{\rm EM}$ must be greater than $0.05$ and less than $0.95$. Jets
  in the forward region ($\mdeta > 2.5$) must satisfy $f_{\rm EM}> 0.04$.
  This requirement is not enforced on jets in ICR.
\item The fraction of energy in the coarse hadronic calorimeter ($f_{\rm CH}$)
  must be less than $0.44$ for jets with $\mdeta < 0.8$, less than $0.46$ for
  jets in the endcap region $1.5 < \mdeta < 2.5$, and less than 0.4
  for all other jets. Exception is the jets in the region $0.85 < \mdeta < 1.25$,
  which are allowed to have $f_{\rm CH}<0.6$, if at the same time the number of cells
  that contain $90\%$ of the jet energy is less than 20. The requirement on $f_{\rm CH}$ is
  aimed at removing jets dominated by noise originating in the coarse
  hadronic part of the calorimeter.
\item The jet must be ``confirmed'' by the independent readout of
  calorimeter energies in the Level~1 trigger, \ie, the energy
  of the trigger towers inside a cone of $\deltar = 0.5$ around the jet axis
  must be at least 50\% of the energy of the jet as reconstructed by the
  precision readout. This condition, progressively loosened to 10\%
  for forward jets ($\deta > 1.4$) and soft jets ($\pT < \GeVc{15}$),
  suppresses spurious jets due to calorimeter readout noise.

\end{itemize}


%% file: overview.tex
\section{Overview of jet energy scale determination}
\label{sec:overview}

The evolution from the colored parton to a jet of hadrons is dominated by low energy processes that
are not calculable perturbatively by Quantum Chromodynamics (QCD),
and lead to large variations in the composition of a jet.  The energy calibration of a jet is 
fundamentally different than for any other object in particle physics, since it does not 
correspond to a single well-defined particle such as an electron or a muon. 
The measured energy of a jet is not fully correlated to energy of its progenitor
parton due to two effects: the parton-to-hadron fragmentation that leads to the
creation of the jet, and the interaction of the final state hadrons with the detector.  

The goal of the jet energy scale correction is to relate, on average, the
jet energy measured in the detector to the energy of the final 
state particle jet. Because jets are composite objects, the algorithm used to
reconstruct them defines the particle jet to which we calibrate.  We employ
a calibration methodology related, to but modified from, \RefCite{JES_run1}.
The particle jet energy \Eptcl can be related to
the measured energy \Emeas of the reconstructed jet via:
\begin{equation}
\Eptcl = \frac{\Emeas-E_{\textnormal{O}}}{R \cdot S} ,
\label{master1}
\end{equation}
\noindent where:
\begin{itemize}
\item $E_{\textnormal{O}}$ represents an offset energy, which includes several contributions.
  Noise arises from
  electronics and radioactive decay of the uranium absorber.  Additional in-time
  \ppbar{} interactions and those from previous crossings, termed ``pile-up'', also
  contribute.
  The underlying event, defined as the energy contributed by the proton
  and antiproton constituents not participating directly 
  in the hard interaction (``spectators''), is considered to be part of the high-\pT event and
  therefore not subtracted. The offset energy depends on the jet cone radius
  ($\rcone$), jet detector pseudorapidity (\deta), number of 
  reconstructed primary vertices ($\npv$), and instantaneous luminosity (\InstLumin).
  \extend{The definition of \eta is: ``one tenth of the \ET-weighted jet 
  position in the calorimeter''. In practice: $0.1\times \texttt{TMBJet:detEta()}$).}
\item $R$ represents the response of the calorimeter to the energy of
  the particles comprising the jets.  Its value is generally
  smaller than unity, primarily because response to hadrons, particularly to
  charged pions, 
  is lower than response to electrons, that is set to unity by the calibration (\cref{calibration}).
  The ratio of responses, $e/\pi$, has a significant dependence on particle energy.
  Significant energy is also lost in non-sampled material before the 
  calorimeter, and in non-instrumented regions between calorimeter modules.
  The ICR region is poorly sampled and this leads to energy scale
  variations. 
  For these reasons, the jet response is a function of jet energy and,
  particularly in the ICR region, of \deta.  A small but 
  non-negligible variation occurs for different cone algorithms, since
  particles near the jet core tend to have higher energy 
  and thus higher response than particles near the jet boundary. 
\item The function $S$ represents corrections for the showering of particles
  in the detector.  Due to the cone size $\rcone$ and the calorimeter cell
  size, energy from particles originating within a jet can spread to cells outside
  the cone radius.  This is not to be confused with parton showering during 
  fragmentation, that is a process occurring prior to interaction with the
  detector.  Conversely, energy may be deposited in cells inside this boundary
  that originated from particles that do not belong to the particle jet
  (\eg, due to showering effects in the calorimeter, or to the
  magnetic field changing the direction of particles outside of the jet cone).
  Typically, the net correction is close to unity. It depends strongly on $\rcone$ and
  \deta, and only mildly on jet energy.
\end{itemize}

We refer to the terms on the right-hand side of \cref{master1} as 
\textit{true} values of the corrections.
In practice, the $E_{\textnormal{O}}$, $R$ and $S$ 
corrections that we measure represent only estimators
of the true corrections and may be affected by a number of biases.  
We explicitly correct for these biases to ensure that 
the mean particle jet energy is recovered.

\subsection{True corrections}
\label{truecorrections}

We first examine the definition of the true corrections discussed in the previous
section. 
Here, we assume that no multiple interactions or pile-up are present, and
only the hard \ppbar interaction produces the jet.
The particle jet energy is defined as the sum of energies of all stable particles 
belonging to the particle jet as defined in \cref{objectid_jets}:
\begin{equation}
\Eptcl = \sum_{i\in \textnormal{ptcljet}} E_i.
\label{eptclj}
\end{equation}
\noindent The measured jet energy receives contributions from particles both inside and outside
the particle jet, as well as the offset energy:
\begin{equation}
\Emeas = \sum_{i\in \textnormal{ptcljet}} \Emeas[$i$] f_{i} + 
              \sum_{i\notin \textnormal{ptcljet}} \Emeas[$i$] f_{i} + E_{\textnormal{O}},
\end{equation}
\noindent where \Emeas[$i$] is the visible calorimeter energy from particle $i$, and $f_{i}$ is the
fraction of such energy contained within the calorimeter jet cone.
%
After subtracting the \emph{true offset energy} $E_\textnormal{O}$, 
we obtain, by definition, the energy inside the calorimeter jet cone in the absence
of any noise, pile-up or multiple interaction effects.

We define the \emph{true response} correction to be the ratio of visible energy
for particles from the particle jet divided by the energy of the incident particle jet (given by \cref{eptclj}):
\begin{equation}
R = \frac{\sum_{i\in \textnormal{ptcljet}} \Emeas[$i$]}{\Eptcl}.
\label{rtrue}
\end{equation}
\noindent This definition includes all the constituents
of the particle jet, regardless of whether their energy is deposited within the 
cone radius of the reconstructed jet.

To satisfy \cref{master1}, the \emph{true showering} correction is 
defined as:
\begin{equation}
S = \frac{\sum_{i\in \textnormal{ptcljet}} \Emeas[$i$] f_{i} + 
                 \sum_{i\notin \textnormal{ptcljet}} \Emeas[$i$] f_{i}}{\sum_{i\in \textnormal{ptcljet}} \Emeas[$i$]}.
\label{strue}
\end{equation}
\noindent 
This represents a correction from the visible energy inside the
calorimeter jet cone, resulting from particles both inside and outside the
particle jet, to the total visible energy resulting from the particle jet,
whose cone may differ from the one of the calorimeter jet.

\subsection{Estimated corrections} 
\label{estimated_corrections}

The jet offset, response and showering corrections
can be estimated in data, and are represented by $\hat{E}_{\textnormal{O}}$, $\hat{R}$
and $\hat{S}$. 
Ideally, the corrected jet energy would be given by
\cref{master1}, with the true corrections replaced by the \emph{estimated corrections}:
\begin{equation}
{\hat E}^{\textnormal{corr}} = \frac{\Emeas-\hat{E}_{\textnormal{O}}}{\hat{R}\cdot\hat{S}}.
\label{master2}
\end{equation}
\noindent Since the estimated corrections suffer from biases, the
corrected jet energy as given by \cref{master2} can differ by several
percent from \Eptcl. We therefore determine additional corrections 
using Monte Carlo (MC) samples to 
remove biases of the estimated corrections.
The final jet energy correction is given by the modified expression:
\begin{equation}
\Ecorr = \frac{(\Emeas-\hat{E}_{\textnormal{O}})\kO}{\hat{R}\;\kR\,\hat{S}},
\label{master3}
\end{equation}
\noindent where \kO{} and \kR{} represent the bias corrections to offset and
response, respectively. As will be discussed in \cref{sec:showering}, $\hat{S}$ is 
\emph{a~priori} an unbiased estimator of the true showering correction, and no bias correction 
is required.
After these corrections, \cref{master3} provides, on average, the unbiased energy of
the particle jet.

\subsection{Biases from the sample composition}
\label{ssec:SampleDependencyIntro}

The corrections to data and MC simulation are extracted independently, although the procedure is similar.
All corrections are determined on average in the sense that they are parametrized on only a few characteristic
properties of the jet.

Jets have different characteristics according to whether they originate from a light quark, 
$c$ quark, $b$ quark, or a gluon
(the ``parton flavor'' of the jet).
The jet energy scale correction outlined above 
considers a mixture of jets with parton flavors
produced by the physical process used in the calibration, namely \photonjet{}.
This correction calibrates samples composed of a mixture 
of jets with parton flavor content similar to \photonjet production processes.
In samples with different composition, this correction will 
generally have a bias depending on the partonic content of the sample.

The method described in this article uses both \photonjet and \dijet events.
This yields the extraction of two energy scales, appropriate for the analysis
of data samples with composition similar to the \photonjet and \dijet processes,
respectively.

Even without knowledge of the precise composition of the data sample,
it is still possible to perform measurements with the available energy scale by comparing
with the MC, provided that the simulation 
describes the features and biases of jets
with different parton flavor, \ie{}, their calorimeter response.
\Cref{sec:fdc} describes an improvement to this description
based on the calibration of the simulated response to single particles inside
a jet using data, and \cref{sec:jssr} presents a further correction to
improve the description of simulated jets.

%% file: response_overview.tex
\section{Overview of response correction}
\label{sec:response_overview}

\newcommand{\qlpar}[1][]{\JESmathSym{c_{#1}}}

The response correction ($R$) is numerically the largest correction in 
the jet energy scale calibration procedure, since it accounts for
a number of sizable instrumental effects that influence the 
jet energy measurement. First, particles emerging from the hard 
scattering interact with the material before the calorimeter
and lose a fraction of their energy, which can be significant
for low momentum particles. Furthermore, charged particles
are deflected in the magnetic field and, depending on their \pT,
can potentially fail to reach the calorimeter 
(\eg, charged particles in the central rapidity region with $\pT < \GeVc{0.3}$).
Most particles reaching the calorimeter (except for muons and neutrinos,
which constitute, on average, a small fraction of the jet energy) are fully 
absorbed and their deposited energy is transformed into a signal. 

The \DZero{} calorimeter is non-compensating: it has
a higher and more linear response to electromagnetic particles 
($e^\pm,\photon$) than to hadrons ($e/h>1$). The energy dependence of the
response to hadrons is nearly logarithmic as a result of 
the slow rise of the fraction of $\pi^0$ mesons produced as a function of the
hadron energy during hadronic shower development 
\cite{wigmans}. Zero suppression can also significantly
contribute to the non-linearity of response to hadrons, especially
at low jet momentum. Finally, calorimeter module-to-module inhomogeneities or
poorly instrumented regions (\eg, the ICR) can result in
significant distortions to the measured jet energy.

Some of these instrumental effects (\eg, the calorimeter response to
hadrons) are difficult to model accurately
in the MC simulation. As a result, data and MC
have a different response to jets, requiring response corrections to be
determined separately for data and MC. 
While in MC it is \emph{a~priori} possible to compute
the response correction exactly by comparing the measured jet energy
to the particle jet energy, this information is not available in data.
The Missing \ET Projection Fraction (MPF) method~\cite{JES_run1,mpf}
has been developed to measure the calorimeter response to jets in data.
We use this method to measure the jet response in both data
and MC. Applying the MPF method to MC, where the true
jet response is known, allows an evaluation of the biases of the
method and development of suitable correction procedures to be applied to data.
In the next sections we give an overview of the MPF method, followed 
by the discussion of the expected biases and the corresponding corrections. 
Finally, we outline the strategy to determine the jet energy response correction.

\subsection{Missing \ET Projection Fraction method}
\label{mpf}

We consider a two-body process \Xjet, where $X$ (\photon, $Z$ boson, or jet) is 
referred to as the ``tag object'', and the jet is the ``probe object'' whose response 
we are estimating. The MPF method can be used to estimate
the calorimeter response of the probe jet relative to the response of the tag object.
This method is also exploited to intercalibrate the response of different calorimeter
regions. 

At the particle level, the transverse momenta of the tag object
($\MakeTransverseExt{\vec{p}}{}{\,tag}$)
and of the hadronic recoil ($\MakeTransverseExt{\vec{p}}{}{\,recoil}$) are balanced due to 
overall transverse momentum conservation in a given event:
\begin{equation}
\MakeTransverseExt{\vec{p}}{}{\,tag} + \MakeTransverseExt{\vec{p}}{}{\,recoil} = 0.
\label{ideal_cal}
\end{equation}
\noindent The probe jet is part of the hadronic recoil, 
but may not constitute the entire hadronic recoil.
In a calorimeter, the responses of the tag object
($R_\textnormal{tag}$) and the hadronic recoil ($R_\textnormal{recoil}$) might be different,
which results in a transverse momentum imbalance as measured by the calorimeter:
\begin{equation}
\MakeTransverseExt{\vec{p}}{\,\textnormal{meas}}{\,tag} + \MakeTransverseExt{\vec{p}}{\,\textnormal{meas}}{\,recoil} = -\vecmet{},
\label{real_cal}
\end{equation}
\noindent where $\MakeTransverseExt{\vec{p}}{\,\textnormal{meas}}{\,tag} = R_\textnormal{tag}\,\MakeTransverseExt{\vec{p}}{}{\,tag}$ 
is the measured transverse momentum
of the tag object, $\MakeTransverseExt{\vec{p}}{\,\textnormal{meas}}{\,recoil} = R_\textnormal{recoil}\,\MakeTransverseExt{\vec{p}}{}{\,recoil}$
is the measured transverse momentum of the hadronic recoil, and \vecmet is the 
missing transverse energy measured in the event (see \cref{met_description}).

From \cref{ideal_cal,real_cal} we derive the following expression:
\begin{equation}
\frac{R_\textnormal{recoil}}{R_\textnormal{tag}} = 1+\frac{\vecmet{}\cdot \MakeTransverseExt{\vec{n}}{}{\,tag}}{\pTtag[\textnormal{meas}]},
\label{mpf_method}
\end{equation}
\noindent which shows that the response of the hadronic recoil relative to the response of the
tag object can be estimated from the projection of \vecmet onto the direction of the tag object 
in the transverse plane, $\MakeTransverseExt{\vec{n}}{}{\,tag}$.

In the ideal case, where the probe jet is identical to the hadronic recoil, 
we can replace $R_\textnormal{recoil}$ in \cref{mpf_method} by the jet response, $R$. 
However, 
the presence of additional jets in the event, some of which might not even
be reconstructed, make this idealized situation impossible to achieve in practice. 
By requiring exactly two
reconstructed objects (tag and probe) back-to-back in azimuthal angle, 
it is possible to significantly improve the
approximation that $R\approx R_\textnormal{recoil}$.
Residual effects at the percent level remain and are subsequently corrected
(see \cref{sec:mpfbiases}). To avoid confusion with
the true response of the particle jet ($R$), we will refer to the jet response estimated with the 
MPF method as \MPF{sample}, where the superscript will be used to indicate which 
sample has been used to estimate the response. This information is important,
since the MPF response 
depends on the sample used (\eg, via the parton flavor
composition, color flow, etc.). It also depends on the corrections applied 
to the energy of the tag object, which are propagated into \MET.

\subsubsection{Resolution bias}
\label{resolbias_overview}

\cref{mpf_method} attributes the average imbalance in transverse momentum in the event, \MET,
to differences in calorimeter response between the tag and probe objects. 
For a precise determination of this relative response it is important to eliminate all
sources of imbalance that are unrelated to calorimeter response.

In particular, when measuring $R_\textnormal{recoil}/R_\textnormal{tag}$ in bins of \pTtag[\textnormal{meas}], 
there is a possibility of a significant imbalance (\MET) arising purely from resolution effects.
The dominant effect arises from the finite calorimeter energy resolution coupled
with a steeply falling jet \pT spectrum. In this case, each \pTtag[\textnormal{meas}] bin tends 
to contain on average more upward fluctuations from lower
\pT than downward fluctuations from higher \pT. As a result, there is positive bias in the average
\pTtag[\textnormal{meas}] that translates into an artificial source of \MET
imbalance in the event. We refer to this effect as ``resolution bias''.
Because this bias depends on the jet energy resolution, its size also depends on
\deta of the jet.

This bias can be precisely estimated if the tag object \pT spectrum and \pT resolution
are known.
The expected resolution bias in the transverse momentum of the tag photon in
\photonjet events is much smaller than $1\%$ \cite{PRD88_072008_PhotonJet}
and can thus be neglected. In contrast,
the expected resolution bias in the transverse momentum of the tag jet in \dijet events
can be much larger~\cite{IncJets} and needs to be explicitly corrected. To evaluate
this correction a detailed numerical calculation is performed taking into account the measured \pT spectrum 
in data \dijet events as a function of \deta of the probe jet and a precise measurement of the
jet energy resolution for a jet in the calorimeter~\cite{IncJets}. 
This correction procedure has been validated in MC and verified to properly
correct the bias, within an uncertainty of $0.5\%-1.0\%$.

\subsubsection{Absolute MPF response}
\label{absolute_mpf_response}

The absolute MPF jet response is estimated from \cref{mpf_method} using \photonjet events
with a jet in the central calorimeter region ($|\deta|<0.4$),
assuming that the measured transverse momentum of the photon
is converted to the particle level (\pTgamma) by the EM energy scale corrections.
In this case the photon response $R_{\photon} = 1$, and \cref{mpf_method} can be rewritten as:
\begin{equation}
\MPF{\photonjet} = 1 +  \frac{\vecmet\cdot \MakeTransverseExt{\vec{n}}{}{\photon}}{\pTgamma}.
\label{mpf_gj}
\end{equation}
\noindent The most important dependence of the jet response 
is on the jet energy. 
As discussed in \cref{resolbias_overview},
the jet energy resolution causes a bias in the
estimated jet response. 
Therefore, to measure the energy dependence of the jet response with minimal
impact from resolution effects we use the jet energy estimator \Eprime, defined as:
\begin{equation}
\Eprime = \pTgamma \cosh(\eta),
\label{eq_eprime}
\end{equation}
\noindent where $\eta$ is the jet pseudorapidity with respect to
the reconstructed 
$p\bar{p}$ collision vertex in the event.
The estimator \Eprime is calculated using the photon transverse momentum and the jet
direction, which are measured more precisely than the jet energy itself.
It is strongly correlated with the particle level jet energy.
We also use the quantity \pTprime, defined as
\begin{equation}
\pTprime = \Eprime / \cosh(\deta),
\label{eq_ptprime}
\end{equation}
\noindent where \deta is the detector pseudorapidity of the probe jet~\cite{d0_coordinate}.

The energy dependence of the jet response is well described by a quadratic logarithmic
function~\cite{IncJets,JES_run1}:
\begin{equation}
\func[R]{\Eprime} = \qlpar[0] + \qlpar[1] \log(\Eprime/E_0)+ \qlpar[2] \log^2(\Eprime/E_0),
\label{quadlog}
\end{equation}
\noindent where $E_0$ is a constant and \qlpar[i] ($i=0,1,2$) are free parameters to be determined.
%
A detailed description of the measurement of the jet absolute response 
is given in \cref{sec:response}.

\subsubsection[Relative MPF response]{Relative MPF response: pseudorapidity dependence}
\label{sec:relresp}

Even after individual cells are calibrated, the \DZero{} calorimeter exhibits a non-uniform
response to jets as a function of $\deta$. The jet response
is rather uniform within the CC region;
however, in data (MC) the EC response is $\sim 15\%$ $(10\%)$ lower than the CC response. 
An important contribution to this non-uniformity arises from the poorly 
instrumented ICR region ($1.1 < \mdeta < 1.5$). As discussed in \cref{calorimeters}, 
 a substantial amount of energy in this region is lost in the solenoid, cryostat walls, module
end-plates, and support structures. 
As a result, the ICR region has the largest deviation in energy dependence
of response respect to the central calorimeter. 
In the $1.2 < \mdeta < 1.4$ region,
the calorimeter lacks an electromagnetic section and the total depth drops below
$6 \lambda_{\textnormal{int}}$. 
The goal of the relative MPF response correction is to
address this effect in such a way that the corrected MPF
response is uniform for the entire calorimeter, independent of \deta.
Since different calorimeter regions have 
different energy dependence of the response, this correction is not
only a function of \deta, but also of energy. 

To express the dependence on \deta, the relative MPF response correction, $\Feta$,
as defined in detail in \cref{ssec:etadep:method},
is estimated using samples of \photonjet and \dijet events (see ). 
The former sample allows a direct and consistent
derivation of the MPF response relative to the central calorimeter,
with a normalization of $\Feta$ to unity for the central jets ($|\deta|<0.4$).
The \dijet sample provides the additional statistics
required to measure this correction in fine bins of \deta and up to much
higher energies than the \photonjet{} sample can reach. By 
combining these two different samples, we reduce both statistical and
systematic uncertainties of the relative response correction.

The $\Feta$ factors are found from a global fit
to data in all $\deta$ bins simultaneously. 
The \dijet sample has a different composition of quark- and gluon-initiated jets
as compared to \photonjet events (\cref{sec:qcd_specific}).
For this reason, the responses in the two samples
differ, albeit having the same form of energy dependence in a given $\deta$
(see \cref{feta_gj_vs_dijet}). From the global fit, the correction factors
for the \photonjet and \dijet samples, $\Feta[\photonjet]$ and $\Feta[dijet]$,
are available separately. The $\Feta[dijet]$ factors are used later 
to derive jet energy scale corrections for \dijet events, described
in \cref{sec:qcd_specific}.

\subsection{MPF response biases}
\label{mpfbias}

As discussed in the previous section, an estimate of the absolute 
jet response  can be obtained by applying the MPF method to selected \photonjet events
(free from the resolution bias), so that
$\hat{R} = \MPF{\photonjet}$. 
However, the MPF response is not a perfectly unbiased estimator of the true jet response, and
explicit bias corrections are required. These corrections are estimated using
MC simulation which models modest relative changes reliably,
despite the fact that they do not correctly predict the absolute jet response.
The nature of these biases and how the corresponding corrections are determined
is discussed below.

The calorimeter calibration yields an overestimation
      of the photon \pT with respect to the particle-level photon \pT,
      due to the smaller energy loss of photons respect to electrons (\cref{calibration}).
      This miscalibration would result in a negative bias to \MPF{\photonjet}.
      Such bias is prevented by correcting photon energy as described in \cref{photon_corrections}.

The selected \photonjet sample in data suffers from a non-negligible background
      contamination (especially at low \pT) from \dijet events, where one of the jets is misidentified as a photon.
      In these background events there is a hadronic energy around 
      the misidentified photon that is undetected, and thus
      the measured \pT of the photon candidate is too low resulting in a positive bias to \MPF{\photonjet}.
      A correction factor, \kR[\photon], is derived in MC to correct the measured 
      MPF response \MPF{mixture} of the mixture (signal+background) sample to the response \MPF{\photonjet}
      of a pure \photonjet sample with the photon at the particle level:
\begin{equation}
\kR[\photon] = \frac{\MPF{\photonjet}}{\MPF{mixture}}.
\label{eq_kRgamma}
\end{equation}
      \noindent This correction is described in \cref{sssec:response:background_corrections}.

Due to the different effects of zero suppression inside and outside the jet,
      the presence of the offset energy in the event introduces 
      a transverse momentum imbalance in the direction
      opposite to the jet, which results in a positive bias to \MPF{\photonjet}. 
     As we describe in \cref{sec:datasets}, the zero bias (ZB) events from data,
     overlaid to our MC events provide a more realistic simulation of the offset energy observed in data.
     A corresponding correction factor, 
      \kRZS, is determined in the \photonjet MC, by comparing the MPF response 
      (using the particle-level photon) in the MC simulation to a special MC simulation with no ZB overlay 
      (\ie, no offset energy):
\begin{equation}
\kRZS = \frac{\MPF{\photonjet,noZB}}{\MPF{\photonjet}}.
\label{eq_kOffset}
\end{equation}
      The evaluation of this correction is described in detail in \cref{zs_bias_corr}.

Finally, the MPF method provides an estimate of the response to the hadronic recoil against
      the photon, which can differ from the true jet response, especially for
      forward jets. This bias also depends on the topological selection applied to 
      the \photonjet events. A corresponding correction factor, \kR[topo], is 
      determined in \photonjet MC without ZB overlay, and defined as the ratio 
      of the true jet response (given by \cref{rtrue}) to the MPF response (using the particle-level photon):
\begin{equation}
k_\textnormal{R}^\textnormal{topo} = \frac{R}{\MPF{\photonjet,noZB}}.
\label{eq_mpfbias}
\end{equation}
      \noindent This last correction is described in more detail in \cref{topo_bias_corr}.

The total correction to the estimated jet response in \cref{master3} is given by:
\begin{equation}
	\kR = \kR[\photon]\,\kRZS\,\kR[topo].
	\label{eq:resp_oview:ResponseCorrection}
\end{equation}
All the corrections are estimated for both cone algorithms $\rcone=0.7$ and 0.5.

\subsection{Estimation of the true response}
\label{response_strategy}

Here we give a brief outline of the procedure used to estimate the true jet response, which will
be discussed in detail in \cref{sec:response,sec:etadep,sec:mpfbiases}. 
The first step is to estimate the MPF response \MPFCC{\photonjet} for a CC jet in a pure sample of \photonjet events 
with the photon corrected to the particle level.
This is straightforward in the case of MC, since there is no \dijet background
contamination and the MPF response can be computed using the particle-level photon 
event-by-event. For the data, the MPF response \MPFCC{mixture} for the selected 
sample of photon candidate and jet (with the jet in the CC region), is computed, and then corrected for the background contamination 
and by the photon energy scale using \kR[\photon] from \cref{eq_kRgamma}.
The estimated \MPFCC{\photonjet} is then parameterized in both data and MC 
as a function of \Eprime using the functional form given in \cref{quadlog}.
A discussion of this measurement and the related uncertainties is the main topic of
\cref{sec:response}.

In a second step, a correction \Feta[\photonjet] is determined to intercalibrate
the MPF response as function of \deta, \MPFeta{\photonjet}, with respect to the central calorimeter.
This $\eta$-dependent correction is defined by
\begin{equation}
\MPFeta{\photonjet} = \MPFCC{\photonjet}\,\Feta[\photonjet].
\label{ideal_feta}
\end{equation}
By combining selected \photonjet and \dijet events, it is possible to determine \Feta[\photonjet] 
with high resolution over a wide energy and rapidity range. Combining the measurements in \photonjet 
and \dijet events is not trivial, due to differences arising from the diverse parton flavor
composition in the two samples.
In addition, it is necessary to correct for the effect of the \dijet background
contamination in the \photonjet data sample. Using this
data-driven approach instead of relying on MC allows a reduction of the dependency
on physics and detector modeling.
A detailed discussion of the procedure 
is given in \cref{sec:etadep}.

Finally, the true response for a jet with detector pseudorapidity \deta is computed as:
\begin{equation}
R = \MPFCC{\photonjet}\,\Feta[\photonjet]
\,k_{\textnormal{R},\eta}^\textnormal{ZS}\,k_{\textnormal{R},\eta}^\textnormal{topo},
\label{master_resp}
\end{equation}
\noindent where $k_{\textnormal{R},\eta}^\textnormal{ZS}$ and $k_{\textnormal{R},\eta}^\textnormal{topo}$ are the bias correction factors
described above, now with $\eta$ dependence.

%% file: datasets.tex
\section{Data and Monte Carlo samples}
\label{sec:datasets}

This section gives an overview of the data and MC
samples used to determine the jet energy scale corrections.

\subsection{Data samples}
\label{data_samples}

Various data samples are required to determine and validate different components
of the jet energy scale corrections. 

\begin{itemize}
\item \textbf{Minimum bias (MB)}: This sample is collected using 
a trigger that requires only hits in the 
luminosity counters, signaling the
 presence of a \ppbar{} inelastic collision. It is used to measure the contribution from
 multiple \ppbar interactions to the offset energy (\cref{sec:offset}).
\item \textbf{Zero bias (ZB)}: This sample is collected during beam crossings without any 
 trigger requirement. 
 It is used to measure the contribution from
 noise and pile-up to the offset energy (\cref{sec:offset}).
\item $\boldsymbol{\photon}$\textbf{+jet}: This sample is collected using triggers 
 that require an isolated EM cluster, with different transverse momentum thresholds.
 It is used to measure the
 calorimeter response to a jet (\cref{sec:response}), intercalibrate the calorimeter response
 as a function of jet pseudorapidity (\cref{sec:etadep}), determine the showering
 correction (\cref{sec:showering}), and to tune the particle response in simulation 
to data (\cref{sec:fdc}).
\item \textbf{Dijet}: This sample is collected using jet triggers that require at least one jet
 with transverse momentum $\pT > \GeVc{~15,\,25,\,45,\,65,\,95, \text{ or }125}$. It is used
 together with the \photonjet sample described above to intercalibrate the calorimeter response
 as a function of jet pseudorapidity (\cref{sec:etadep}).
\item $\boldsymbol{Z(\to \mu^+\mu^-)}$\textbf{+jet}: This sample is used
 to derive corrections for the relative energy scale shift and 
 resolution effects  for MC jets to better match experimental data (\cref{sec:jssr}).
\end{itemize}

These samples have been extracted from the full \II dataset, which corresponds to
an integrated luminosity of approximately \invfb{9.7}.
Due to 
changes in the detector configuration (\cf{} \cref{ssec:detector:overview}),
instantaneous luminosity, object reconstruction, and trigger selections,
\II{} is split into 5 data taking periods, 
corresponding to an integrated luminosity of \invfb{1.1} for \IIa,
and of 1.2, 3.0, 2.0 and \invfb{2.4} for \IIbOne, 2, 3 and 4, respectively.
Jet energy calibration has been performed separately
for each of these periods.
Data are required to satisfy the quality requirements developed in the \DZero{} Experiment.
Photon and jet selection criteria are described in \cref{sec:objectid}.


\subsection{Monte Carlo samples}
\label{mc_samples}

Since jet energy scale corrections are 
determined for MC separately, the following samples have been used:

\begin{itemize}
\item $\boldsymbol{\photon}$\textbf{+jet}: This sample includes the $2\to 2$ direct photon production processes 
  $qg\to \photon q$ and $\qqbar \to \photon g$ simulated using \PYTHIA~\cite{pythia} 
  with \textsc{cteq6l1}~\cite{cteq6l} parton distribution functions (PDFs)
  and with the transverse momentum of the outgoing partons 
  ranging from \GeVc{5\text{ to }980}.
\item \textbf{Dijet}:  This sample includes the inclusive parton processes 
  used for modeling the inclusive jet production
  (\eg, $gg\to gg, qq\to qq, qg\to qg, gg\to q\bar{q}$, \etc) and is simulated with \PYTHIA.
\item $\boldsymbol{\photon}$\textbf{-like jets}: This sample includes the same inclusive \dijet processes as above, 
  with a specific selection applied at the particle level in order to enrich the sample
  with jets having a photon-like signature due to fluctuations in jet fragmentation~\cite{IncGam}. 
  This sample is mainly used to study and correct for the contamination from the \dijet 
  background in data.
 \item $\boldsymbol{Z(\to \mu^+\mu^-)}$\textbf{+jet}: This sample has been simulated
  by the \ALPGENPYTHIA{} MC~\cite{alpgen} with a matrix element allowing 
  real emissions of up to five light partons.
%
%
\end{itemize}

\PYTHIA{} is used to compute the leading-order matrix elements for each of
 the above samples except \Zjet, and to simulate the underlying event, which includes
 the contribution from beam remnants and additional parton interactions. Only phenomenological
 models exist for these processes. We use the ``\PYTHIA{} tune A'' model~\cite{tuneA}, which has been optimized
 to describe CDF data~\cite{jetshapes}. Fragmentation, hadronization and particle decays are also
 handled by \PYTHIA.
 Comparisons to other \PYTHIA{} tunes are described in sections devoted to corrections
 for the topology bias (\cref{topo_bias_corr}) and showering effects (\cref{sec:showering}).

 Generated events are processed through the \GEANT-based~\cite{geant} simulation
 of the \DZero \II detector. To achieve a more realistic simulation of noise, pile-up,
 and additional \ppbar interactions, the digitized signals from ZB data events are overlaid
 on the simulated MC processes.
 The default MC production at \DZero uses overlaid ZB events with the symmetric $1.5\,\RMSped$
 zero-suppression (\cref{calibration}) applied at the calorimeter cell level (``suppressed ZB overlay'').
 To study the impact of this selection, additional \photonjet and \dijet samples 
 have been generated without ZB overlay (``no ZB overlay''), as well as with ZB overlay from data without
 the $1.5\,\RMSped$ zero-suppression requirement (``unsuppressed ZB overlay'').
 Finally, the events are processed through the same reconstruction program as for collider data.

 \extend{Unfortunately, the ZB samples used for MC overlay do not span the full run range 
 in data, but are typically restricted to relatively short data taking periods. Furthermore,
 the overlay sample not always contain a luminosity spectrum representative of the full
 Run~IIa dataset. To illustrate these points, \cref{datasets_plots} in Appendix~\ref{datasets_appendix} 
 presents a comparison of the distributions of run number, instantaneous luminosity
 and primary vertex multiplicity for selected $\gamma$+jet events 
 (following \cref{photonjet_selection}, except for the cut on $\npv$, which
 is relaxed to $\npv\geq 1$) in data, MC with unsuppressed ZB overlay and MC with
 suppressed ZB overlay. As it can be appreciated, the unsuppressed ZB overlay sample, which
 is used to derive some of the corrections for the data JES, provides a sufficiently
 realistic description of the luminosity profile in data.}



%% file: offset.tex
\section{Offset correction}
\label{sec:offset}


The goal of the offset correction is to subtract the energy not associated
with the $p\bar{p}$ collision producing the high-$\pt$ interaction. 
Hence, the energy included in a jet that originates from soft interactions 
involving the spectator partons constituting the colliding proton and antiproton 
(underlying event) 
is not subtracted.
The excess energy to be subtracted includes contributions from electronic noise, 
pile-up, and additional \ppbar collisions (multiple interactions) within the same bunch crossing.


The shaping time of the calorimeter preamplifier is longer than the time between bunch crossings
(\ns{396}). It is therefore possible that the signal 
may be on top 
of energy from a previous bunch crossing, resulting in an overestimation of the energy.
This effect is called pile-up and it depends on the instantaneous luminosity of the
previous bunch crossings, as well as the location of the present bunch crossing
with respect to the beginning of the bunch train.

A hard-scatter event with multiple interactions can be modeled as the superposition of one hard
parton scattering and one ZB event at the same instantaneous luminosity. 
The number of additional \ppbar inelastic interactions 
in the ZB event follows a Poisson distribution with average given by
$\sigma_\textnormal{inel} \InstLumin_\textnormal{bunch}$, where $\sigma_\textnormal{inel}$ is the 
total \ppbar inelastic  cross section and $\InstLumin_\textnormal{bunch}$ is the
luminosity of the colliding bunches~\cite{LumiNIM}.

The energy contribution from noise, pile-up, and multiple interactions is estimated 
using ZB and MB data samples, which are described in the next 
section. However, this estimate can differ substantially from the true offset energy
(\cref{truecorrections}), due to the different impact of zero suppression 
inside the jet as compared to the ZB and MB data samples. 
Corrections for this effect, estimated in MC to be 1--5\%, are described in \cref{offset_bias_correction}.

\subsection{Sample selection}

The components of the offset energy from noise, pile-up and multiple interactions,
are estimated using samples of MB and ZB events (\cref{mc_samples}).
%
The MB sample is dominated by soft interactions and is used to estimate
the contribution from multiple \ppbar interactions to the offset energy.
%
The ZB events represent a truly
unbiased measurement of the energy in the calorimeter regardless of the nature of
the \ppbar interaction.
This sample, depleted of multiple
interactions by rejecting events with hits on both sides of the luminosity detector
(LD veto) and with reconstructed \ppbar collision vertices,
is then used to estimate the contribution from noise and pile-up to the offset energy.

%



\subsection{Method}

The average offset energy, $\hat{E}_\textnormal{O}^\textnormal{ring}$, 
is estimated for each calorimeter ring in \ieta
(summing over all towers in \iphi), and as a function of the number of reconstructed
\ppbar collision vertices, \npv, and instantaneous luminosity \InstLumin
by adding the estimated contributions from noise and pile-up (NP),
$\hat{E}_\textnormal{NP}^\textnormal{ring}$,
and multiple interactions (MI),
$\hat{E}_\textnormal{MI}^\textnormal{ring}$: 
\begin{equation}
\hat{E}_\textnormal{O}^\textnormal{ring}(\ieta,\npv,\InstLumin) =
\hat{E}_\textnormal{NP}^\textnormal{ring}(\ieta,\InstLumin)+
\hat{E}_\textnormal{MI}^\textnormal{ring}(\ieta,\npv,\InstLumin). 
\label{eoring}
\end{equation}
\noindent The NP contribution is expected to depend on \InstLumin via the pile-up component. 
The contribution from multiple \ppbar interactions depends mainly on \npv,
assuming that every additional interaction contributes a reconstructed
vertex in the event. It is also parameterized as a function of
$\InstLumin$ in order to take into account a possible luminosity dependence of the primary
vertex reconstruction efficiency.
To maximize the efficiency to 
identify multiple interactions, no requirement is applied on the number of tracks 
in an event nor on the location of the vertices.

\subsubsection{Noise and pile-up}
\begin{figure}[t]
	\centerline{\includegraphics[width=\columnwidth]{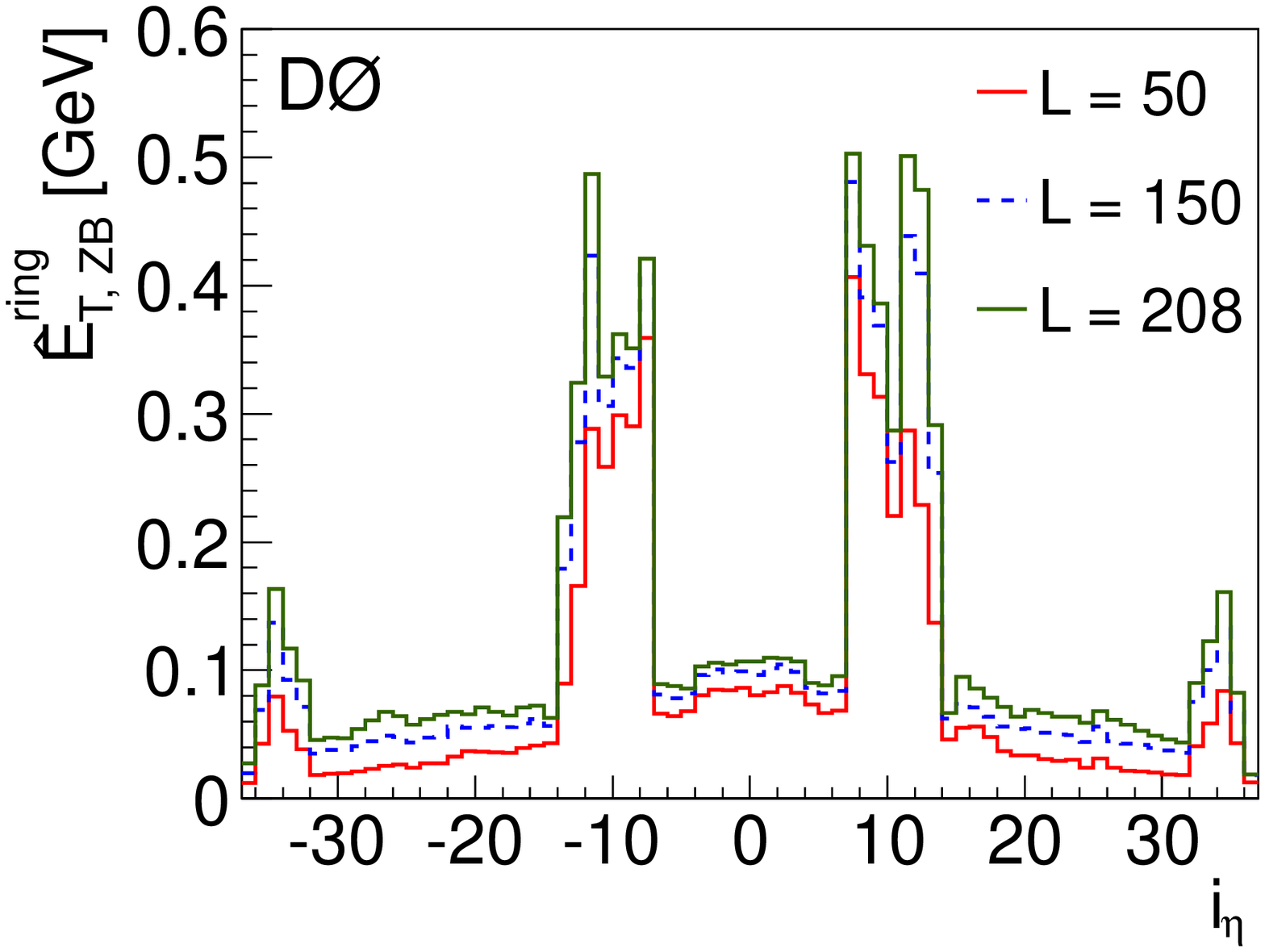}}
	\vspace{-5mm}
	\caption{
		(color online) Average transverse energy per \ieta ring 
		in ZB events selected as discussed in the text.  
		Lines with various colors correspond to $\InstLumin = 
		\InstLumTimes{50, 150 \text{ and }208}$.
	}
	\label{zb}
\end{figure}

The average energy per \ieta ring due to noise and pile-up 
is measured in ZB events requiring the LD veto to reject  inelastic activity.
Since the luminosity monitor is not 100\% efficient, 
we also exclude events with any reconstructed \ppbar collision vertex.
The average transverse energy \ET, where
\begin{equation}
	\ET = E / \cosh(\eta),
	\label{eq:ET}
\end{equation}
is parameterized for each \ieta ring as a function of \InstLumin.
\Cref{zb} shows the average \ET per \ieta ring,
$\MakeTransverseExt{\hat{E}}{\textnormal{ring}}{,ZB}$, 
for four different 
values of \InstLumin. 
The structure in the $8 \leq \abs{\ieta} \leq 15$ range corresponds to the poorly instrumented ICR region, 
where the noise fluctuations are amplified by large weight factors applied to convert 
ADC counts into energy, while at $\abs{\ieta}>32$ (as described in \cref{calorimeters}),
the cell size grows by a factor of two or more, resulting in a larger transverse energy per $\abs{\ieta}$ ring.

\subsubsection{Multiple interactions}

\newcommand{\npvTrue}[1][]{\JESmathSym{n_{\textnormal{PV\,true}}^{#1}}}

\begin{figure}[t]
	\centerline{\includegraphics[width=\columnwidth]{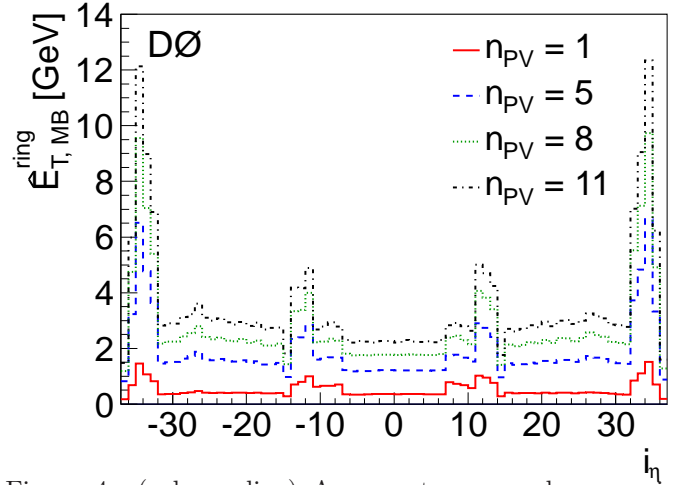}}
	\vspace{-5mm}
	\caption{
		(color online) Average transversed energy in minimum bias events as a function of \ieta.
		Lines with various colors correspond to
		$\npv=1$, $\npv=5$, $\npv=8$, and $\npv=11$.
	}
	\label{mb}
\end{figure}

The average energy per \ieta ring due to multiple interactions 
is estimated from the average energy per ring measured in MB events. 
\Cref{mb} shows the average transverse energy per \ieta ring 
$\hat{E}^{\textnormal{ring}}_{\textnormal{T,MB}}$, 
for MB events with different $\npv$, and corresponding to $\InstLumin=\InstLumTimes{200}$.
\begin{figure}[h]
	\centerline{\includegraphics[width=\columnwidth]{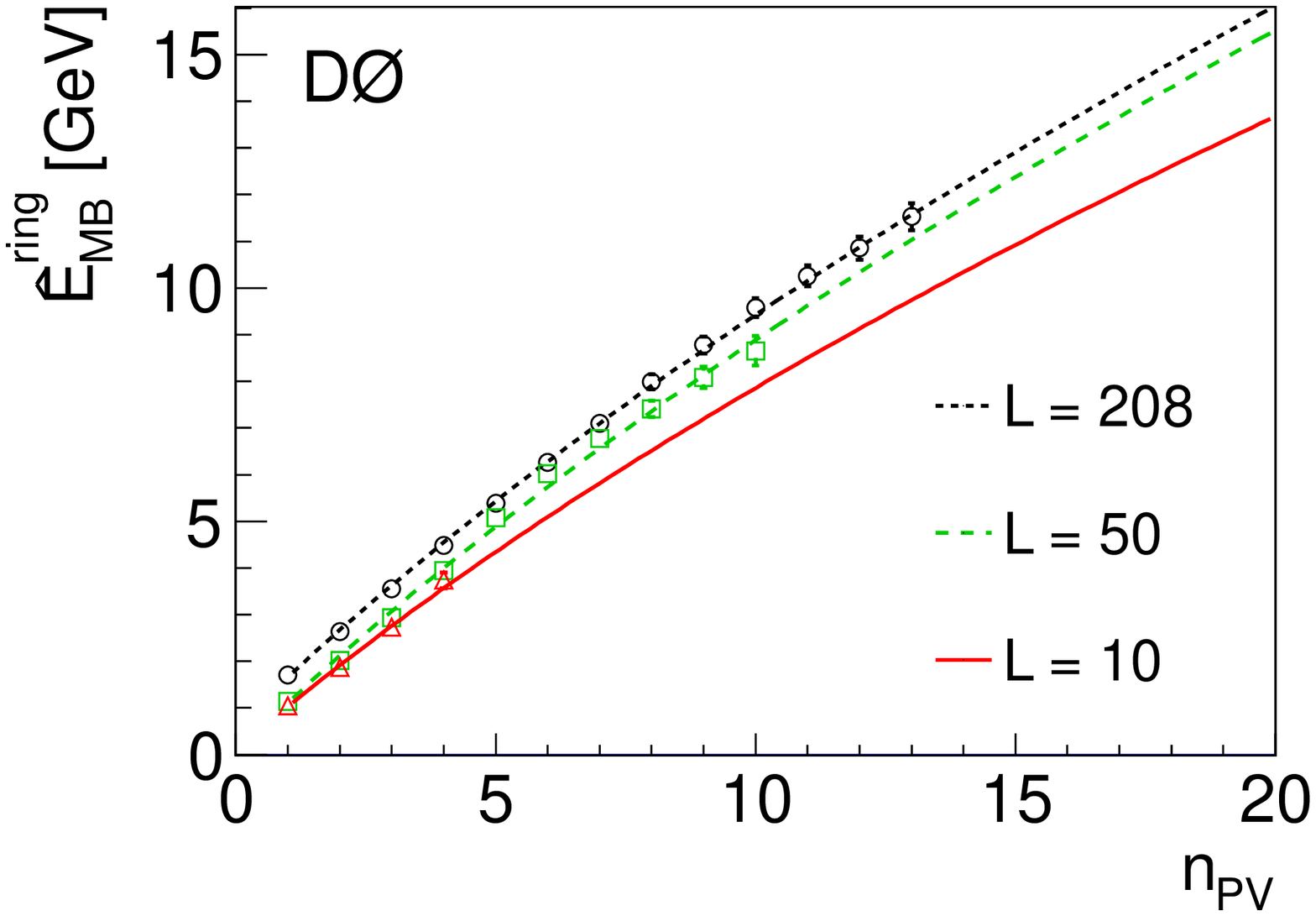}}
	\vspace{-5mm}
	\caption{
		(color online) Minimum bias energy as a function of $\npv$ 
		for one particular ring $\ieta=20$. Lines correspond to different
		luminosities $\InstLumin = \InstLumTimes{10, 50 \text{ and } 208}$.
	}
	\label{npv}
\end{figure}

For each \ieta and \InstLumin bin, the average MB
energy is measured as a function of \npv, for $\npv\leq14$, and 
extrapolated up to $\npv\leq20$ using function:
\begin{equation}
\hat{E}^{\textnormal{ring}}_{\textnormal{MB}}{(\npv)} = ( a + b\,\sqrt{1 + 4 c\,\npv} - 1) / 2c ,
\label{eq:TrueNPV}
\end{equation}
with empirically determined constants $a$, $b$, and $c$.
The form of the function assumes that the
offset energy depends on the number of \ppbar collision, \npvTrue,
linearly as $a + b\,\npvTrue$, while the observed number of primary vertices is
$\npv = \npvTrue + c\,\npvTrue[2]$, 
due to fake tracks and vertices~\cite{TrackNIM}. 
\Cref{npv} illustrates the average offset energy for MB events,
$\hat{E}^{\textnormal{ring}}_{\textnormal{MB}}$, as a function of
$\npv$ for $\ieta=20$ (taken as an example) collected at different luminosities. 
This simple model accurately describes the observed dependency
of the minimum bias energy on \npv.

We define
the average energy per \ieta ring due to multiple interactions
as the difference between the MB energy $\hat{E}_\textnormal{MB}^\textnormal{ring}$
for events with $\npv$ \ppbar collision vertices and with exactly one \ppbar collision vertex:
\begin{multline}
\hat{E}_\textnormal{MI}^\textnormal{ring}(\ieta,\npv,\InstLumin) \\
= \hat{E}_\textnormal{MB}^\textnormal{ring}(\ieta,\npv,\InstLumin) 
- \hat{E}_\textnormal{MB}^\textnormal{ring}(\ieta,\npv=1,\InstLumin) .
\label{mi}
\end{multline}

\subsubsection{Total offset energy}

The estimated total offset energy for a jet, $\hat{E}_\textnormal{O}$, is calculated using
the average energy for each ring ($\hat{E}_\textnormal{O}^\textnormal{ring}$),
taking into account the fraction of towers ($f^\textnormal{twr}$) in each \ieta
ring within the jet cone:
\begin{multline}
\hat{E}_\textnormal{O}(\deta,\npv,\InstLumin) =
\sum_{\ieta \in \rcone}
\hat{E}_\textnormal{O}^\textnormal{ring}({\ieta,\npv,\InstLumin})\times \\
\times f^{\textnormal{twr}}({\ieta,\deta}),
\label{mastereq_offset}
\end{multline}
where \deta is the detector pseudorapidity of the cone axis.

\subsection{Results}

\begin{figure}[t]
	\includegraphics[width=0.45\textwidth]{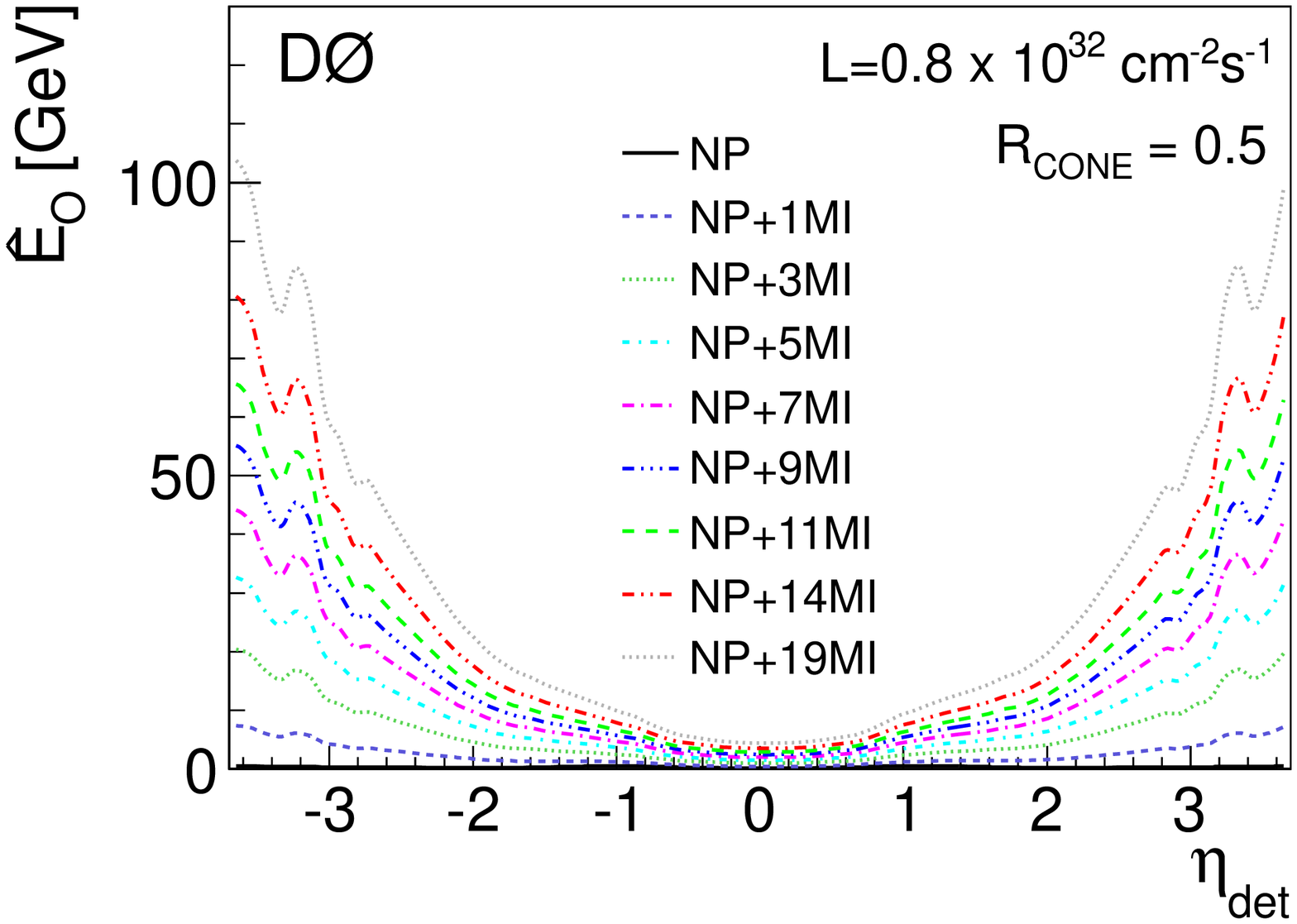}
	\caption{
		(color online) Estimated total jet offset energy, $\hat{E}_\textnormal{O}$, 
		as a function of \deta, for jets with $\rcone=0.5$.
		The different lines show the prediction
		for noise and pile-up (NP) only ($\npv=1$), as well as NP and multiple interactions (MI) ($\npv>1$).
	}
	\label{jetoffset05}
\end{figure}

\begin{figure}[t]
	\includegraphics[width=0.45\textwidth]{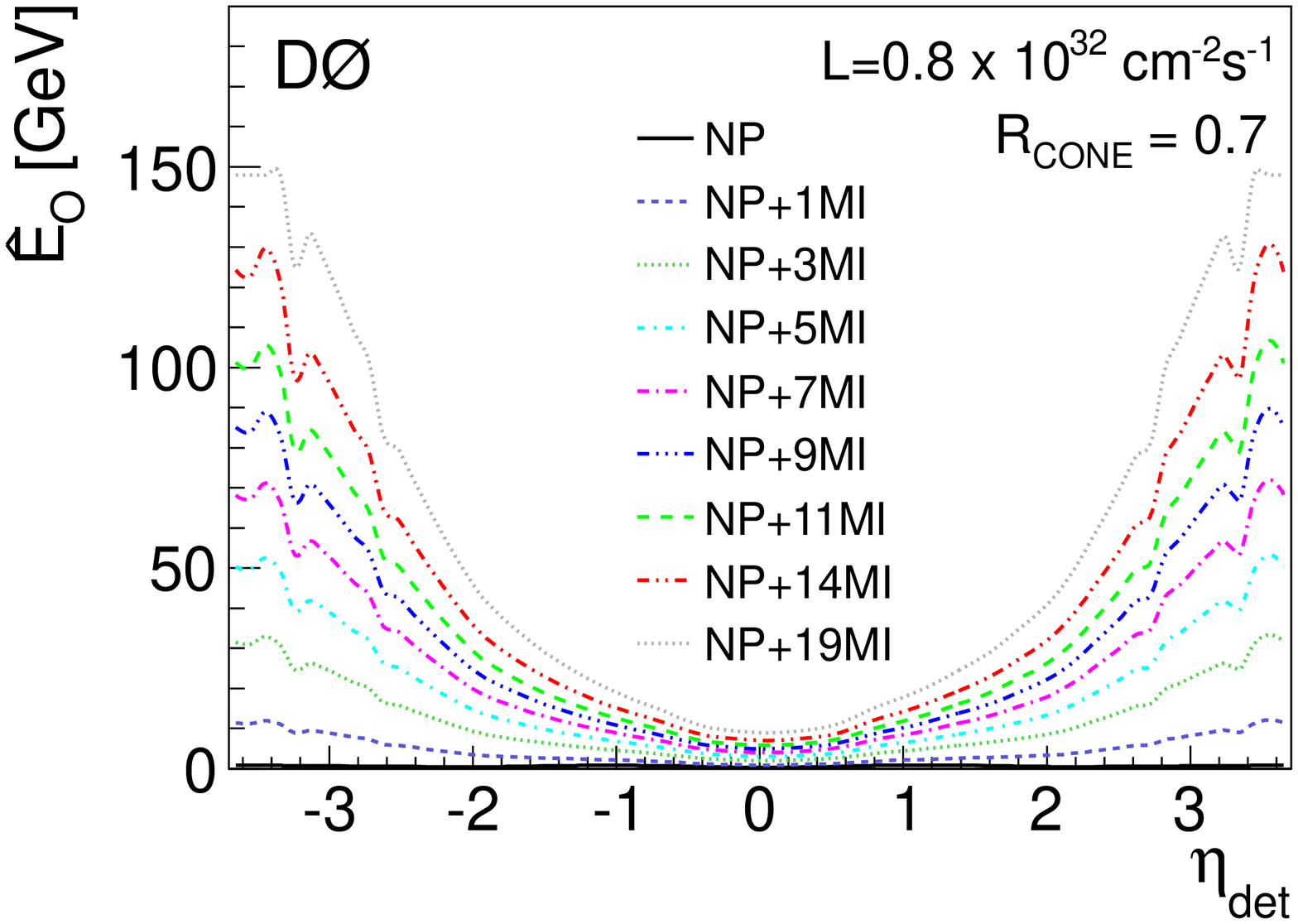}
	\caption{(color online) Same as on \cref{jetoffset05} for jets with  $\rcone=0.7$.}
	\label{jetoffset07}
\end{figure}

\Cref{jetoffset05,jetoffset07} show the estimated jet offset energy as a function of
\deta for events with different number of reconstructed \ppbar collision vertices.
This estimate has been obtained using \cref{mastereq_offset}, separately 
for jets with $\rcone=0.7$ and 0.5, and $\InstLumin=\InstLumTimes{80}$,
which represents the average instantaneous luminosity of the MB sample. 
The offset energy for $\rcone=0.5$ jets is approximately a factor of two
smaller than for $\rcone=0.7$ jets, in agreement with the expectation based
on the ratio of their geometrical areas.


%
%
%
%

\subsection{Zero-suppression bias correction}
\label{offset_bias_correction}

The total offset energy estimated from MB and ZB events can differ substantially
from the true offset energy inside the jet cone. This is because the
calorimeter cells inside the jet cone already contain energy from the hard 
interaction and therefore they are more likely to be above 
threshold compared to the cells outside the jet. 
As a result, the actual offset energy deposited inside the
jet cone is higher than that estimated using the MB and ZB events with a lower cell occupancy. 
We thus derive an average correction factor from the
offset-corrected jet energy to the actual jet energy in the absence of 
noise, pile-up, and multiple interactions. This correction factor can 
be estimated in MC by comparing the measured jet energy from 
the same high-$\pt$ events processed with and without offset energy added.

The factor which corrects for this effect (\kOZS) is estimated 
comparing the measured energy of the leading jet
from the same high-\pT \photonjet event with and without
offset energy added, denoted by \Emeas and $E^\textnormal{meas,noZB}$,
respectively:
\begin{equation}
	\kOZS = \frac{E^\textnormal{meas,noZB}}{\Emeas - \hat{E}_\textnormal{O}}.
	\label{eq:kOZS}
\end{equation}

For this purpose, we consider the same \photonjet MC events
processed in three ways (see \cref{mc_samples}):
\begin{enumerate}
\item no ZB overlay, \ie, no offset energy from noise, pile-up, and
multiple interactions. This provides the reference level to which to correct 
($E^\textnormal{meas,noZB}$).
\item ZB overlay (providing \Emeas):
	\begin{enumerate}[(a)]
	\item Using zero-suppressed overlay the derived correction factor
	will be applicable to the jet energy scale calibration in MC since 
	the standard MC simulation
	uses zero-suppressed ZB overlay,
	\item Using ZB overlay without zero-suppression, the derived correction factor
	will be applicable to the jet energy scale calibration in data since they
	provide the most realistic description of the per-cell energy spectrum
	arising from noise, pileup, and multiple interactions.
	\end{enumerate}
\end{enumerate}

Only matched jets contribute to \cref{eq:kOZS},
\ie, only events where a reconstructed jet in the case of ZB overlay 
is unambiguously matched within $\Delta{\cal R}<\rcone/2$
with a jet in the case of no ZB overlay are considered.
Furthermore, we have the same set of physical events (with common partonic origin)
in the samples without ZB overlay and with ZB overlay, both suppressed and unsuppressed.
The correction is measured separately for jets with $\rcone=0.7$ and 0.5,
in intervals of 0.4 of jet \mdeta, and as a function of \pTprime (defined in \cref{eq_ptprime})
for suppressed and unsuppressed ZB overlay. 

\begin{figure}
	\includegraphics[width=0.45\textwidth]{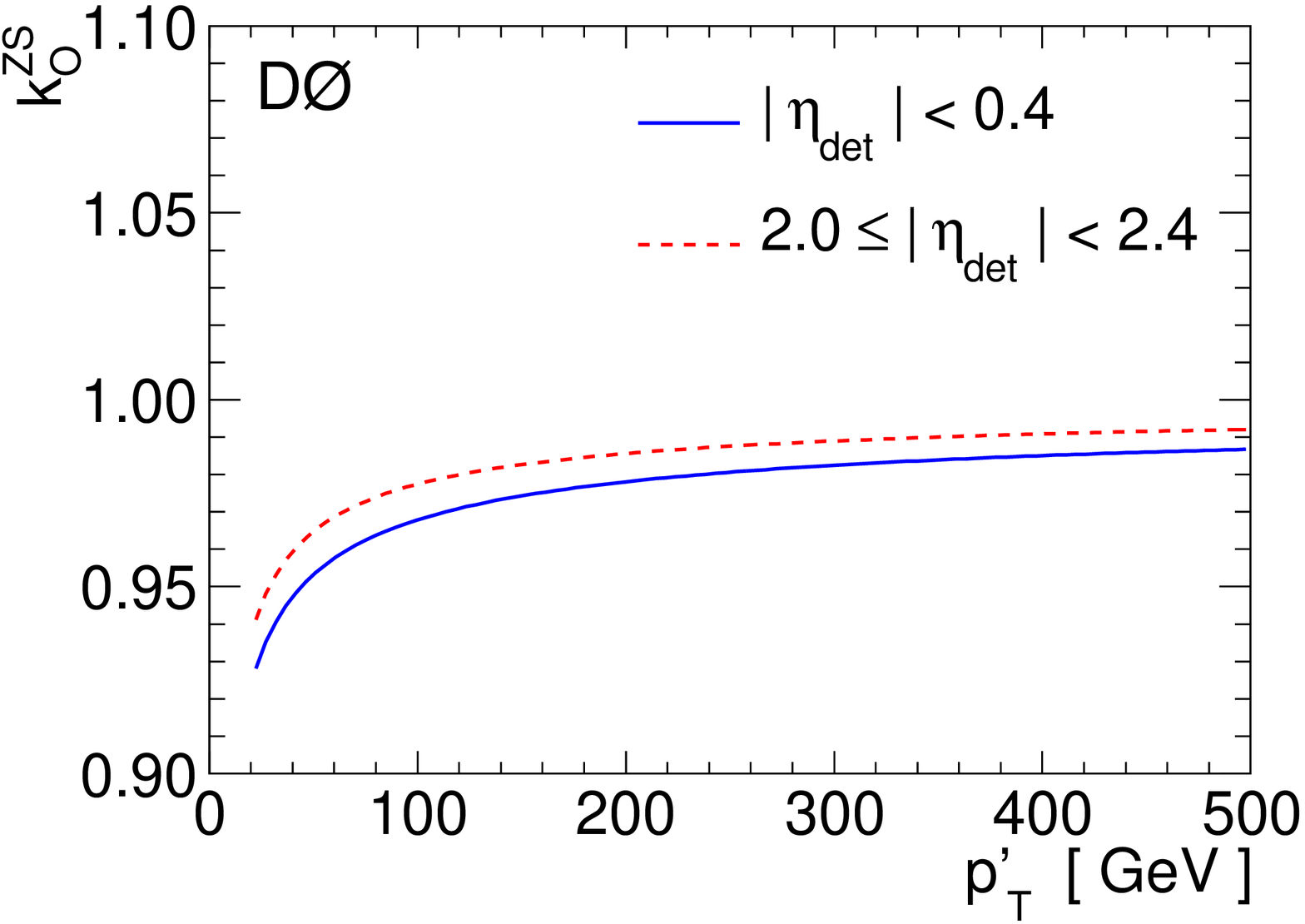}
	\caption{
		Correction factor for the zero-suppression bias  \kOZS
		for  $\rcone=0.7$ jets in the unsuppressed ZB overlay sample
		within $\mdeta < 0.4$,
		and  $2.0 \le \mdeta < 2.4$ ($\average{\npv} = 2.6$, \IIbTwo data). 
	}
	\label{jcca_kO_bin0}
\end{figure}

The \kOZS factor depends on \deta and \npv
and it is extracted for the average number of \ppbar collision vertices, 
$\langle\npv\rangle$.
\Cref{jcca_kO_bin0} 
illustrates the extracted \kOZS factor for two
\deta intervals.

\subsection{Uncertainties}
\label{offset_uncertainties}

The offset correction measurement in data as given by \cref{mastereq_offset}
has a high statistical precision.
Statistical uncertainties do not exceed 2\%. 

%
%

The systematic uncertainty originates from the fitting procedure for
$\hat{E}_\textnormal{MB}^\textnormal{ring}$ and is estimated for each \ieta ring (see \cref{npv})
from the residual difference between fit and data.
The uncertainty is found to be mildly dependent on \deta and 
varies between \MeV{50\text{ and }200} for $\rcone=0.5$
and between \MeV{100\text{ and }350} for $\rcone=0.7$.

\begin{figure}
	\centerline{\includegraphics[width=0.45\textwidth]{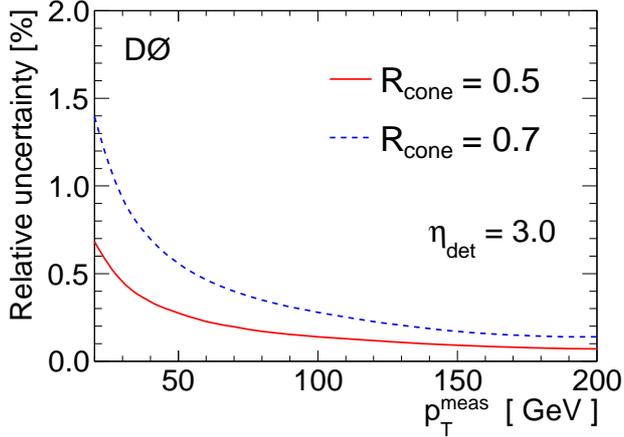}}
	\caption{
		Relative systematic uncertainty on the offset correction as a function of
		the measured transverse momentum \pT[meas] of the jet,
		for jets with $\rcone=0.5$ and $0.7$, and $\deta=3.0$.
	}
	\label{fig:ofs:syst3}
\end{figure}

\Cref{fig:ofs:syst3} shows the relative systematic uncertainty as a function of 
the measured transverse momentum of the jet, for jets with $\rcone=0.5$ and $0.7$, 
and $\deta = 3.0$, which has the largest systematic uncertainty.
The systematic uncertainties of the jet transverse momenta due to the offset correction
are typically less than $1\%$.

%% file: response.tex
\section{Absolute MPF response}
\label{sec:response}


This section describes the determination of the response correction 
for central calorimeter jets using the MPF
method as described in \cref{sec:response_overview}. 
The response from the central region provides the main correction factor for jet energy calibration.
The calibration of forward jets, relative to the response in
the central region, is described in \cref{sec:etadep}. 

\subsection{Sample selection}
\label{photonjet_selection}

The \photonjet samples described in \cref{sec:datasets} are used
for the determination of the absolute response correction in both data and MC.
Further selection criteria are applied to extract a subset
of events with suitable characteristics for the measurement of the jet
response via the MPF method. These requirements are:
\begin{itemize}
	\item Events are rejected unless they have exactly one or two reconstructed
	\ppbar{} collision vertices. The main vertex associated 
	with the hard interaction must satisfy the vertex selection criteria 
	discussed in \cref{pvsel}. The inclusion of events with two vertices doubles
	the size of the sample and has been shown not to introduce any bias.
	\item Each event must have 
	exactly one photon candidate with measured transverse momentum,
	$\pTgamma[meas] > \GeVc{7}$, satisfying the 
	tight photon identification criteria (see \cref{photonid}).
	The photons must be in the central calorimeter corresponding to 
	$\abs{\deta[\photon]} < 1$. The momentum \pTgamma[meas] does not include
	the photon calibration described in \cref{photon_corrections}.
	\item To avoid a possible bias caused by trigger inefficiency,
	\pTgamma[meas] is required to be in the high efficiency
	range of the particular trigger used to collect the event. In addition,
	the directions of the photon candidate and the electromagnetic trigger tower at Level~1
	trigger must match within $\deltar < 0.4$.
	\item Each event has to have exactly one reconstructed jet 
	(with $\rcone=0.7$ or 0.5, as appropriate) 
	satisfying the jet selection criteria described in \cref{jetid}. 
	This jet is referred to as the ``probe jet''. No additional 
	jet is allowed in the event, except if its direction matches the photon candidate 
	within $\deltar < 0.2$, since the photon candidate can also be reconstructed
	as a jet.
	\item The probe jet must have $\mdeta < 0.4$, so that its core is well 
	contained inside the central calorimeter.
	\item The photon and jet are required to be back-to-back in the $r$-$\phi$ plane:        
	the difference of their azimuthal angle should be 
        $\Delta\azim(\photon,\textnormal{jet})>3.0$~rad.
	\item Data events with cosmic muon candidates, indentified using muon system
	timing information, are rejected.
	\item To further eliminate cosmic rays and other physics backgrounds,
	an upper limit is imposed on the ratio $\MET/\pTgamma[meas]$
	in the range of 0.65 to 1.1, where a looser cut corresponds to the lower
	photon \pT 
	~\cite{JES_run1,gj_PRD}.
\end{itemize}

\subsection{Backgrounds in the \photonjet sample}
\label{sssec:PhotonJetPurity}

Two types of background contaminate the
\photonjet sample:
events with electrons or multiple photons from electroweak interactions that are
misidentified as a single photon,
and events where strong interactions produce a jet misidentified as photon.

Background processes of the first type are \Xjet[W(\to e\nu)],
\Xjet[Z/\photon^{*}(\to{\fpfm[e]})], and diphoton production.
The contributions from these backgrounds are estimated from MC simulation. 
In the case of \Xjet[W(\to e\nu)] events, with the electron misidentified as a photon, the neutrino will contribute additional
missing transverse energy \MET. 
The combination of the track veto (part of the photon identification criteria)
and the capping of the ratio $\MET/\pTgamma[meas]$ reduces
the contribution from these processes to a negligible level, less than $0.5\%$. 
Contributions from \Zjet and diphoton events are found to be even smaller.
The total expected bias on the MPF response is studied in MC and is estimated to be below $0.1\%$.

The second type of background is represented by \dijet events, 
where one of the partons showers to produce a well isolated, energetic $\pi^0$ or $\eta$ meson,
decaying into a multi-photon final state.
The probability for a jet to be misidentified as a photon depends on the
photon identification criteria but is typically very small. Nevertheless 
this background contamination remains sizable, particularly for
photons with low transverse momentum \pTgamma, due to the high rate of \dijet production.

The photon purity is estimated using the
\photonjet and \dijet(\photon-like) MC samples described in \cref{mc_samples}.
To estimate background from the \dijet events remaining after the photon selection, 
we use the scalar sum of the transverse momenta of all tracks in a hollow cone
of $0.05<\deltar<0.7$ around the direction of the photon candidate
(see \cref{photonid}).
The distributions for the simulated photon signal and \dijet background samples are fitted to 
the data for each $\pTgamma[meas]$ bin using a maximum likelihood fit \cite{HMCMLL} to obtain 
the fractions of signal and background components in the data.
The systematic uncertainties on the purity measurement are estimated
from the uncertainties on the fit result
and from a comparison with alternative fitting functions.
An additional contribution is included due to the dependencies on the
fragmentation model implemented in \PYTHIA{}.
The overall systematic uncertainty is found to be $5\%$ at $\pTgamma\approx \GeVc{30}$, 
$3\%$ at $\pTgamma[meas] \approx \GeVc{50}$, 
and $2\%$ at $\pTgamma[meas] \gtrsim \GeVc{70}$~\cite{IncGam}.
\extend{More detailed information
about the purity determination of the \photonjet sample can be found
in Appendix~\ref{response_appendix}\ref{syst_purity}.}

\Cref{f_rjet_purity} illustrates the estimated purity of the
selected \photonjet sample with central jets ($\mdeta<0.4$, as an example) 
as a function of \Eprime (defined in \cref{eq_eprime}). 
Individual points represent purity determined from the data. 
The purity improves for higher \Eprime, as the probability for production of
isolated EM showers through the fragmentation process decreases.
 
The presence of this instrumental background leads to a positive bias in 
the measured MPF response, since the photon candidate is usually surrounded by 
hadronic activity resulting from the fragmentation of the original parton.
This effect can be suppressed by using more stringent photon identification criteria, 
but it cannot be completely eliminated. Therefore, we explicitly correct
the measured MPF response for this effect. 

\begin{figure}
	\centerline{
		\includegraphics[width=0.9\columnwidth]{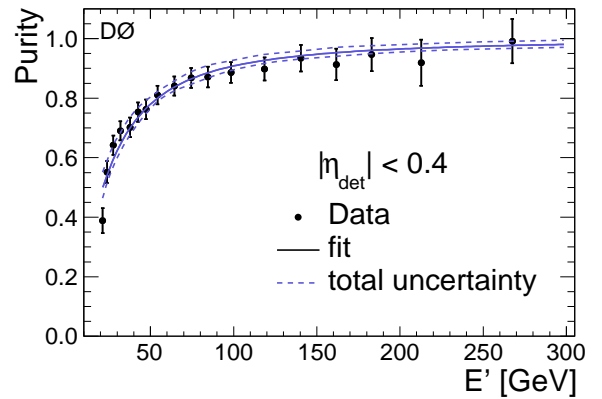}
	}
	\caption{
		Photon purity measured in the \photonjet
		data sample selected using tight photon identification criteria,
		for $\mdeta<0.4$ as a function of \Eprime. 
		Also shown is the total uncertainty band.
	}
	\label{f_rjet_purity}
\end{figure}

\subsection{Method}
\label{absolute_response_method}

The measurement of the absolute MPF response is discussed 
in \cref{absolute_mpf_response}. The goal is to estimate the MPF response for $\gamma$+jet events 
with the photon at the particle level. In the case of MC, this is achieved
by using a modified version of \cref{mpf_gj}:
\begin{equation}
	\MPFCC{\photonjet} = 1 +  \frac{\vecmet \cdot \MakeTransverseExt{\vec{n}}{}{\photon}}{\pTgamma[ptcl]},
	\label{eq_rmc}
\end{equation}
\noindent where, on an event-by-event basis, the particle
level photon transverse momentum, \vecpTgamma[ptcl], is used for the tag,
and \vecmet is corrected accordingly, similarly to \cref{eq:METcorrEM}:
\begin{equation}
	\vecmet \enskip \Rightarrow \enskip \vecmet + \vecpTgamma - \vecpTgamma[ptcl].
	\label{eq:correctedMET}
\end{equation}
In the case of data, as discussed in \cref{mpfbias},
the application of \cref{mpf_gj} results in a measurement of the MPF response which
is affected by the bias in measured photon transverse momentum,
as well as the presence of the \dijet background. 
Explicit corrections for these biases are discussed below.


\subsubsection{Photon energy scale correction}
\label{photon_corrections}

The first correction is related to the calibration of the photon energy scale.
As discussed in \cref{calibration}, the absolute energy calibration of the electromagnetic
calorimeter is obtained using electrons from $Z \to \fpfm[e]$ 
decays with about 0.5\% accuracy. Corrections for the energy loss of electrons in the 
material in front of the calorimeter as a function of \deta and 
\pT are determined in MC and applied to electromagnetic objects in data.
However, photons interact less with the material of the detector than electrons,
and as a result the electron energy scale correction overcorrects the 
photon energy (\Emeas[\photon]) relative to the particle level (\Eptcl[\photon]).
This effect is particularly sizable at low energy.

\begin{figure}
	\centerline{\includegraphics[width=0.9\columnwidth]{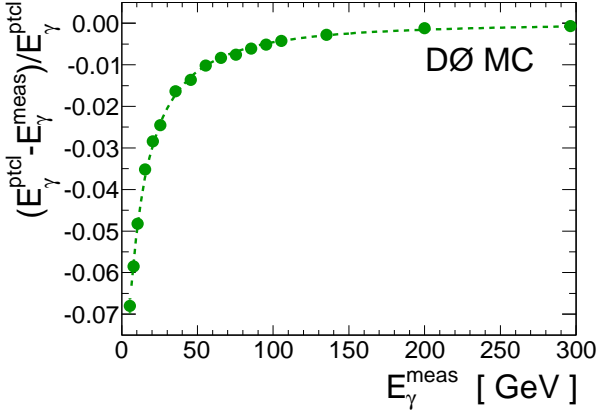}}
	\caption{
		Photon energy scale correction as estimated from dedicated MC.
		This correction is applied to photons after the calibration of the EM
		calorimeter described in \cref{calibration}.
	}
	\label{f_rjet_photoncorr}
\end{figure}

   The difference in the response of the calorimeter for electrons and photons
is evaluated in dedicated MC with an improved \GEANT{} description
of electromagnetic showers~\cite{D0WMass2012},
which is not used for standard simulation of physics processes 
due to its low execution speed.
Calorimeter response is simulated for single photons and electrons entering the
\DZero detector at different angles and positions.
At low energies ($\Emeas[\photon] \approx \GeV{20}$),
the photon energy overcorrection (\cref{f_rjet_photoncorr}) is estimated to be about $3\%$. The difference 
between electrons and photons becomes smaller, but still remains sizable, at high energies.
This photon energy scale correction is applied to the reconstructed EM object,
and the missing energy is corrected accordingly (see \cref{eq:correctedMET}).

   Three main sources of photon energy scale systematic uncertainties are considered:
the electron energy calibration, the difference between photon 
and electron energy scale, and the contamination by \photon-like jets.
The first is estimated to be about $0.5\%$ and it is
mostly connected with long-term stability of the calorimeter response. 
The second is due to the different nature of photon and electron interactions with the material
in front of the calorimeter. This effect is estimated by varying 
the amount of this material in the simulation within its uncertainty.
Finally, the energy calibration of the candidate photons is affected by
the presence of misidentified \photon-like jets. 
The size of this effect is found to be smaller than $0.2\%$, and included in the uncertainty.

\subsubsection{Background correction}
\label{sssec:response:background_corrections}


The \photonjet sample selected according to criteria of \cref{sec:objectid} is a mixture of \photonjet signal events
and \dijet background. The photon candidate in the latter sample is caused by \photon-like jets.
The measured MPF response for this mixed sample
can be expressed as a linear combination of the MPF responses for signal 
and background  weighted by the respective sample fractions:
\begin{equation}
	\MPFeta{mixture} = \rho_\eta \MPFeta{\photonjet}+
		(1-\rho_\eta)\MPFeta{\dijet},
\end{equation}
\noindent where both MPF responses are with respect to the photon \pT, 
and $\rho_\eta$ is the \photonjet sample purity (see, \eg, \cref{f_rjet_purity}), as
function of the jet pseudorapidity \deta.
Since the same approach is used later for the relative calibration of forward jets,
the dependence on jet \deta is explicitly kept in this formula.
The relative difference between 
the MPF response of the mixed sample and the MPF response of the pure sample 
is then:
\begin{equation}
	c_{\textnormal{bckg},\eta} \equiv 
		\frac{\MPFeta{mixture}}{\MPFeta{\photonjet}}-1
		= (1 - \rho_\eta) \left(\frac{\MPFeta{\dijet}}{\MPFeta{\photonjet}}-1\right),
	\label{eq_bkgrcorr1}
\end{equation}
and the correction factor \kR[\photon] described in \cref{eq_kRgamma}
can therefore be written as
\begin{equation}
	\kR[\photon] = \frac{1}{1+c_{\textnormal{bckg},\eta}},
	\label{eq:kappa_eta}
\end{equation}
\noindent
where $\eta$ refers to the pseudorapidity of the jet recoiling from the central EM object.

\begin{figure}[t]
	\centerline{\includegraphics[width=0.90\columnwidth]{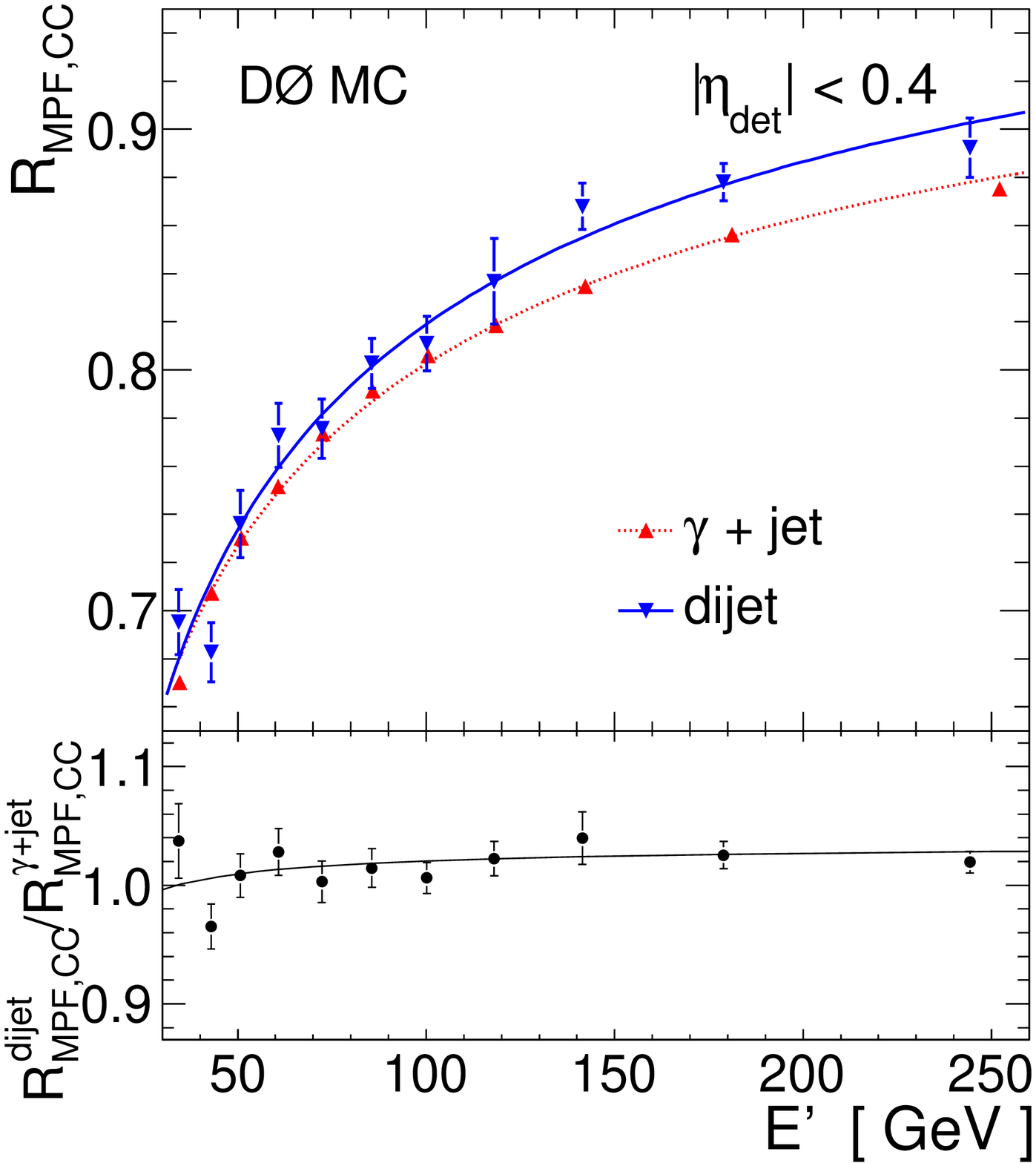}}
	\caption{
		Comparison of MPF responses in \photonjet and \dijet MC samples
		selected as described in \cref{photonjet_selection} in the events with central jets. 
		The lines in the upper panel show
		fits to the responses in the \photonjet and \dijet events, respectively.
		The line in the bottom plot shows a ratio of fits of those responses, while
		points correspond to the ratio $\MPFCC{\dijet}/\MPFCC{\photonjet}$ in each bin.
	}
	\label{f_rjet_bkgr}
\end{figure}

   Jet response in a pure \photonjet sample and in \dijet
background is estimated from MC.
The response from MC does not accurately reproduce the jet response in the data.
Therefore the background correction
is determined with the corrected MC simulation of \cref{sec:fdc}.
\Cref{f_rjet_bkgr} compares \MPFCC{\dijet}, defined similarly to \cref{eq_rmc} 
(with a photon candidate energy corrected according to \cref{f_rjet_photoncorr}), and
\MPFCC{\photonjet} as predicted by the MC, for events with a jet within $\mdeta < 0.4$.
%
%
The ratio of fits of responses in the \photonjet and \dijet events
appearing in the right side of \cref{eq_bkgrcorr1}, also shown in \cref{f_rjet_bkgr},
is $1\%-3\%$ above unity, due to additional hadronic activity around the misidentified 
photon in the \dijet sample. This activity reduces \MET in the direction of the jet,
increasing the measured MPF response relative to that for the \photonjet sample. 
The tight
photon criteria, which are applied for the final jet response measurements, 
suppress much of this additional hadronic activity, yielding a MPF response 
for the \dijet sample which is not more than 2\% larger than for the \photonjet sample. 

\subsection{Results}
\label{response_results}

The MPF response as a function of \Eprime for $\rcone=0.7$ jets
is shown in \cref{f_rjet7} for MC and data.\extend{Plots corresponding to 
$\rcone=0.5$ can be found in \cref{f_rjet5} in Appendix~\ref{response_appendix}.}
In the case of MC, the MPF response is obtained directly using \cref{eq_rmc}. In the
case of data, the MPF response for the mixture sample is first computed
using \cref{mpf_gj} and then corrected using the following equation,
\begin{equation}
	\MPFCC{\photonjet} = \MPFCC{mixture}\,\kR[\photon],
\label{eq_rdata}
\end{equation}
where \kR[\photon] is defined in \cref{eq:kappa_eta}.

In both data and MC, the tight photon identification criteria are used.
Since jet energies do not enter directly into the calculation of the MPF response, the dependence
on $\rcone$ is expected to be very small. As an example, the MPF response for $\rcone=0.5$
is about $0.5\%$ higher at $\Eprime\approx \GeV{100}$ than for $\rcone=0.7$, 
in both data and MC.
The measured MPF response is fitted using the parameterization in \cref{quadlog}.

\begin{figure}
	\subfloat{
		\includegraphics[width=\columnwidth]{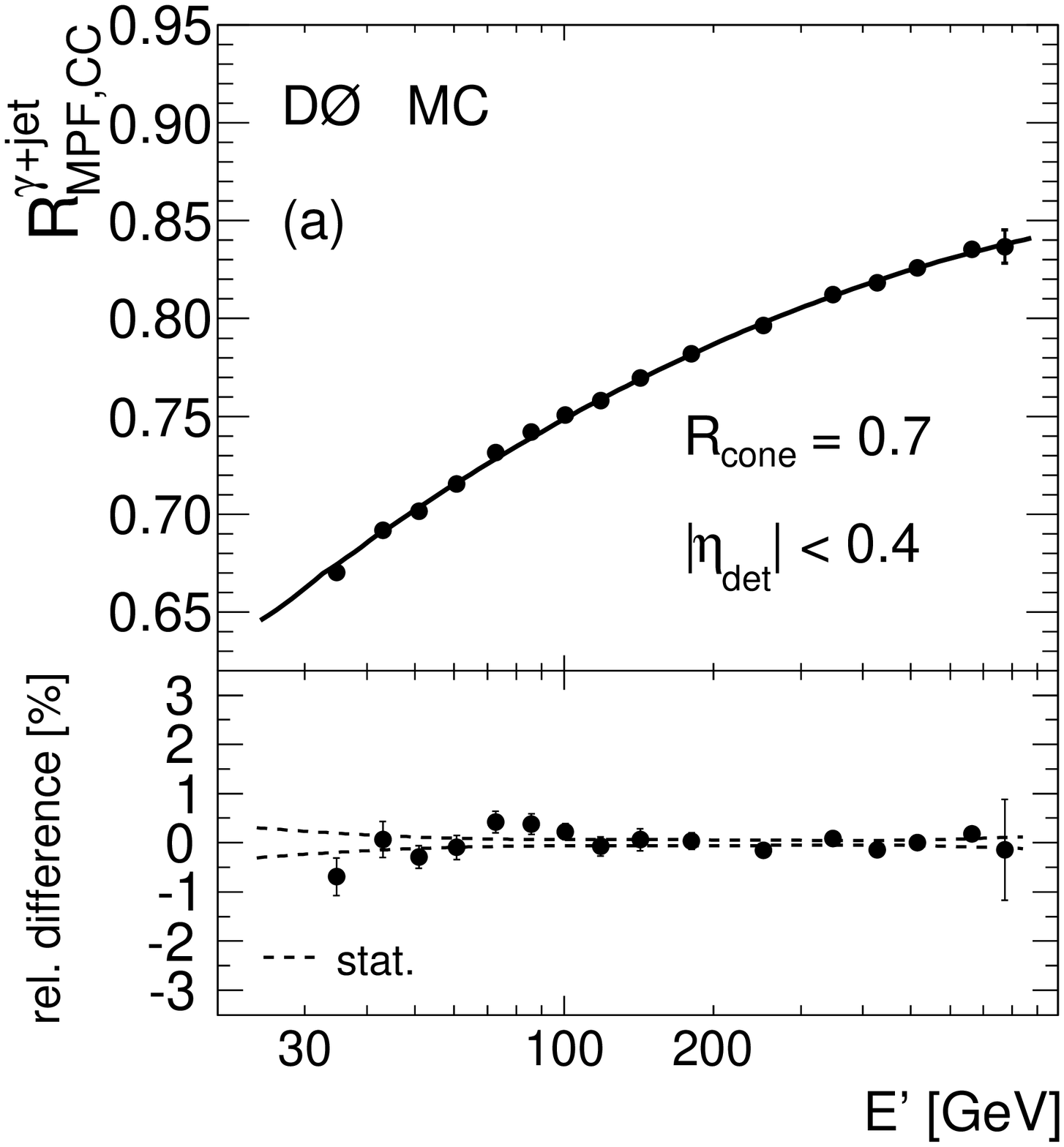}
		\label{f_rjet7_MC}
	}
	\\
	\subfloat{
		\includegraphics[width=\columnwidth]{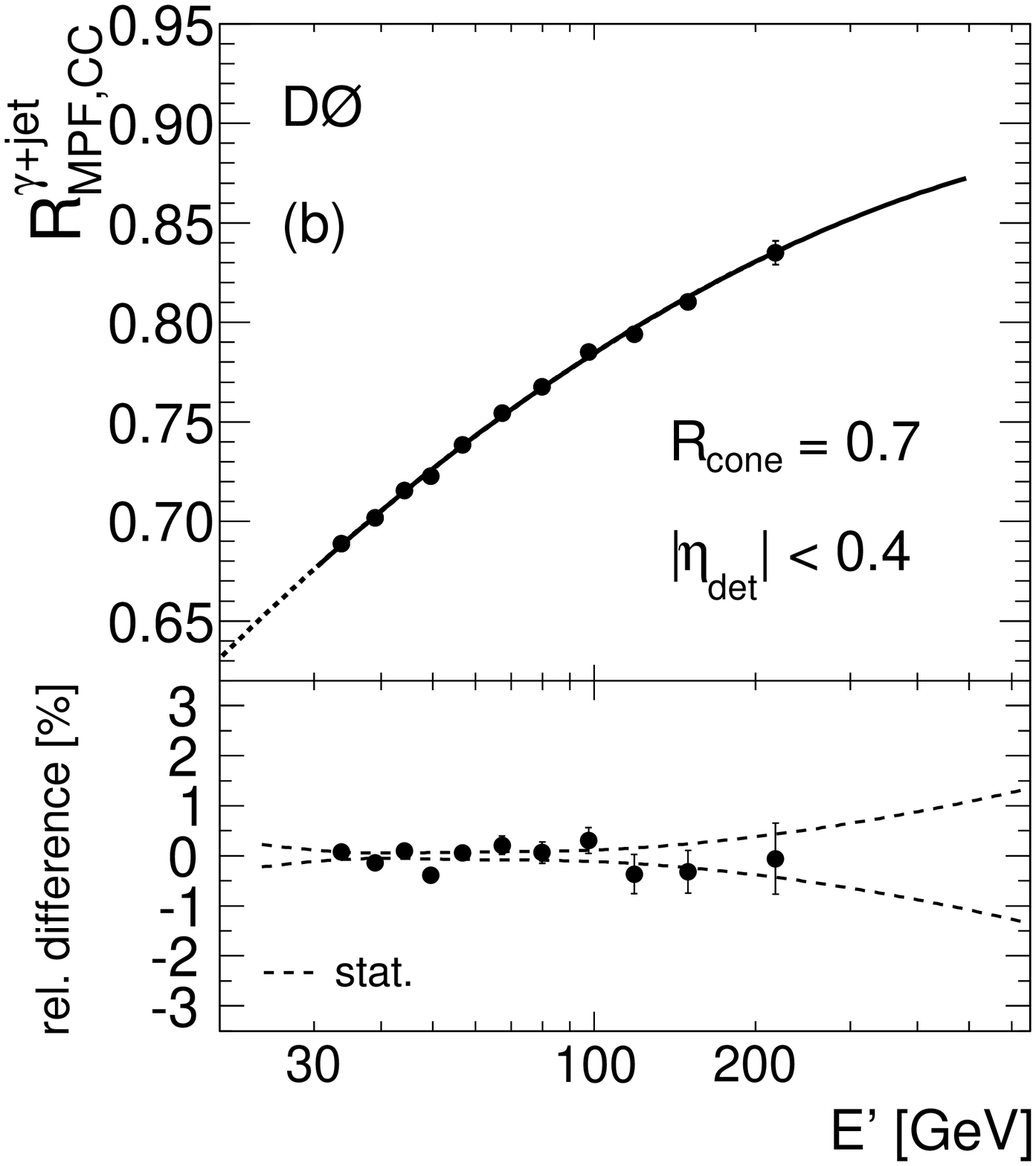}
		\label{f_rjet7_data}
	}
	\caption{
		Absolute MPF response for $\rcone=0.7$ jets in
		\protect\subref{f_rjet7_MC} MC and \protect\subref{f_rjet7_data} data
		as a function of \Eprime.
		The solid line indicates the fit to the function in
		\cref{quadlog}. The lower plots show the relative difference
		of the points with respect to the fitted function, along with the 
		statistical uncertainty from the fit.
	}
	\label{f_rjet7}
\end{figure}

\extend{In general, the response follows the quadratic-logarithmic dependence 
quite well. However, there is a residual structure at $\eprime\sim 40-50$ GeV of the order $\pm1\%$ which 
is not present in MC and whose origin is currently not understood. Nevertheless, the 
shape distortion is within the claimed systematic uncertainty on the photon energy scale 
and it could possibly be explained by the difference of the real photon energy scale and the
one obtained from MC. The following potential sources were investigated and ruled out: 
trigger bias,  bias due to \MET cut and \Xjet[W] contamination. None of them
was found to be responsible for the apparent deviation from the quadratic-logarithmic fit 
at $\eprime\sim 40-50$ GeV. A possible trigger bias was studied by 
increasing the values of \pT thresholds for the selected triggers. The cut on $\met$
for $p_{T\gamma}>50$ GeV was found to affect only the highest $\eprime$ bins
as intended, and the effect of the cut is below $0.2\%$ at $\eprime\sim50$ GeV. 
MC studies suggest that the $W$+jet contamination is negligible.
This was further confirmed in data by releasing and tightening the cut 
on the spatial track-match probability. Varying the cut from
$10^{-6}$ up to $0.1$ did not lead to any significant change
in response at $\eprime\sim50$ GeV.}

\subsection{Uncertainties}
\label{response_uncertainties}

\begin{figure}
	\centerline{\includegraphics[width=0.90\columnwidth]{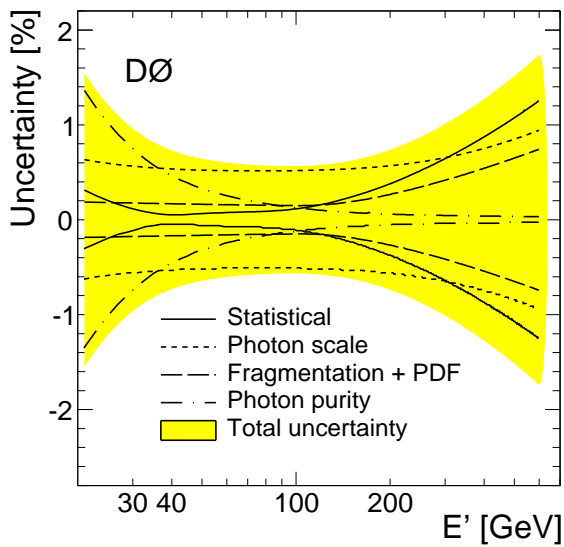}}
	\caption{(color online) Relative uncertainties on the MPF response measurement 
in data for $\rcone=0.7$ jets.
        The shaded region shows the total uncertainty, while the statistical uncertainty is shown by the solid line.
        Uncertainties related to photon energy scale and purity, 
        parton-to-hadron fragmentation and PDF effects are also shown.
        Uncertainties for $\rcone=0.5$ jets are very similar.
}
	\label{f_rjet_err}
\end{figure}

In the case of the MPF response measurement in MC, the only uncertainty is from the
statistical uncertainty of the fit (including the full covariance matrix)
shown in \cref{f_rjet7_MC}.
The main sources of uncertainty in the MPF response measurement in data
are shown in \cref{f_rjet_err} for $\rcone=0.7$ jets. 
They include the statistical 
uncertainty of the fit, the uncertainties on the photon energy scale,
on the correction for the \dijet background contamination,
on the high energy extrapolation procedure (see below), and
an uncertainty to account for the stability of response versus time.
The uncertainties for
$\rcone=0.5$ jets are almost identical except the statistical uncertainty and
those connected with the high energy extrapolation procedure which is
performed only for $\rcone = 0.7$ jets.
\extend{See Appendix~\ref{response_appendix}\ref{time_dep_response} for a discussion
on the time-dependence of response. STILL NEED
TO INCLUDE THIS UNCERTAINTY IN THE PLOTS, but it is already implemented in jetcorr.
In the central region, it is anyway quite small, 0.1\%, and it will not 
change the picture.}

The main source of uncertainty in data is the photon
energy scale in almost the entire range of accessible energies.
At low jet energies (below \GeV{30}), the uncertainty
due to the \dijet background correction dominates.

The uncertainty on the \dijet background correction
is related to the uncertainty on $c_\textnormal{bckg}$ (Eq.~\ref{eq_bkgrcorr1}),
with two independent components:
purity and relative response between \photonjet and \dijet MC events.
\extend{See Fig.~\ref{f_emid_sys} in Appendix~\ref{response_appendix}.}
Ideally, the corrected MPF response in data should be independent of the photon identification
criteria, despite differences in purity. 
We have compared the MPF response in data for the different photon criteria, before 
and after 
the background correction.
The observed small residual differences after background correction are consistent with
the assigned systematic uncertainty.
Part of the observed difference between medium and tight criteria 
is unrelated to the background and can already be observed in \photonjet MC with the photon at 
the particle level. \extend{See Fig.~\ref{f_rjet_mcdiff} in Appendix~\ref{response_appendix}.}
This effect is believed to be caused by distortions in the hadronic activity in the photon hemisphere, which
propagates to \MET, as a result of tightening the photon isolation. 
This effect will be corrected by the topology bias correction (see \cref{topo_bias_corr}), 
and therefore it does not represent an additional source of systematic uncertainty.
The \dijet background correction (see \cref{eq:kappa_eta}) can be verified using data. 
Jet responses with different photon selections should be equal once they are corrected for 
the background admixture.
To cover potential mismodeling in the MC simulation, 
an additional $1\%$ systematic uncertainty is assigned to the relative difference
between the response in the \photonjet and background \dijet samples.
This is roughly half the size of the background correction
in case of the tight photon selection (\cref{f_rjet_bkgr}).

Measurements from events with only two or three jets, typically using
jets with $\rcone = 0.7$, include jets with very large energy (up to \TeV{1}).
The MC is used to constrain the response for such high energy jets in data.
The uncertainty on this high energy extrapolation
includes the following two sources of systematic uncertainty: 
parton distribution functions (PDFs) and fragmentation model.
These uncertainties are related to the dependence of the predicted hadron spectra at high
energy on the parton flavor of jets as well as the modeling of the fragmentation.
\extend{Figure~\ref{f_highEErr} in Appendix~\ref{response_appendix}\ref{syst_highe} summarizes
the uncertainties affecting the high energy extrapolation of response.}
More details about the high energy extrapolation can be found in Ref.~\cite{IncJets}.

%% file: etadep.tex
\section{Relative MPF response}
\label{sec:etadep}


\newcommand{\SFeta}[1]{\JESmathSym{SF_{\eta}^{\textnormal{#1}}}}

\newcommand{\pTCC}{\JESmathSym{\MakeTransverseExt{p}{}{\textnormal{CC}}}}

In the previous section we derived the absolute response for jets in the very
central part of the calorimeter.
The relative MPF response normalizes the response for jet energy as a function of pseudorapidity,
allowing the description of the response for jets in any part of the detector.
The derivation of this correction relies on events from two different processes,
\photonjet and \dijet production (see \cref{sec:relresp}).

The selection criteria for \photonjet{} events are identical to those used for
the measurement of absolute response (\cref{photonjet_selection}),
with the exception that the probe jet is not restricted to the central calorimeter.
The samples used for the determination of the \dijet{} correction in MC
are described in \cref{mc_samples}.

The selection of \dijet{} events closely follows that of \photonjet{} events,
with one of the jets playing the role of the photon. 
Events with no reconstructed PV or with more
than four PVs are rejected. 
Events must have exactly two reconstructed jets with $\rcone=0.7$ or $0.5$ as appropriate, 
satisfying the jet selection criteria described in \cref{jetid}.
Events with additional reconstructed jets that do not satisfy those criteria are rejected.
At least one of the jets must be within $\mdeta \leq 0.4$, so that its core is well 
contained inside the central calorimeter. This jet is referred to as the ``tag'' 
whereas the other jet is referred to as the ``probe''. If both jets have $\mdeta < 0.4$
both possibilities for tag and probe assignments are considered.
To avoid a bias from the trigger, the uncorrected transverse momentum
of the tag jet is required to be large enough so that the efficiency of
the trigger for such jets is above $98\%$.
The jets are required to be back-to-back in the $\radius$-$\azim$ plane, \ie,
the difference of their azimuthal angles, $\func[\Delta\azim]{\textnormal{probe},\textnormal{tag}}$, 
must be larger than \rad{3.0}.
Events with muons that are cosmic ray candidates are rejected. 
To further reduce cosmic rays, the ratio of the measured \MET{} over 
the \pT{} of the most energetic jet is required to be $\MET / \pT < 0.7$.

\subsection{Method}
\label{ssec:etadep:method}


The relative response correction, \Feta, is evaluated for jets up to $\mdeta \leq 3.6$.
The samples are split according to the pseudorapidity of the probe jet,
each group typically spanning $\Delta\deta = 0.1$,
with the exception that the range is narrower (0.05) for the ICD region,
where the response varies rapidly with \deta, and wider (up to 0.4) starting with $\mdeta > 2.0$
in order to compensate for decreased statistics.

Although similar in spirit, the treatment of \photonjet{} and \dijet{} events differs
in some details due to the different nature of the samples.
The following sections describe their treatment separately.

%
%

\subsubsection{Relative response in the \photonjet{} sample}
\label{feta_gj_procedure}

The MPF response \MPFeta{} (see \cref{mpf_method}) is estimated as the average
over all the events in each \Eprime{} bin,
similar to the procedure described in \cref{absolute_response_method}, 
in each $\deta$ region independently.
In the case of data,
the estimated response does not correspond exactly to the \photonjet{} response
because of the contamination by \dijet{} events
and the imperfect calibration of the photon energy,
as discussed in \cref{absolute_response_method}.
In the case of MC, the measured MPF response for a pure \photonjet{} sample
is known at the particle level directly from the MC
information. This response is denoted as \MPFeta{\photonjet}.

The relative MPF response correction is computed as:
\begin{equation}
	\Feta[\photonjet] = \frac{\func[\MPFeta{\photonjet}]{\Eprime}}{\func[\MPFCC{\photonjet}]{\Eprime}}.
	\label{feta_gj_points}
\end{equation}
The denominator, measured in \cref{response_results}, represents the MPF response of a jet with 
approximately the same energy as the probe,
but measured in the central calorimeter.

\begin{figure}[th]
	\subfloat{
		\includegraphics[width=\columnwidth]{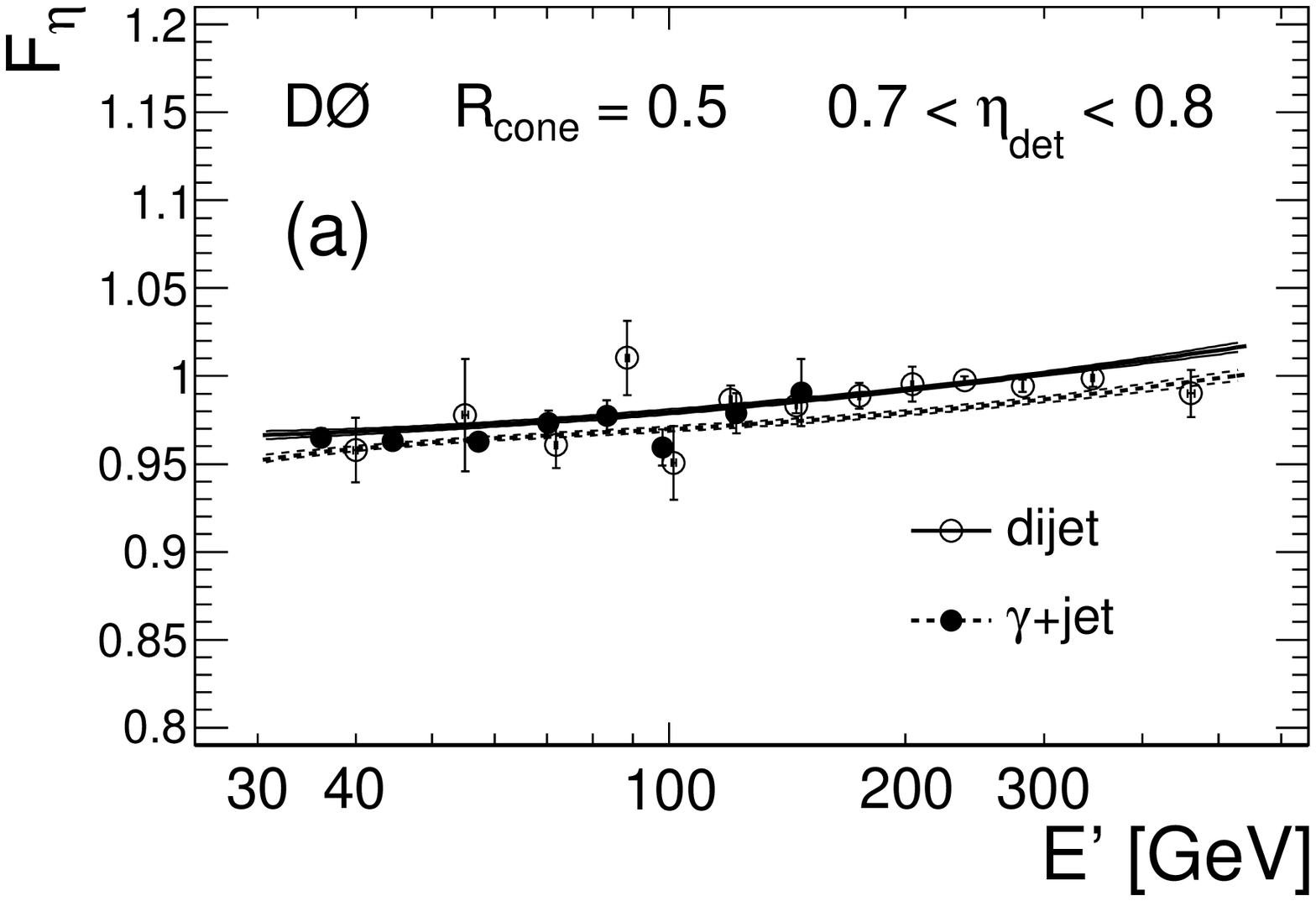}
		\label{example_feta_central}
	}
	\\
	\subfloat{
		\includegraphics[width=\columnwidth]{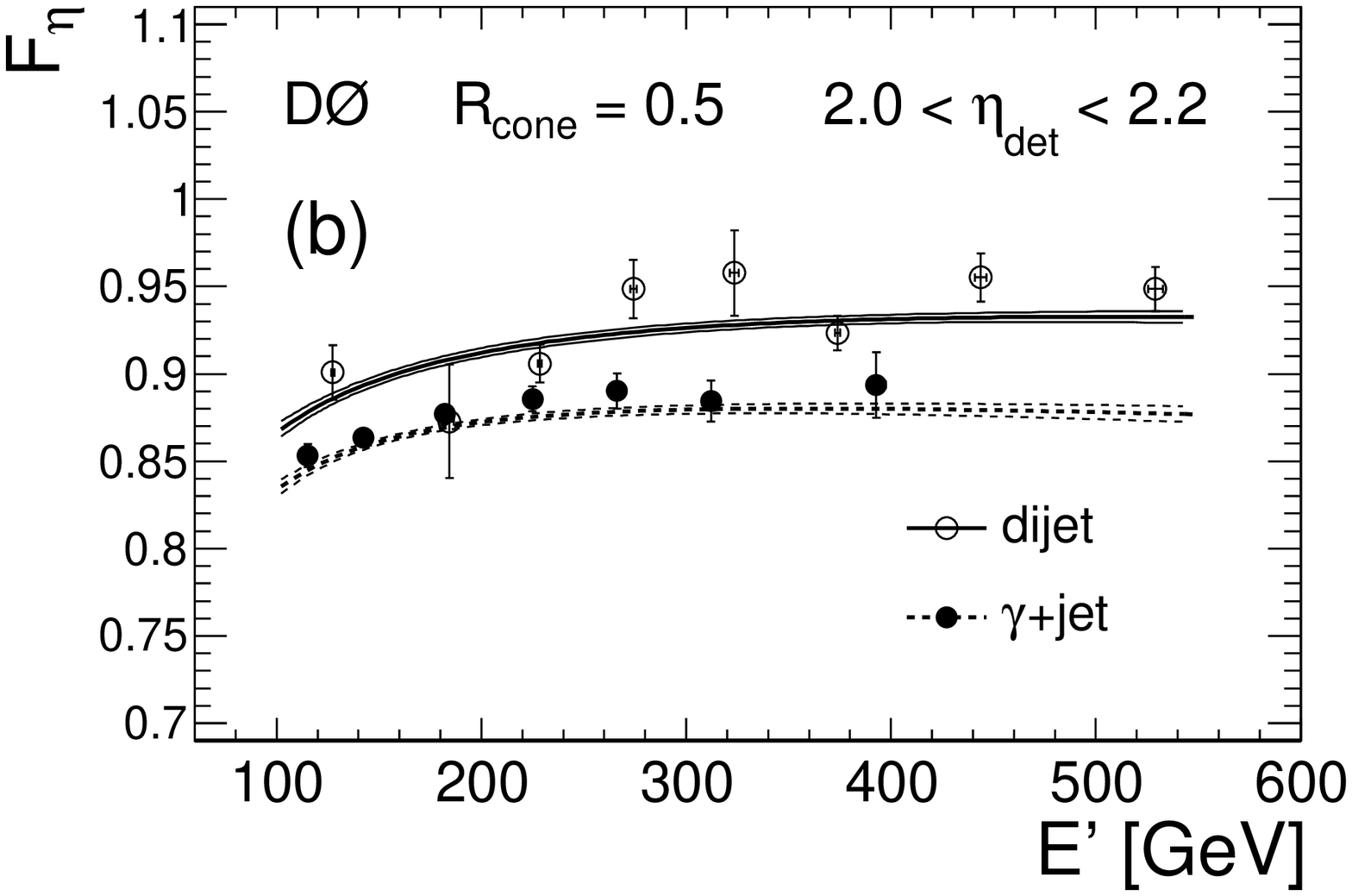}
		\label{example_feta_forward}
	}
	\caption{
		Relative MPF response for $\rcone = 0.5$ jets in 
		data as a function of \Eprime{} for jets with
		\protect\subref{example_feta_central}~$0.7 < \deta <0.8$
		and \protect\subref{example_feta_forward}~$2.0 < \deta < 2.2$
		(note the different \Eprime{} scale), measured in the \photonjet{} and \dijet samples.
		The curves represent the result from the global fit discussed 
		in \cref{global_fit} with statistical uncertainty shown by the error bands.
	}
	\label{example_feta}
\end{figure}

\Cref{example_feta} shows the \Feta[\photonjet] values measured from
\photonjet events as a function of \Eprime{} in two different $\deta$ regions.
The measured \Feta[\photonjet] differs from \cref{feta_gj_points} only in that it is not
corrected for background contamination
(\cref{sssec:response:background_corrections}),
for the reason that will be explained in \cref{global_fit}.

\subsubsection{Relative response in the dijet sample}
\label{feta_dijet_procedure}

The procedure to extract the relative response \Feta{} from the \dijet{} sample is complicated by the
presence of an uncalibrated jet as the reference object.
For a \dijet event, we measure the MPF response of a probe jet located at $\deta$ relative to the response of
the central tag jet ($\abs{\deta[tag]} < 0.4$), following \cref{mpf_gj}, as:
\begin{equation}
R_{\textnormal{relMPF},\eta}^\textnormal{\dijet} = 1 + \frac{\vecmet \cdot \MakeTransverse{\vec{n}}{\textnormal{tag}}}{\pT[tag]} ,
\label{mpfdijets}
\end{equation}
\noindent where \vecpT[ tag] is the measured transverse momentum of the tag jet,
corrected only for offset energy subtraction.
This observable can also be interpreted according to \cref{mpf_method} as:
\begin{equation}
	R_{\textnormal{relMPF},\eta}^\textnormal{\dijet} \simeq
		\frac{\func[\MPFeta{\dijet}]{\Eprime}}{\func[\MPFCC{\dijet}]{\pTprime}}, 
\label{eq:MPFrelInterpretation}
\end{equation}
\noindent where \func[\MPFeta{\dijet}]{\Eprime} and \func[\MPFCC{\dijet}]{\pTprime}
denote the MPF response of the probe and tag jets, respectively.
The transverse momentum of the tag jet \pT[tag], the response $R_{\textnormal{relMPF},\eta}^\textnormal{\dijet}$, 
and also \Eprime and \pTprime,
are corrected for the jet resolution bias (\cref{resolbias_overview}).

%

\Feta[\dijet] is defined similarly to \cref{feta_gj_points} for the \photonjet events:
\begin{equation}
	\Feta[\dijet] = \frac{R_{\textnormal{relMPF},\eta}^\textnormal{\dijet}\:\func[\MPFCC{\photonjet}]{\pTprime}}
		{\func[\MPFCC{\photonjet}]{\Eprime}} \cdot r_{\Eprime} ,
	\label{feta_dijet_points}
\end{equation}
where the response of the probe jet in the numerator is compared to the response of the tag jet in the denominator.
If the \dijet responses were used consistently in the \Feta[\dijet] definition, its average
in the $\mdeta \leq 0.4$ region would be unity by construction.
Since we do not measure the absolute response from \dijet events,
the calibration of the tag jet is instead based on the
\photonjet response, \MPFCC{\photonjet}.
The factor $r_{\Eprime}$ corrects for the difference between the two types of
response in \photonjet and \dijet events.
It is evaluated by enforcing \Feta[\dijet] to be on average unity
for the central region $\mdeta \leq 0.4$, 
thus ensuring $F_{\eta \approx 0}^{\textnormal{\dijet}} \approx 1$.
The deviation of the factor $r_{\Eprime}$ from unity is found to be always smaller than $0.5\%$.
%
%
\Cref{example_feta} shows the measured \Feta[\dijet] 
(\cref{feta_dijet_points}) as a function of \Eprime{} in two different $\deta$ regions.

\subsubsection[Sample dependence]{Sample dependence of the relative MPF response}
\label{feta_gj_vs_dijet}

The measured correction \Feta{} is significantly 
different for the \photonjet{} and \dijet{} samples at large pseudorapidities,
as illustrated in \cref{example_feta_forward}.
The main contribution to the discrepancy, 
particularly for large $\deta$, 
is from the different parton flavor composition of the \photonjet{} and \dijet{} samples. Whereas the
leading jet in \photonjet{} events originates predominantly from quarks at low energy
and gluons at high energy, this trend is reversed in the \dijet{} sample.
The different fragmentation of quarks and gluons results in a lower expected
response for jets from gluons, owing to their softer spectrum of particles. 


In a given $\deta$ range, the ratio of relative responses
in \photonjet{} and \dijet events is found to be nearly independent of \Eprime{}
over the range where both samples overlap. We therefore define the 
scale factor \SFeta{} as:
\begin{equation}
	\SFeta{} = \frac{\func[{\Feta[\dijet]}]{\Eprime}}{\func[{\Feta[\photonjet]}]{\Eprime}},
	\label{sf_eta}
\end{equation}
\noindent which will be the key to combine both sets of measurements in \cref{global_fit}.

\label{photoncorr_vs_eta}

The photon energy corrections for the selected \photonjet{} events in data
with the jet in the central calorimeter are discussed in \cref{absolute_response_method}.
For a jet at a given \deta, the correction is given by \cref{eq:kappa_eta},
which only depends on $\eta$ because of the expected sample purity
$\rho_{\eta}$ and the ratio $\MPFeta{\dijet}/\MPFeta{\photonjet}$
in $c_{\textnormal{bckg,$\eta$}}$. 
The estimation of the purity is described in \cref{sssec:PhotonJetPurity}.
For the ratio of $\MPFeta{\dijet}/\MPFeta{\photonjet}$, large MC
samples are required for a stable determination as a function of $\deta$.
Instead, a different approach is followed which,
in addition to statistical stability, reduces the dependence on MC modeling.

Under the assumption that the response of the recoil against the central tag object in
\dijet{} events is independent on
whether such an object is a jet or a $\gamma$-like jet, the following relation holds:
\begin{equation}
	\frac{\func[\MPFeta{\dijet}]{\Eprime}}{\func[\MPFeta{$\photon$+jet}]{\Eprime}} =
	\SFeta{}\,\frac{\func[\MPFCC{\dijet}]{\pTprime}}{\func[\MPFCC{$\photon$+jet}]{\pTprime}},
	\label{ratio_pred}
\end{equation}
\noindent where the ratio $\MPFCC{\dijet} / \MPFCC{\photonjet}$ is estimated
in \linebreak[4] \cref{sssec:response:background_corrections},
and \SFeta{} is defined by \cref{sf_eta}.
The validity of this approximation has been verified in MC by comparing the measured 
$\MPFeta{\dijet} / \MPFeta{\photonjet}$ to the prediction
given by \cref{ratio_pred}.

\subsubsection{Global fit to \photon{}+jet and dijet samples}
\label{global_fit}

In each $\deta$ range, the relative MPF response correction in the \photonjet{} sample is modeled as:
\begin{multline}
		\func[{\Feta[\photonjet]}]{\Eprime;\{\qlpar[i]\}}
			= \frac{\func[\MPFeta{\photonjet}]{\Eprime;\{\qlpar[i]\}}}
			{\func[\MPFCC{\photonjet}]{\Eprime}}\\
			= \frac{\qlpar[0]+\qlpar[1]\log(\Eprime/E_0)+\qlpar[2]\log^2(\Eprime/E_0)}{\func[\MPFCC{\photonjet}]{\Eprime}},
	\label{feta_gj}
\end{multline}
\noindent where $\func[\MPFCC{\photonjet}]{\Eprime}$ has been determined in \cref{sec:response},
$E_0$ is a constant, and \qlpar[i] ($i=0,1,2$) are coefficients 
to be estimated from the measurements in each $\eta$ bin. 
These coefficients correspond to the MPF response for
the pure \photonjet{} sample,
whereas
the measurements in the \photonjet{} sample do not have
the background contamination corrections applied yet.

Following \cref{sf_eta}, the relative MPF response correction in the \dijet{} sample is modeled as:
\begin{equation}
	\func[{\Feta[\dijet]}]{\Eprime;\{\qlpar[i]\},\SFeta{}} = \SFeta{}\cdot\func[{\Feta[\photonjet]}]{\Eprime;\{\qlpar[i]\}}
	\label{feta_dijet}
\end{equation}
\noindent with the additional \SFeta{} coefficient to be determined.

In each of the 66 $\deta$ subsamples, only four $\eta$-dependent parameters $\braces{\qlpar[0], \qlpar[1], \qlpar[2], \SFeta{}}$ are required to 
define \Feta[\photonjet] and \Feta[\dijet].
All the parameters can be estimated from a
simultaneous fit (``global fit'') to the measurements in the \photonjet{} and \dijet{} samples
(see \cref{example_feta}).
The \photonjet{} and \dijet{} samples together provide several hundred \Feta{}
measurements to constrain the parameters. The simultaneous fit extends
the coverage of the measurement to high energies 
and to large pseudorapidities exploiting the advantages of the \photonjet{} and \dijet{} events, respectively.
The larger coverage yields higher precision for the measurement of the response
in these regions.

The fitting function is given by:
\begin{equation}
	\Feta =
	\begin{cases}
		\Feta[\photonjet] / k_{\textnormal{R},\eta}^{\photon} & \text{if \photonjet,} \\
		\Feta[\dijet]  & \text{if \dijet}
	\end{cases}
	\label{global_feta}
\end{equation}
\noindent where $k_{\textnormal{R},\eta}^\photon$ is a correction factor
which takes into account the background contamination in the \photonjet{} measurements.
The factor $k_{\textnormal{R},\eta}^\photon$ depends (via \cref{eq:kappa_eta,ratio_pred}) 
on the actual \SFeta{} estimate.
This procedure reduces the MC modeling dependence of the photon corrections applied to the data measurements.

\begin{figure}
	\includegraphics[width=\columnwidth]{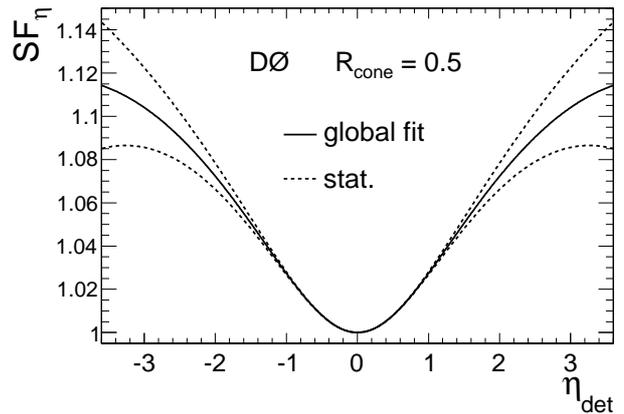}
	\caption{
		Parameterization for the scale factor \SFeta{} resulting from the global fit to the
		relative MPF response measurements in 
		data for $\rcone=0.5$ jets. The dashed lines illustrate the statistical
		uncertainty band.
	}
	\label{global_fit_param_data_jccb}
\end{figure}

Following the discussion above, a total of 264 parameters would have to be determined.
The very fine $\deta$ binning has the advantage of an accurate determination of the relative MPF response
correction in regions where the energy dependence changes quickly with $\deta$ (\eg, in the ICR region).
On the other hand, the limited statistics available in each of the bins can introduce
large fluctuations in the extracted parameters. To reduce the fluctuations and to ensure a smooth parameterization of the
relative MPF response correction in the $(\Eprime,\deta)$ plane, each of the four parameters (\qlpar[0], \qlpar[1], \qlpar[2] and \SFeta{}) is
expressed as a function of $\deta$, whose coefficients now become 
the actual parameters to be determined. For instance, \SFeta{} is found to be well described
by the following parameterization (\cref{global_fit_param_data_jccb}):
\begin{equation}
	\SFeta{} = 1 + b\,\func[\log]{\cosh \deta} + c\,\func[\log^{2}]{\cosh \deta} .
\end{equation}

This procedure reduces the total number of free parameters in data from 264 to less than 60, 
which are determined from the global fit.

\Cref{example_feta} shows two examples of the global fit result in data 
for $\rcone=0.5$ jets. The dashed and continuous lines represent
\Feta[\photonjet] (including background contamination)
and \Feta[\dijet], as shown respectively in the upper and lower
branch of \cref{global_feta}. As in the case of the absolute response, the
relative response is very similar for jets with $\rcone = 0.5$ and $0.7$.



\subsection{Results}
\label{etadep_results}

The relative response \Feta[\photonjet] as derived by the global fit 
describes the average response of the jets as produced by the \photonjet processes.
The correction \Feta[\dijet] can be used for samples where the parton final state is closer to that of \dijet processes.

\begin{figure}[th]
	\subfloat{
		\includegraphics[width=\columnwidth]{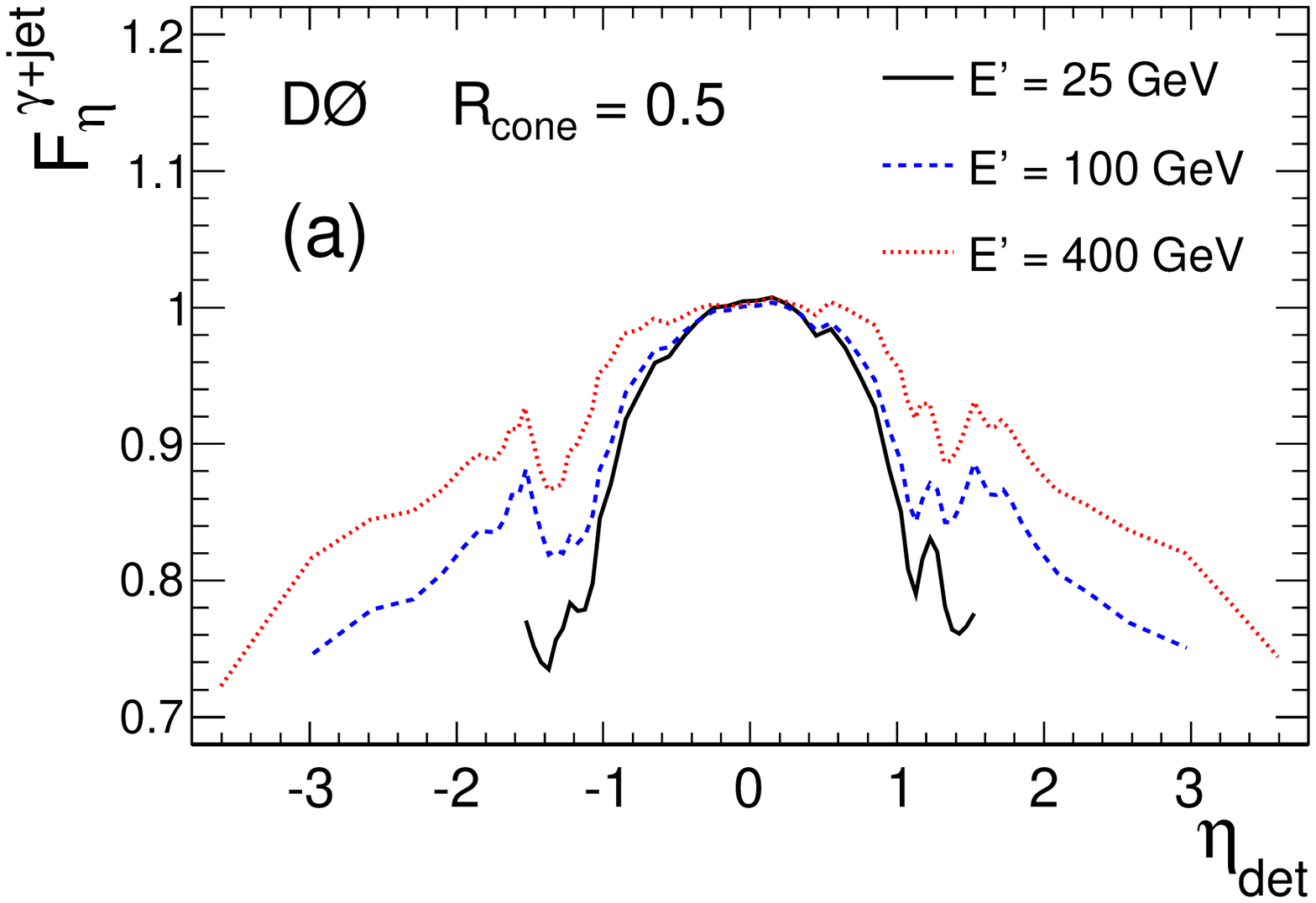}
		\label{feta_esummary_data_jccb_gamjet}
	}
	\\
	\subfloat{
		\includegraphics[width=\columnwidth]{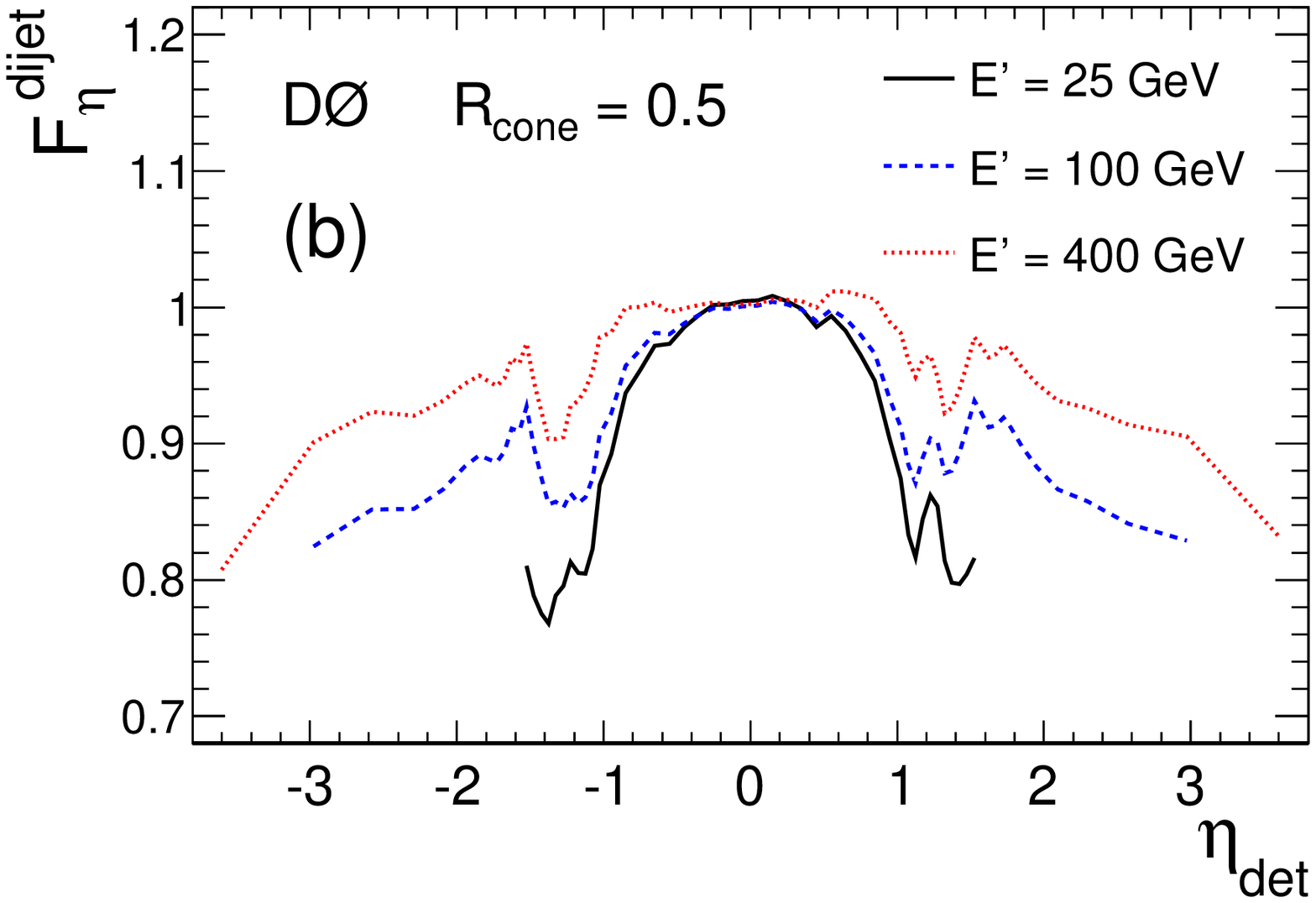}
		\label{feta_esummary_data_jccb_dijet}
	}
	\caption{
		Relative MPF response correction in data 
		for $\rcone=0.5$ jets as function of \deta{} and separately
		for \protect\subref{feta_esummary_data_jccb_gamjet}~\photonjet (\Feta[\photonjet])
		and \protect\subref{feta_esummary_data_jccb_dijet}~\dijet (\Feta[\dijet]) events,
		as given in \cref{global_feta}.
		Each line corresponds to a different value of \Eprime as indicated.
	}
	\label{feta_esummary_data_jccb}
\end{figure}

\Cref{feta_esummary_data_jccb} presents the final relative MPF response correction
in data for $\rcone=0.5$ jets 
for selected values of \Eprime.
This figure illustrates the non-uniform response of the calorimeter as a
function of $\deta$, especially in the ICR, with its complex geometry and rapid
variation of amount of inactive material in front of the calorimeter
(see \cref{sec:relresp}).
The measured relative MPF response for $\rcone=0.7$ jets is very similar.

\label{feta_closure_test}

\begin{figure}
	\includegraphics[width=\columnwidth]{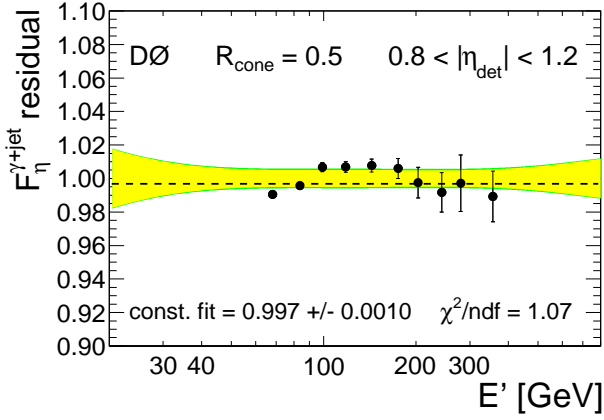}
	\caption{
		(color online) Residual \Feta{} for $\rcone=0.5$ jets in \photonjet data as a function of \Eprime{}
		for $0.8 \leq \mdeta < 1.2$.
		The yellow shaded band represents the statistical uncertainty from the global fit,
		while the green band shows the additional uncertainty described in \cref{feta_closure_test}.
	}
	\label{Fig:Eta_FetaBin_JCCB_Closure_Example}
\end{figure}

A self-consistency test is performed to verify the effectiveness
of the parametrized $\eta$-dependent jet response in correcting the jet energies.
The same samples used in the derivation of \Feta{}, \photonjet{} and \dijet{},
are split into $\Delta \deta = 0.4$ subsamples, wider than in the derivation,
in order to achieve higher statistical precision.
For the test, the \MET{} used for the MPF (\cref{mpf_method}) is corrected
using \Feta{} from the global fit, and then \Feta{} is measured again.
The result is a residual \Feta{} correction that is expected to be consistent
with unity.
An example of the consistency test with $\rcone=0.5$ jets is displayed in
\cref{Fig:Eta_FetaBin_JCCB_Closure_Example}.
An additional uncertainty is added to the 
relative MPF response
to cover the presence of the small residual \Feta.
This uncertainty is also shown in \cref{Fig:Eta_FetaBin_JCCB_Closure_Example}.

%

\subsection{Uncertainties}
\label{etadep_uncertainties}

The systematic uncertainties on the relative response for the \photonjet sample (\Feta[\photonjet])
include the residual \Feta correction (\cref{feta_closure_test}) and the uncertainty on the
correction for background contamination (\cref{sssec:response:background_corrections}).
The systematic uncertainties on \Feta[dijet] also include
the residual \Feta correction, plus uncertainties from resolution bias correction
and high energy extrapolation (described in \RefCite{IncJets}).

\begin{figure}
	\includegraphics[width=\columnwidth]{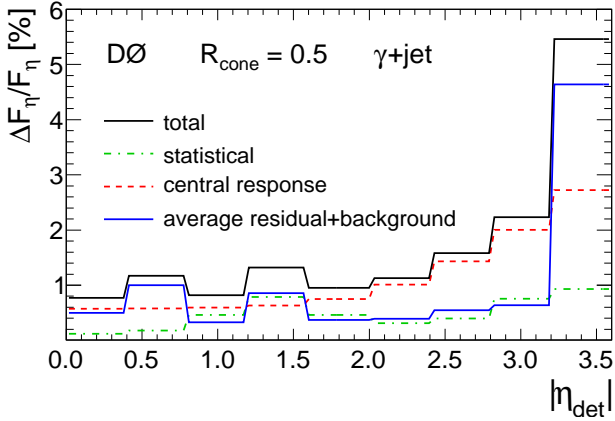}
	\caption{
		Uncertainties on the relative MPF response correction in \photonjet data for $\rcone=0.5$ jets,
		as a function of $\mdeta$
		for $\pTprime = \GeVc{50}$.
		The bold solid line shows the total uncertainty, split into individual contributions:
		statistical, average residual and background correction combined,
		and central response correction.
	}
	\label{feta_error_summary_data_jccb}
\end{figure}

\Cref{feta_error_summary_data_jccb} shows as an example the summary of the uncertainties
on the response correction for \photonjet data,
including contributions from the absolute response (\cref{response_uncertainties}),
the statistical uncertainty on \Feta[\photonjet] from the global fit,
and the combination of the systematic uncertainties on \Feta[\photonjet].
Uncertainty on \Feta[dijet] are typically of similar size as for
\Feta[photonjet], being a little smaller for high \mdeta and larger energy,
where larger \dijet statistics is available.


%% file: mpfbiases.tex
\section{MPF response bias corrections}
\label{sec:mpfbiases}

The determination of the MPF response suffers
from three main biases (see \cref{mpfbias})
that need to be corrected to recover the true jet
response. The first bias affects the response measurement in data only, and is
related to the imperfect calibration of the measured photon \pT{} as well as the
presence of \dijet{} background contamination in the selected \photonjet{} sample.
The correction for this bias was discussed in \cref{sssec:response:background_corrections}.
This section presents the correction for the other two biases:
from zero suppression and from event topology, that affect the
determination of response in both data and MC.

\subsection{Zero-suppression bias correction}
\label{zs_bias_corr}

Zero suppression, which introduces the bias on the offset energy estimator
described in \cref{offset_bias_correction}, also biases the estimator of the
MPF response. Since the cells in the jet cone are more likely to pass the
suppression threshold than those outside the cone, the \MET{} in the
direction of the jet is reduced, artificially increasing the estimated MPF response.

In analogy with \kOZS in \cref{offset_bias_correction}, a correction factor for the
MPF response, \kRZS, is also estimated in MC
using \photonjet samples where the same generator-level events
are processed with and without ZB overlay. 
The correction factor is defined in \cref{eq_kOffset}.

This factor is measured using the events selected with the same kinematic cuts 
as for \kOZS{} determination with the exception that 
the events should contain exactly one jet and exactly 
one photon back-to-back to the jet with 
$\abs{\Delta\azim(\photon,\textnormal{jet})} > \rad{3.0}$,
and either one or two reconstructed PVs,
following the requirements used in the derivation of the MPF response.

\begin{figure}[t]
	\includegraphics[width=0.45\textwidth]{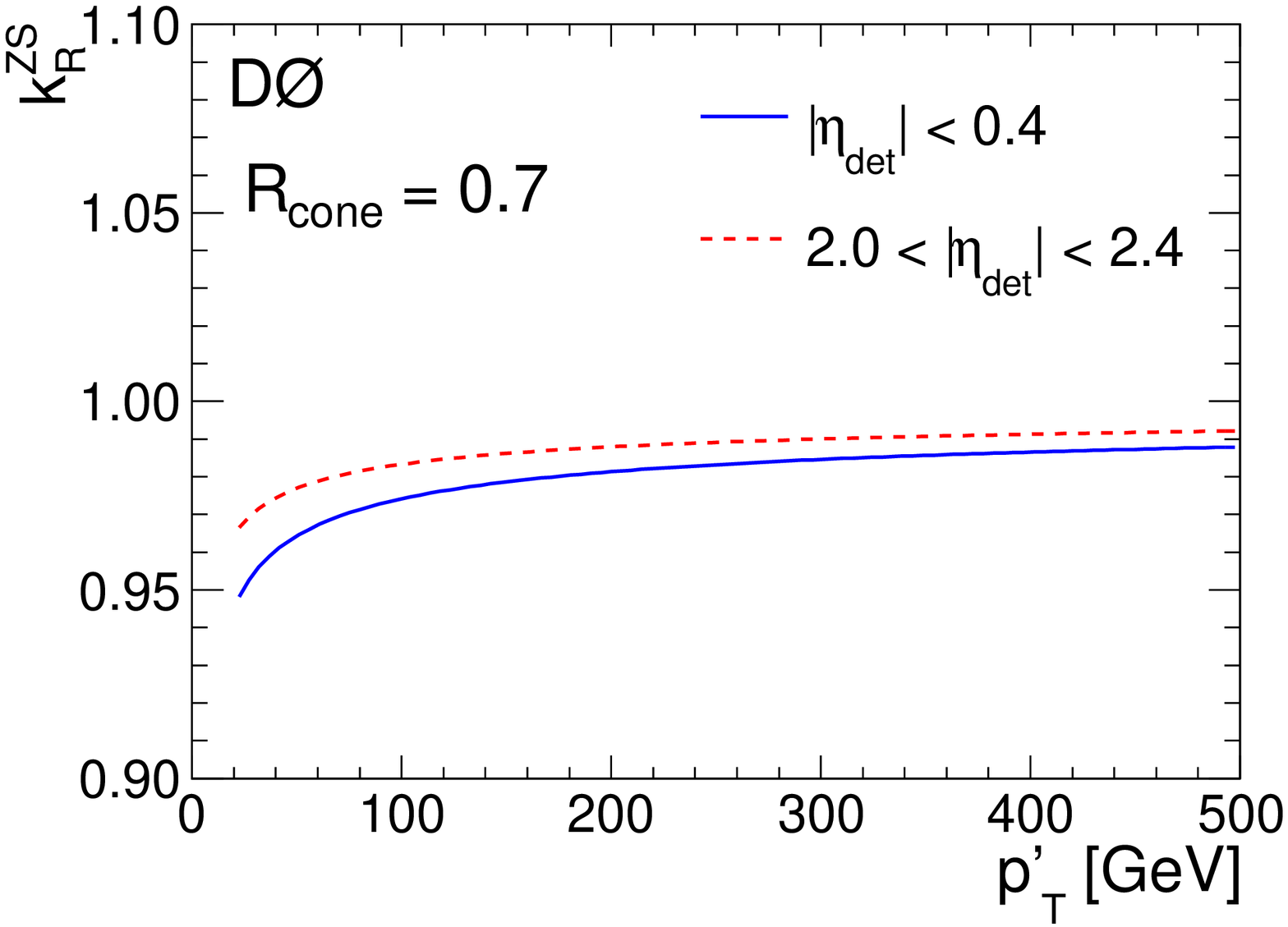}
	\caption{
		\kRZS{} correction factor as a function of \pTprime{}
		for $\rcone=0.7$ jets 
		in two different \mdeta{} regions.
	}
	\label{kRunsuppJCCA}
	\label{kRunsuppJCCAbin0}
	\label{kRunsuppJCCAbin5}
\end{figure}

\begin{figure}[t]
	\includegraphics[width=0.45\textwidth]{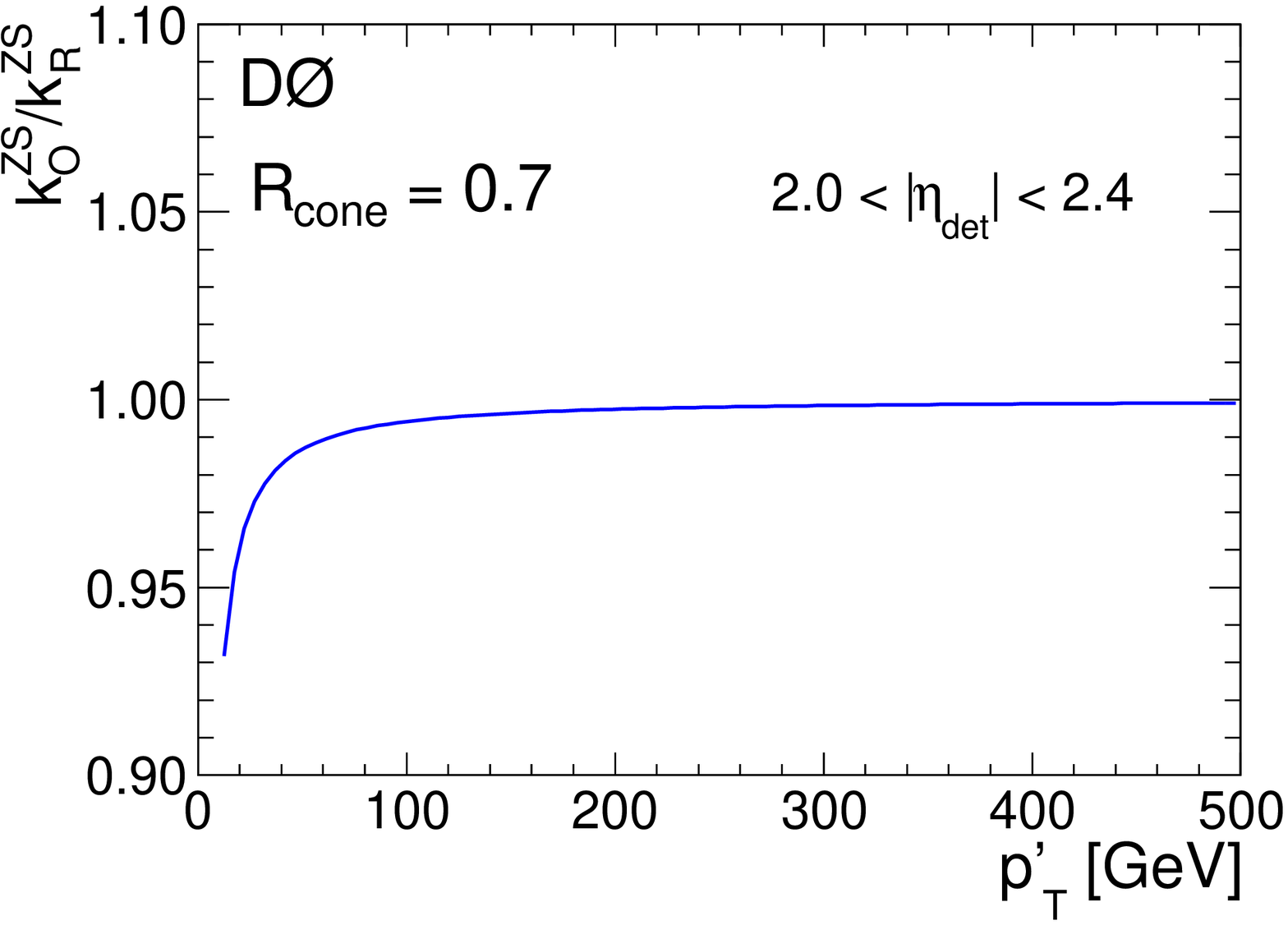}
	\caption{
		$\kOZS/\kRZS$ correction factor as a function of \pTprime{}
		for $\rcone=0.7$ jets in $2.0 < \mdeta < 2.4$.
	}
	\label{kOkRunsuppJCCA}
	\label{kOkRunsuppJCCAbin0}
	\label{kOkRunsuppJCCAbin5}
\end{figure}

An example of the response correction factor for $\rcone=0.7$ jets 
is shown in \cref{kRunsuppJCCAbin0} for two different \mdeta regions. 
The correction factor for $\rcone=0.5$ jets is almost identical. 

The bias arising from the zero-suppression effect on the MPF response is highly correlated with 
the bias on the offset estimation 
discussed in \cref{offset_bias_correction}.
Both are smaller than unity and have the same magnitude (\eg{}, compare 
\cref{jcca_kO_bin0} and \cref{kRunsuppJCCAbin0}),
and therefore they partially cancel in the overall response correction.
%
It is convenient to parametrize the $\kOZS/\kRZS$ ratio, 
which enters \cref{master3} directly.
As an example, \cref{kOkRunsuppJCCA} illustrates the extracted $\kOZS/\kRZS$ factor 
as a function of \pTprime for $2.0 < \mdeta < 2.4$.
The variation of the $\kOZS/\kRZS$ correction is within 1\% over the e\mdeta range.


\subsection{Topology bias correction}
\label{topo_bias_corr}

After applying the two bias corrections above, the MPF 
estimate of the response to the hadronic recoil against the particle-level
photon may still differ from the true jet response
due to a number of physics and instrumental effects.
The goal of the topological bias correction 
is to compensate for the net effect of all the remaining  contributions. 

An example of physics-related bias is soft radiation below
the jet reconstruction threshold of $\pT=\GeV{6}$, despite the stringent
\photonjet{} selection requiring exactly one jet and $\abs{\Delta\azim(\photon,\textnormal{jet})} > \rad{3.0}$.
This can spoil the \pT balance between the jet and the photon.
Depending on whether such radiation
populates the photon or jet hemispheres, the estimated MPF response can be higher or lower
than the true jet response. Another example is the fact that, owing to the
shrinkage of the rapidity space
shrinking in angle of constant intervals of rapidity
(especially for forward jets), the hadronic recoil can cover significantly larger 
physical space than the reconstructed jet. Since the particles outside the jet cone are of lower energy 
than in the core, the estimated MPF response is \emph{a~priori} lower than the actual response to the jet. 
This difference can also be increased by the larger effect of zero suppression on low 
energy calorimeter deposits. Finally, the MPF method inherently relies on \pT balance and therefore is in principle more suitable
for jet \pT, rather than energy, calibration. The difference between jet \pT and energy calibration
is the largest for low energy jets, where jets are wider and mass effects can be sizable. 
The MPF method can absorb
instrumental effects unrelated to energy calibration, such as the rapidity bias in the ICD region 
(see \cref{ssec:qcd:RapidityCalibration}).
\begin{figure}
	\centerline{\includegraphics[width=0.85\columnwidth]{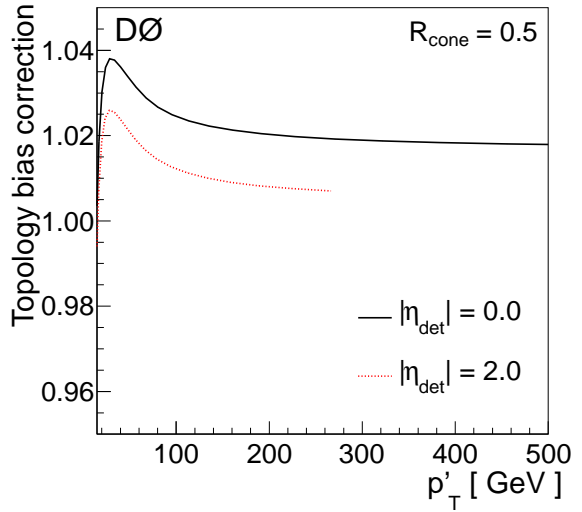}}
	\caption{Example of the topology bias correction \kR[topo] for $\rcone=0.5$ jets at
		$\mdeta = 0$ and $2$. The lines represent
		the result of a smooth parameterization  of the correction.
	}
	\label{f_mpfbias_example}
\end{figure}

The net bias correction factor, denoted by \kR[topo],
is estimated from \photonjet{} MC samples without ZB overlay, selected using the
same criteria as for the absolute response measurement (\cref{photonjet_selection}).
As indicated in \cref{eq_mpfbias}, it is defined as the ratio of the true jet response (\cref{rtrue}) 
and the MPF response with respect to the particle-level photon (\cref{eq_rmc}). 
The true jet
response is estimated as the ratio of the average visible energy in the calorimeter from particles belonging to the
particle jet to the average particle jet energy
(the particle jet is required to match the direction of the reconstructed jet within $\deltar < \rcone/2$).
\Cref{f_mpfbias_example} shows an example of the topology bias corrections for $\rcone=0.5$ jets at $\mdeta = 0$ and $2$.
The corrections for $\rcone=0.7$ are closer to unity.
These corrections are also derived from a MC simulation
using a tuned single particle response (see \cref{sec:fdc}), to be applied to jets in data.
We assign a systematic uncertainty that covers
the difference between the topology bias correction for the standard MC and
the MC with the tuned single particle response.
This difference is smaller than $0.5\%$ for central jets and up to $\approx 1\%$ for forward jets.
%
The list of uncertainties also includes the statistical
uncertainty ($\lesssim0.5\%$) and systematic uncertainty from
varying the matching criterion between the
reconstructed and particle jets ($0.2\%-0.5\%$), varying the hadron response,
and considering alternative physics models, as described below.



The particle-level \pT balance in \photonjet{} events can be modified by 
initial and final state radiation
or by additional soft radiation caused by spectator parton interactions
(``soft underlying event''). 
%
To estimate model dependence, the \kR[topo] correction is obtained using
three sets of \PYTHIA{} parameters, called Tune~A, Tune~B and Tune~DWT.
Tune~A and B are tuned to the CDF~\I{} data~\cite{tuneA}.
Tune~A allows for more initial state radiation than Tune~B. Consequently, the 
contribution of the soft underlying event is smaller in Tune~A than in Tune~B.
\DZero{} \II{} data on \dijet{} azimuthal decorrelations~\cite{dphi} show 
a lack of sufficient initial state radiation in Tune~B while there is too much radiation 
in Tune~A. 
Tune~DWT~\cite{tuneDWT} has been developed to provide an improved description of this observable.
The systematic uncertainty due to physics modeling has been estimated as
the maximum observed difference with respect to \PYTHIA{} Tune~A.
It is smaller than $0.4\%$ for central jets, increasing up to $2\%-3$\% for forward jets.

%% file: showering.tex
\section{Showering correction}
\label{sec:showering}

\newcommand{\Strue}{\ensuremath{S}}

After implementing the offset and full set of response corrections discussed in 
\cref{sec:offset,sec:response,sec:etadep,sec:mpfbiases},
the corrected jet energy does not yet correspond to the particle-jet 
energy. Indeed, after offset subtraction, not
all the energy contained inside the jet cone originates from particles belonging
to the particle jet, so the response correction can not recover the
original particle jet energy.
Particles not belonging to the particle jet 
(\eg, from the underlying event) may contribute to the energy inside the jet
cone, due to effects such as the shower development
from interactions with the detector material, the granularity and pseudo-projective
arrangement of the calorimeter towers, as well as the bending of low momentum 
charged particles in the magnetic field.
The same instrumental effects also cause some of the energy 
from particles belonging to the particle jet to leak outside the jet cone. Therefore,
a ``showering correction'' is required 
to compensate for the
net energy flow through the jet cone boundary. Such a correction must be defined
in a way consistent with 
the rest of corrections to ensure
that the particle jet energy is recovered. The definition of this showering
correction is given as \Strue{} in \cref{strue}.

\subsection{Method}

The showering correction is determined both in data and MC using
\photonjet{} events selected using the same criteria as for the
absolute response measurement (see \cref{photonjet_selection}), 
with the exception that the probe jet is not restricted to be in 
the central calorimeter. The procedures to estimate 
the showering correction in data and MC are different. In the case of MC, it is
possible to directly obtain an unbiased estimator of the true showering correction.
In the case of data, an observable sensitive to the jet showering must be defined, 
resulting in a potentially biased estimator of the showering correction.
The estimator must be calibrated to remove the bias.
The following sections present an overview of both procedures.
%

\subsubsection{Monte Carlo method}
\label{mc_method}

In the case of MC, the showering  correction is estimated in simulated \photonjet{}
events without ZB overlay (\ie{}, with no offset energy). Since detailed information 
is available in the simulation regarding the amount of energy deposited in each calorimeter 
cell by each  particle, it is possible to directly estimate \Strue{} 
according to \cref{strue}. In the absence of offset effects, the numerator of the showering correction
represents the uncorrected jet energy as determined by the jet algorithm. The denominator is
estimated by adding the visible energy in all the calorimeter cells from 
the particles originating
from the particle jet. 
Therefore, the measurement of \Strue{}
in MC requires a spatial matching between the calorimeter probe jet 
and the particle jet that 
is required to be within $\deltar<\rcone/2$.  
 
\subsubsection{Template-based method}
\label{template_method}

The measurement of the showering correction in data is based on examining the energy 
distributions in the calorimeter in annuli of increasing radius $\deltar$
with respect to the jet axis. We refer to such distributions as the ``jet energy profile''.
These are obtained by combining cells into towers 
following exactly the same procedure as the jet algorithm (see Par.~\ref{objectid_jets}),
and then adding the energy from all towers within a particular $\deltar$ annulus.

\begin{figure}
	\includegraphics[width=\columnwidth]{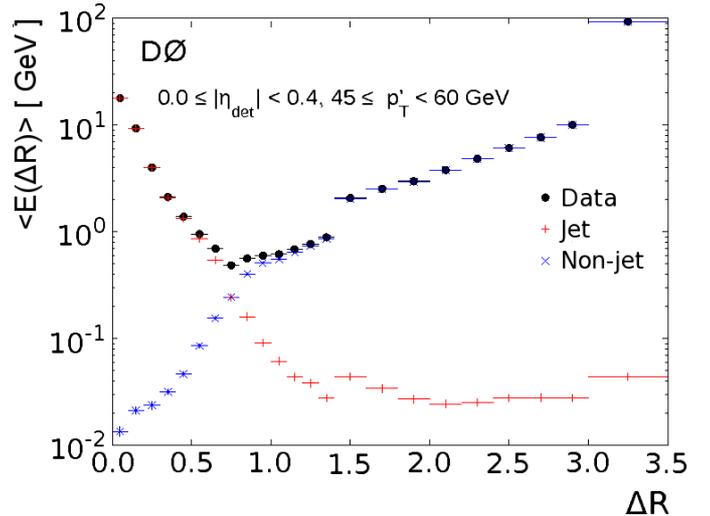}
	\caption{
		Jet energy profiles as a function of distance from the jet axis $\deltar$
		for MC and data.
		The jets in data are corrected for offset energy from noise and additional \ppbar{}
		collisions and are compared to MC jets without offset (called ``jet'' on the plot) 
		and contributions from the underlying event (``non-jet'').
		The MC templates are weighted with their fractions obtained from the fit to data.
	}
	\label{jetprofile_mc_example} 
\end{figure}

\Cref{jetprofile_mc_example} shows an example of the jet energy profile for central jets 
in \photonjet{} data and MC without ZB overlay. 
Exploiting the available MC information, it is possible
to compute the energy profiles corresponding to the particles belonging to the particle jet 
matching the reconstructed jet (``particle-jet profile'') and the rest of the particles
(``non-particle-jet profile''). The latter receives contributions from the underlying event as
well as particles resulting from large-angle gluon radiation during the parton shower evolution.
This figure helps to visualize the need for the showering correction.
The integral of the total reconstructed jet profile up to $\deltar=\rcone$, \Emeas,
represents the uncorrected jet energy as reconstructed by the jet algorithm, which receives 
contributions from both particle-jet and non-particle-jet profiles: 
\begin{equation}
\Emeas \!=\! \Emeas[ptclj](\deltar\!\!=\!\!\rcone) +
\Emeas[non-ptclj](\deltar\!\!=\!\!\rcone).
\label{emeas_jet}
\end{equation}
On the other hand, the integral of the particle-jet profile, 
$\Emeas[ptclj](\text{all}\,\deltar)$, 
represents the total visible energy 
from the particle jet, a small fraction of which is deposited beyond the jet cone boundary.
The ratio of the two integrals represents an estimator of the true showering correction:
\begin{equation}
	\hat{S} 
	= \frac{\Emeas}{\Emeas[ptclj](\text{all}\,\deltar)},
	\label{strue_hat}
\end{equation}
where $\Emeas$ is defined in \cref{emeas_jet}.
The distinct spatial distribution of energy around the jet
centroid for each of these two contributions, as shown in \cref{jetprofile_mc_example}, suggests 
that the showering correction can be estimated from a fit to the total jet profile,
using the particle-jet and non-particle-jet profiles extracted from MC as templates. 
This requires a good description of the jet profile in the MC, including proper modeling
of both the physics and instrumental effects. 
The \photonjet{} MC samples used are generated using \PYTHIA{} Tune~A~\cite{tuneA}, which has 
been verified to successfully describe the jet shapes in inclusive jet production~\cite{jetshapes}.
The level of agreement observed in this measurement confirms this is also the case in \photonjet{} events.

It is also necessary to include  a template describing the offset energy profile.
Such an offset template is estimated in \photonjet{} MC by subtracting from each template in the sample
including unsuppressed ZB overlay the corresponding template in the sample without ZB overlay. 
This takes into account distortions to the template shape related to the interplay between zero-suppression
and the presence of offset energy. Since the overlay in MC is based on ZB data events, the estimated offset profile 
is expected to closely match the one in data. 
The  three profiles obtained using the particle-jet, non-particle-jet and offset templates
are fitted to data to determine the contributions of
$E_\textnormal{ptclj}^{\textnormal{meas}}(\deltar=\rcone)$
and $E_\textnormal{non-ptclj}^{\textnormal{meas}}(\deltar=\rcone)$.
\Cref{jetprofile_mc_example} compares the spatial energy profiles in data with subtracted offset 
contribution and the MC templates.
The MC templates are weighted with their fractions obtained from the  maximum likelihood fit to data.
We see that MC describes the data well when both the energies inside and outside the jet are considered.

The procedure has been validated in full MC, where the estimated showering correction is found
to closely match the true showering correction estimated following the procedure described in
\cref{mc_method}. 
To take into account any potential bias in the method, 
the final value of the showering correction in data is computed as
\begin{equation}
S_\textnormal{data} = \hat{S}_\textnormal{data}\cdot
\frac{S_\textnormal{MC}^\textnormal{true}}{\hat{S}_\textnormal{MC}},
\end{equation}
where the true showering correction $S_\textnormal{MC}^\textnormal{true}$ is directly 
available in MC and estimated following the procedure described in
\cref{mc_method}, and $\hat{S}_\textnormal{MC}$ and $\hat{S}_\textnormal{data}$ are the template-based showering corrections in MC and data, respectively.
The resulting bias is small (typically less than $0.5\%$), and a correction is made
to ensure a properly calibrated estimator in data.

\subsection{Results}
\label{showering_results}

\begin{figure}
	\centerline{\includegraphics[width=0.9\columnwidth]{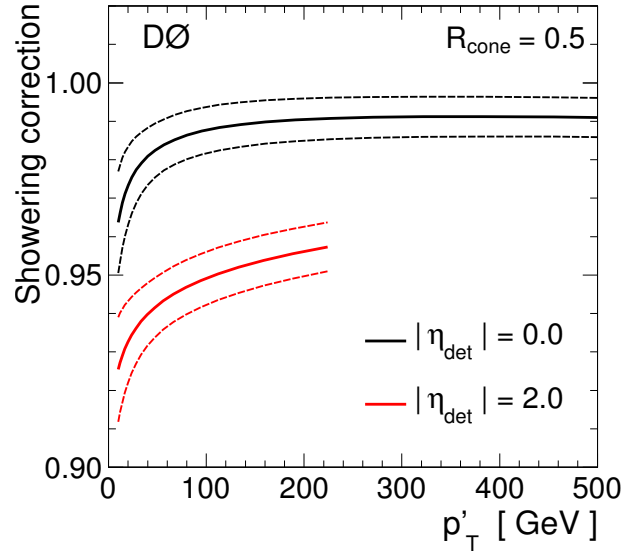}}
	\caption{
		Showering correction for $\rcone=0.5$ jets in data, as
		a function of \pTprime and for two \deta values, $\mdeta = 0$ and $\mdeta=2$. 
		The lines represent the results of a smooth parameterization of the correction. 
	}
	\label{correction_mc_jcca}
\end{figure}

\Cref{correction_mc_jcca} presents the estimated showering corrections 
for $\rcone=0.5$ jets in data, as
a function of \pTprime and for different \deta bins. 
%
The showering corrections in data are in good agreement with MC.
This motivates parameterizing the correction in data using the same energy dependence as observed in MC.

\extend{Figure~\ref{correction_data_jccb} in Appendix~\ref{showering_appendix}\ref{showering_appendix_correction_data} 
presents the showering correction for $\rcone=0.5$ jets in data.}

The uncertainties on the showering correction are typically less than $1.5\%$ (slightly larger for smaller \pT)
of the overall correction factor. The main sources of
uncertainty come from the difference between data and MC in the
single particle response at low \pT, the quality of the fits of MC templates to data, and the description of
the underlying event determined by using alternative \PYTHIA{} tunes.

%% file: summary.tex
\section{Summary of corrections and uncertainties}
\label{sec:summary}

%
%

\subsection{Mapping of measured energy to \Eprime}
\label{summary_mapping}

\newcommand{\Sphys}{\JESmathSym{C}}

The individual corrections discussed in previous sections have all been
parametrized and evaluated as a function of \Eprime or \pTprime (see Eq.~\ref{eq_ptprime}).
Using the uncorrected jet energy (\Emeas), the mapping to $\Eprime$ is found by solving
the following equation (\cf \cref{master3}) for \Eprime:
\begin{gather}
	\begin{aligned}
		\Ecorr &=
			\frac{\Emeas - \hat{E}_{\textnormal{O}}}
			{\func[\MPF{\photonjet}]{\Eprime} \func[\kRtopo]{\Eprime} \func[S]{\Eprime}}
			\frac{\func[\kOZS]{\Eprime}}{\func[\kRZS]{\Eprime}} \\
			&= \Eprime\,\func[\Sphys]{\Eprime},
		\label{eq_eprime_mapping}
	\end{aligned}
\end{gather}
\noindent implying the dependence on $\eta$.
After \Eprime is iteratively extracted from the last two members of the equation,
the corrected jet energy \Ecorr, that represents the particle jet energy \Eptcl,
can be computed directly.
The quantity $\Sphys = \Eptcl / \Eprime$ is
estimated in \photonjet{} MC as a function of \Eprime and \deta,
separately for $\rcone=0.7$ and $0.5$ jets.
This factor is
interpreted as the fraction of energy lost from out-of-cone radiation (physics showering), since it compares the 
particle-jet energy, confined by the geometry of the jet, to \Eprime,
the estimated energy of the parton recoiling against the photon in an ideal
$2 \to 2$ process. As expected, $\Sphys < 1$, especially for $\rcone=0.5$ and/or forward jets,
where values as low as $0.85-0.9$ are reached. Since the different correction
components depend logarithmically on the energy,
a precision of $5\%$ or better is sufficient in this mapping from \Emeas to \Eprime .

To verify the precision of the mapping,
\Eprime, as estimated via \cref{eq_eprime_mapping}, is compared to 
the true \Eprime in \photonjet MC.
The mapping is found to be precise to better than $3\%$ 
over the full \deta range.

\subsection{Results: corrections and uncertainties}

%
%

This section presents a few representative examples of the total correction factors
obtained with the methods
described in \cref{sec:offset,sec:response,sec:etadep,sec:mpfbiases,sec:showering}
for \DZero{} data and MC simulation.

\begin{figure*}[p]
	\subfloat{
		\includegraphics[width=0.9\columnwidth]{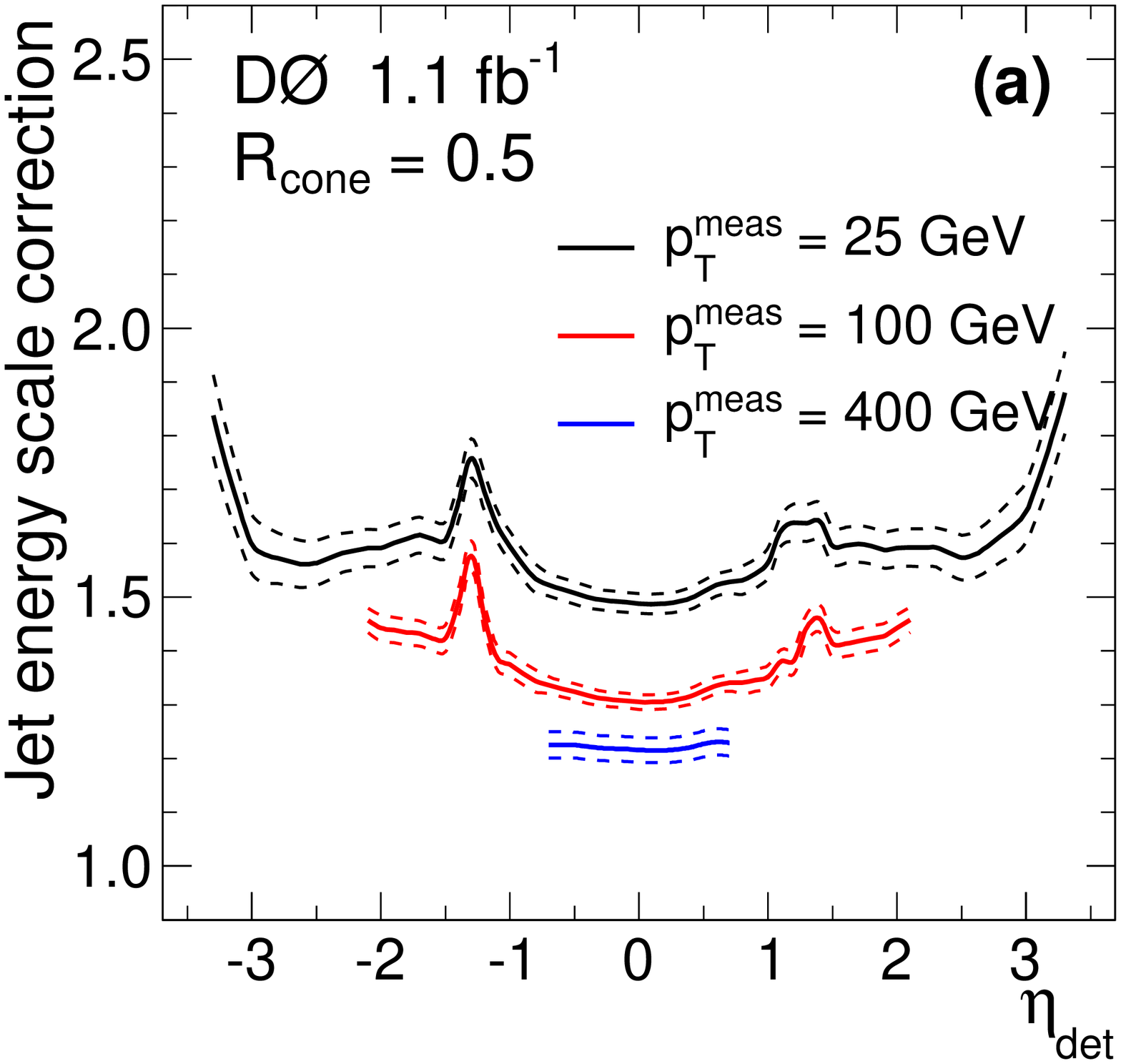}
		\label{fig:Summary_CorrVsEta_JCCB_DATA_IIA}
	}
	\subfloat{
\hspace{5mm} \includegraphics[width=0.9\columnwidth]{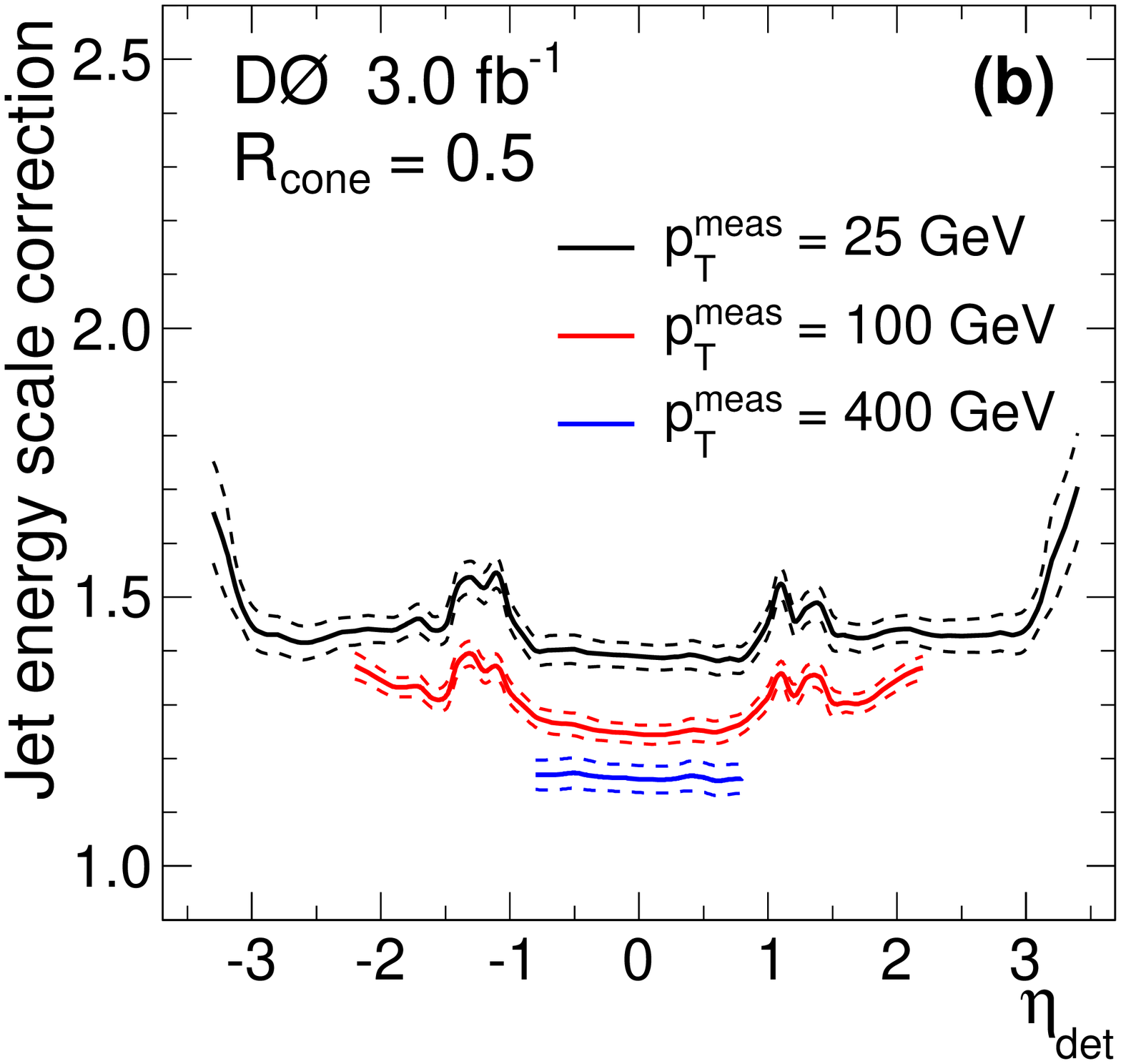}
		\label{fig:Summary_CorrVsEta_JCCB_DATA_IIB2}
	}
	\caption{
		Jet energy scale corrections, $\Eptcl/\Emeas$, for data jets with $\rcone=0.5$
		in \protect\subref{fig:Summary_CorrVsEta_JCCB_DATA_IIA} \IIa and
		\protect\subref{fig:Summary_CorrVsEta_JCCB_DATA_IIB2} \IIbTwo
		as a function of \deta for different uncorrected jet \pT values (\pT[meas]).
		Dashed lines show the total systematic uncertainty on the corrections. 
	}
	\label{fig:Summary_CorrVsEta_JCCB_DATA}
\end{figure*}

Separate corrections and uncertainties are extracted for data and simulation, 
for each of the jet cone sizes $\rcone=0.5$ and $0.7$, and for the five different run periods (see \cref{data_samples}). 
\Cref{fig:Summary_CorrVsEta_JCCB_DATA} shows the typical JES correction factor $\Eptcl / \Emeas$ 
for jets with $\rcone=0.5$ in two run periods, \IIa{} (\invfb{1.1}) and \IIbTwo{} (\invfb{3.0}),
taken as examples throughout this section.
The value of the correction spans a range of less than 10\% in most of the detector. The structures in the
ICD region of the calorimeter ($1.0 < \mdeta < 1.6$)  reflect the reduced coverage and additional
passive material, and the different response in the forward region of the calorimeter ($\mdeta > 3$) is evident.
The overall difference between the two periods is a consequence of the modified jet-finding algorithm 
(see \cref{objectid_jets}) and calorimeter calibration.


\begin{figure*}[p]
	\subfloat{
		\includegraphics[width=0.9\columnwidth]{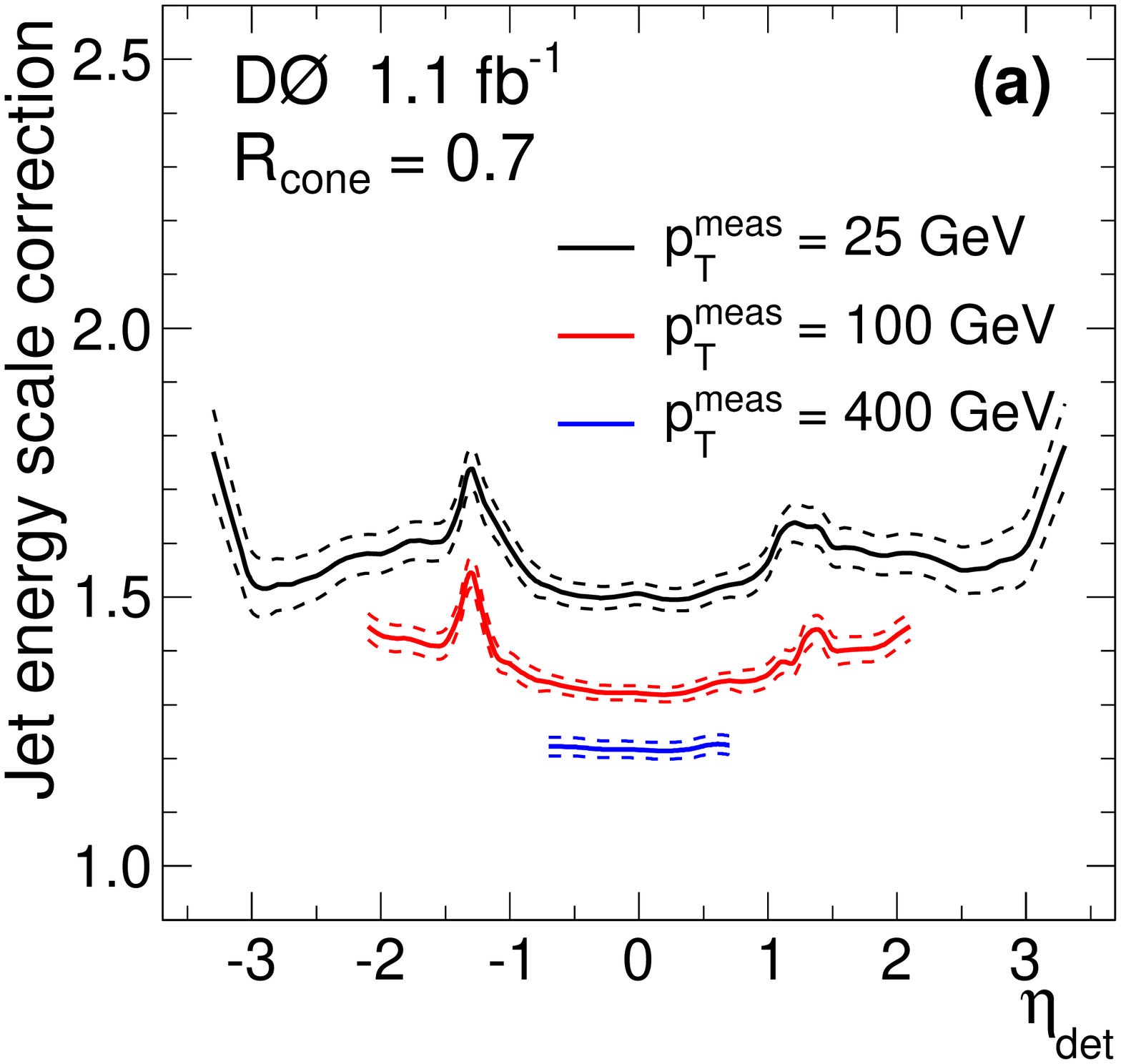}\\
		\label{fig:Summary_CorrVsEta_JCCA_DATA_IIA}
	}
	\subfloat{
\hspace{5mm} \includegraphics[width=0.9\columnwidth]{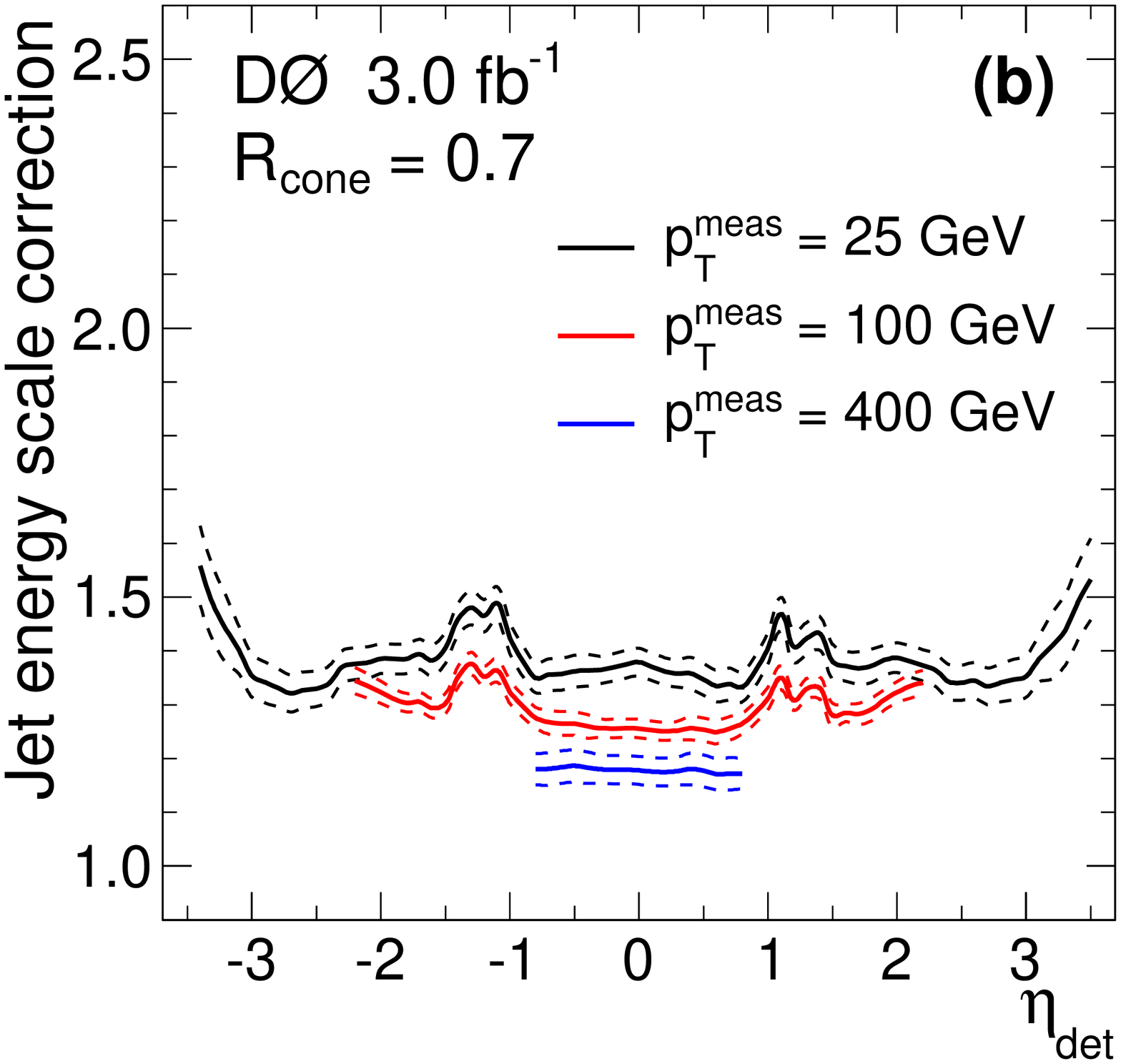}\\
		\label{fig:Summary_CorrVsEta_JCCA_DATA_IIB2}
	}
	\caption{
		Same as in \cref{fig:Summary_CorrVsEta_JCCB_DATA}, but for $\rcone=0.7$.
	}
	\label{fig:Summary_CorrVsEta_JCCA_DATA}
\end{figure*}

\begin{figure}
	\subfloat{
		\includegraphics[width=\columnwidth]{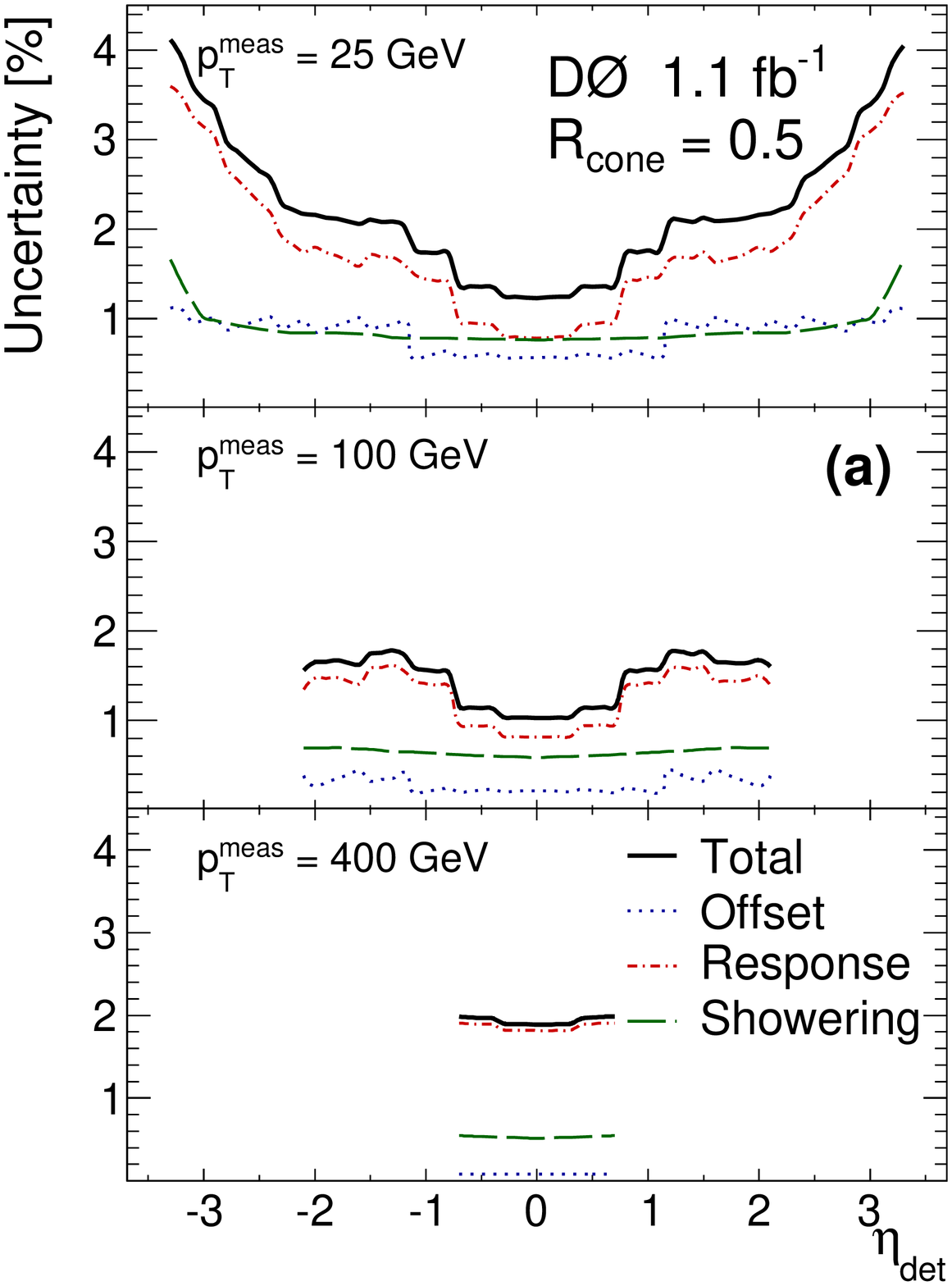}\\
		\label{fig:Summary_ErrVsEta_JCCB_DATA_IIA}
	}
	\\
	\subfloat{
		\includegraphics[width=\columnwidth]{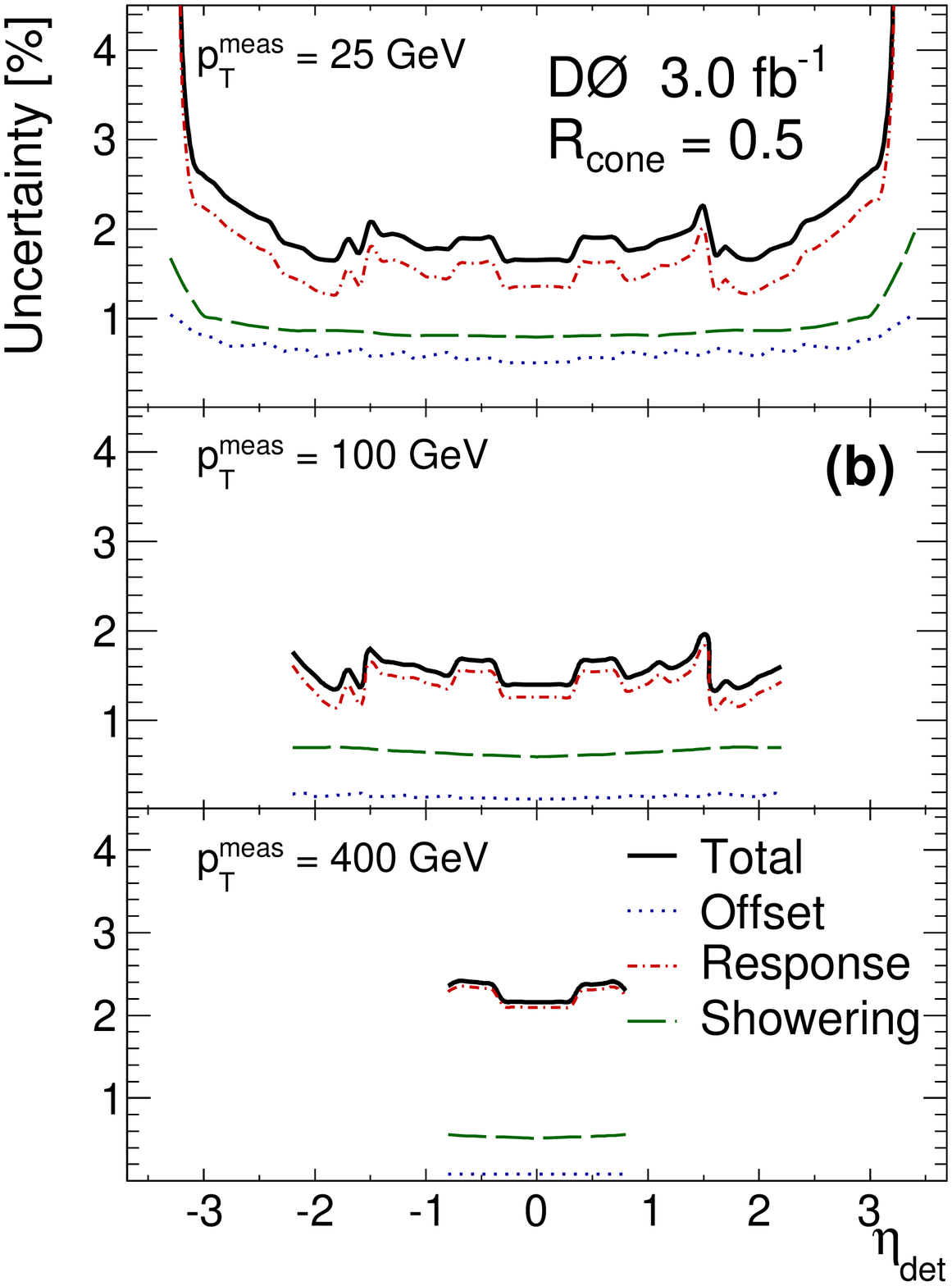}\\
		\label{fig:Summary_ErrVsEta_JCCB_DATA_IIB2}
	}
	\caption{
		Jet energy scale uncertainty for data jets with $\rcone=0.5$,
		in \protect\subref{fig:Summary_ErrVsEta_JCCB_DATA_IIA} \IIa and
		\protect\subref{fig:Summary_ErrVsEta_JCCB_DATA_IIB2} \IIbTwo as a function of \deta
		for different uncorrected jet \pT (\pT[meas]).
	}
	\label{fig:Summary_ErrVsEta_JCCB_DATA}
\end{figure}

The correction factor for $\rcone=0.7$ jets (\cref{fig:Summary_CorrVsEta_JCCA_DATA}) exhibits the same features as for $\rcone=0.5$.
The absolute value of the correction is slightly smaller for $\rcone=0.7$ jets,
reflecting the better geometric coverage provided by the larger cone radius.
This is more evident for low energy jets, whose particle showers are 
less boosted and correspondingly less well collimated. 
\Cref{fig:Summary_ErrVsEta_JCCB_DATA} shows the uncertainty on the correction for $\rcone=0.5$ jets
in the same two run periods, split into their independent contributions.
The uncertainties for $\rcone=0.7$ are similar.

The leading contribution to the uncertainty comes from the correction on the jet response.
At low energy, the offset energy subtraction and the out-of-cone correction contribute significantly.
As the energy of the jet increases, the offset energy becomes a smaller fraction
of the measured energy and the corresponding uncertainty less relevant.
The largest contribution to the uncertainty on the response comes from the absolute response,
which suffers from large statistical uncertainties from extrapolation into regions
not covered by \photonjet data (very low and high energies).


\begin{figure}
	\subfloat{
		\hspace{0.04\columnwidth} 
		\includegraphics[width=0.9\columnwidth]{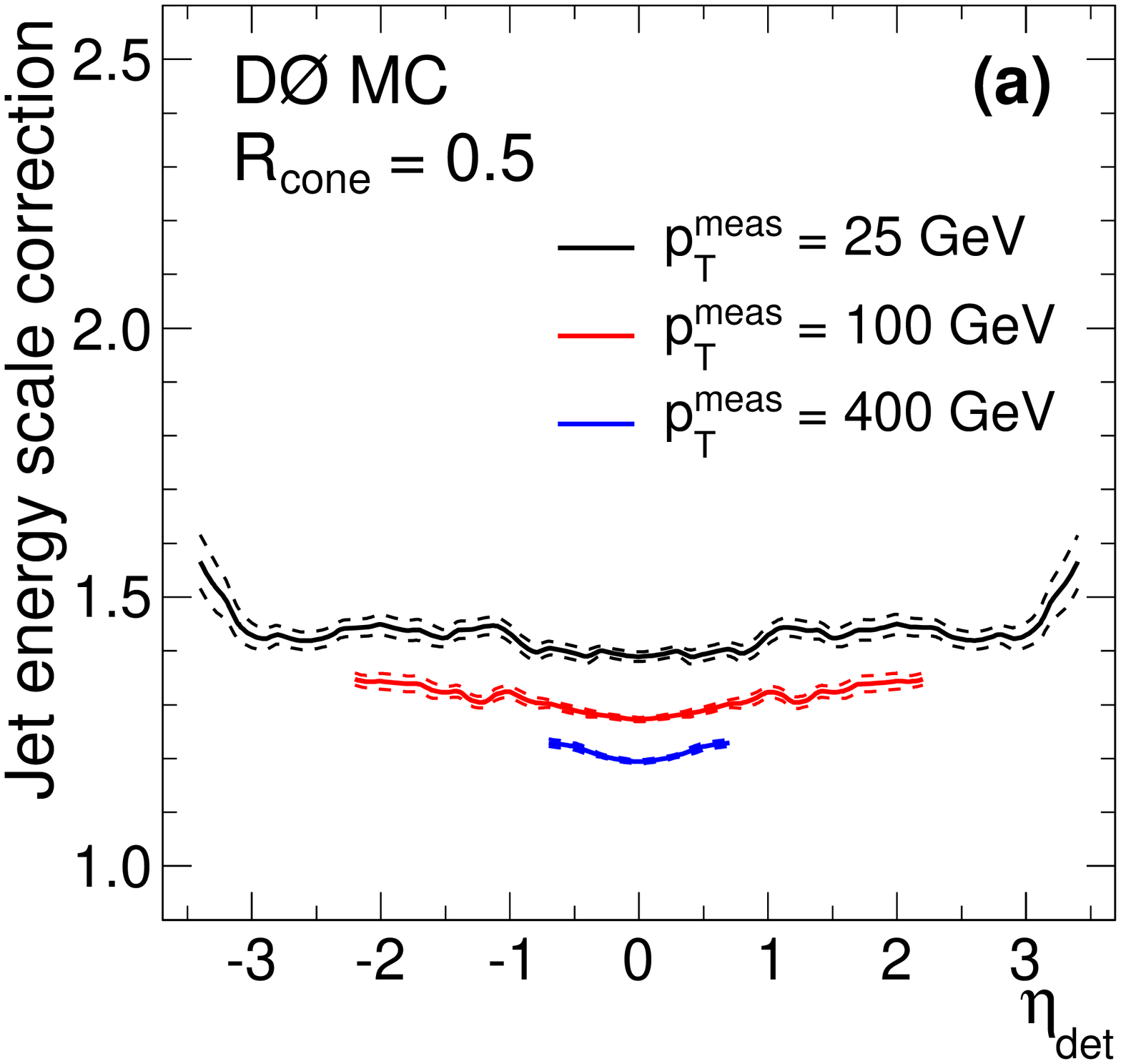}\\
		\label{fig:Summary_CorrVsEta_JCCB_MC_IIB2}
	}
	\\
	\subfloat{
		\includegraphics[width=\columnwidth]{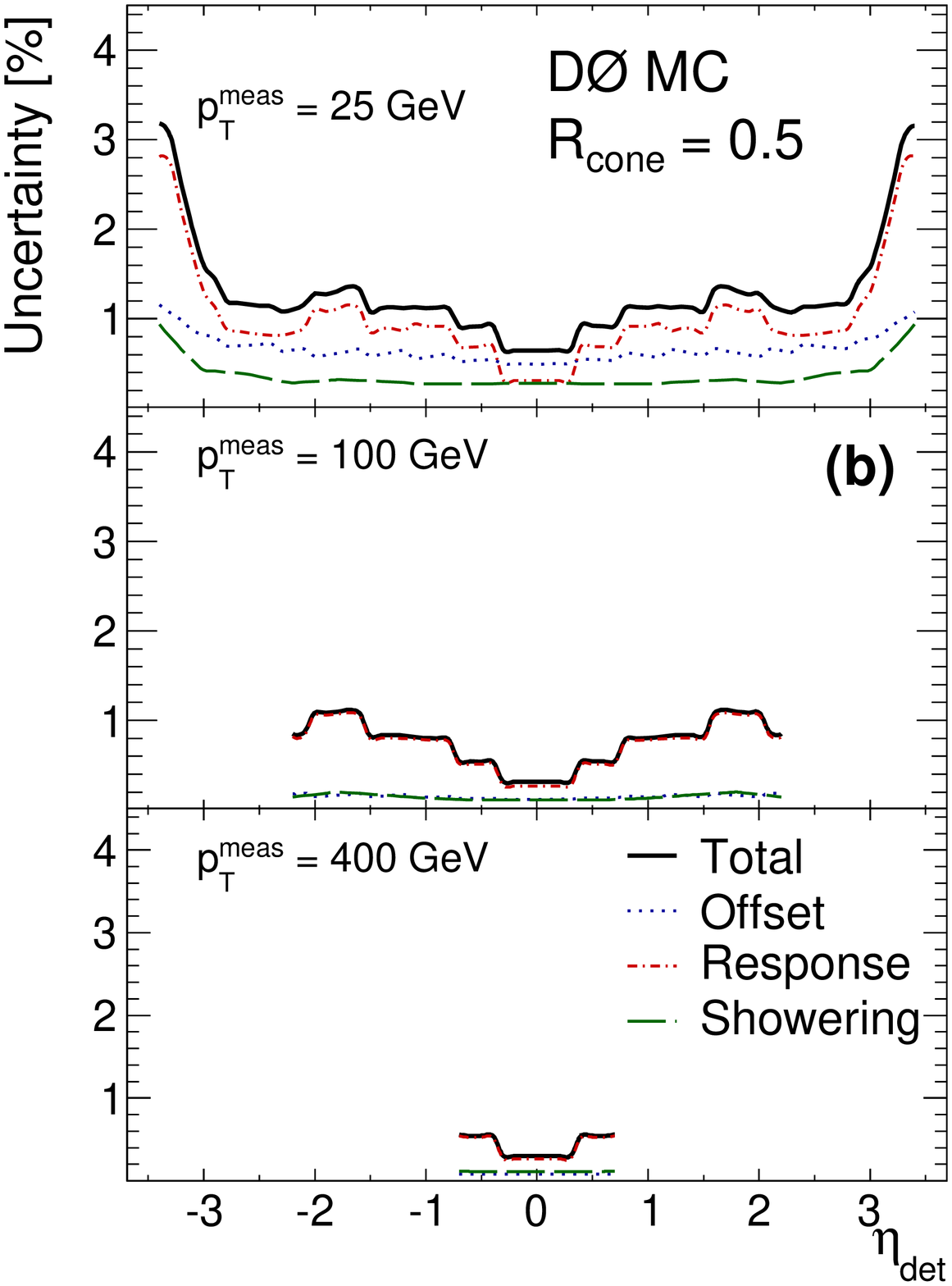}\\
		\label{fig:Summary_ErrVsEta_JCCB_MC_IIB2}
	}
	\caption{
		Jet energy scale \protect\subref{fig:Summary_CorrVsEta_JCCB_MC_IIB2} correction, $\Eptcl/\Emeas$,
		and \protect\subref{fig:Summary_ErrVsEta_JCCB_MC_IIB2} uncertainty
		for jets in MC simulation with $\rcone=0.5$,
		in \IIbTwo{} as a function of \deta
		for different uncorrected jet \pT (\pT[meas]).
	}
	\label{fig:Summary_CorrErrVsEta_JCCB_MC}
\end{figure}

\Cref{fig:Summary_CorrErrVsEta_JCCB_MC} shows an example of jet energy scale correction
and uncertainty for MC events. 
The simplified model in the MC simulation yields 
a more uniform correction versus \deta as compared with data.
The overall uncertainty for simulation is smaller than for data,
while the uncertainty for the offset energy correction, being completely data-driven,
is very similar.

%

%% file: closure.tex
\section{Closure tests}
\label{sec:closure}

This section presents results on the validation of the jet energy scale
corrections and thei uncertainties. These validation tests are referred to as
``closure tests,'' and their goal is to assess whether the corrections  calibrating jet
energy back to the particle level are within the quoted uncertainties. 

To assess whether closure of the corrections is achieved, observables able to
probe the relationship between the particle and (calibrated) calorimeter jet energies must be
defined. In the case of MC 
this can be straightforward. In the case of data, the connection is not direct and effects 
not related to jet energy calibration (\eg, background contamination, photon energy scale) 
must be taken into account.

\subsection{Sample selection}
\label{closure_selection}

Since closure tests are mainly designed to probe the absolute energy scale calibration, 
a natural sample to use is \photonjet events.
The event selections used for the closure tests closely follow those used for the absolute response
measurement (\cref{photonjet_selection}), except that no limit is imposed on the number of
vertices or jet multiplicity. Considering events with $\npv \geq 1$ enforces
consistency with the determination of the offset bias correction \kOZS, which 
is estimated for the inclusive ($\npv \geq 1$) sample.
There are two reasons for allowing events with more than one jet.
First, typical event selections in physics analyses consider final states inclusive in the number of jets.
Second, keeping events with extra jets minimizes biases
in the average offset, since some of those jets will likely arise from
additional parton interactions~\cite{g3j_PRD}. 

Closure tests are performed separately for $\rcone=0.7$ and $0.5$ jets, in
different $0.4$-wide bins of $\mdeta$ (up to $\mdeta < 3.6$) and as a function of \pTprime
(defined in \cref{offset_bias_correction}). 


Due to a finite jet energy resolution around the uncorrected \pT reconstruction threshold of \GeVc{6},
the $E/\pT$ of the jet increases as compared with the particle level jet. This effect
is especially large for lower jet \pT (the ``low-\pT bias''), and can be reduced
by the additional requirement of $\pTgamma \geq \GeV{30}$, applied in the closure test.
However, even for events with exactly one jet, the low-\pT bias with a jet located in or around the ICD region
is still present above \GeVc{30}, mainly due to poorer jet energy resolution as compared with other rapidity regions.

\subsection{Direct closure tests in MC}
\label{closure_direct}

In the case of MC, the availability of the particle jet information allows the definition
of a ``direct'' closure variable $\average{\Ecorr}/\average{\Eptcl}$
%
where \Ecorr is the corrected jet energy (\cref{master3}) and 
\Eptcl is the energy of the closest particle jet matching the reconstructed
jet within $ \deltar < \rcone/2$. The averages in the ratio $\average{\Ecorr}/\average{\Eptcl}$
are taken for the set of events
within the particular \parentheses{\pTprime,\mdeta} bin under consideration.


\begin{figure*}[t]
	\includegraphics[width=\columnwidth]{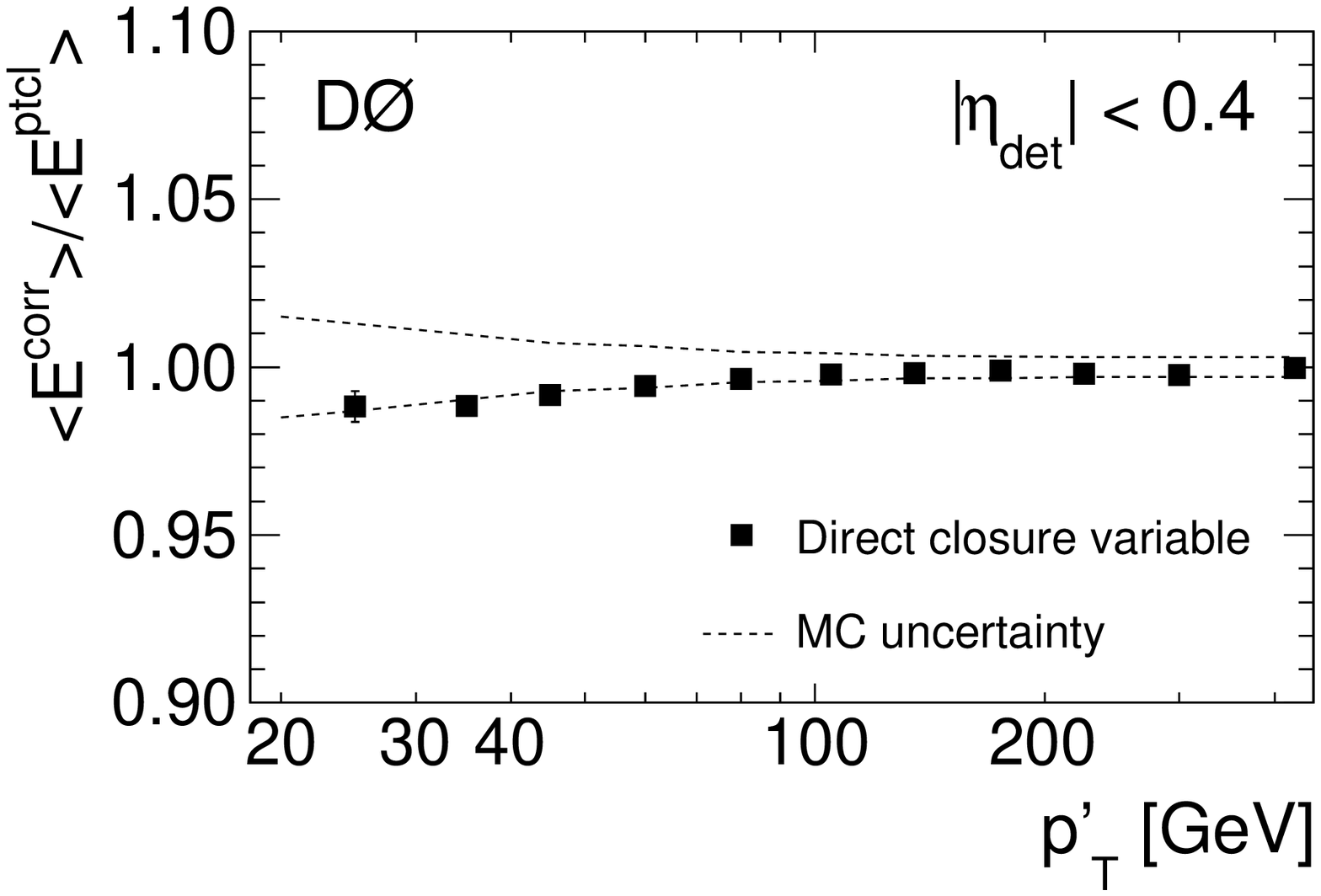}
	\includegraphics[width=\columnwidth]{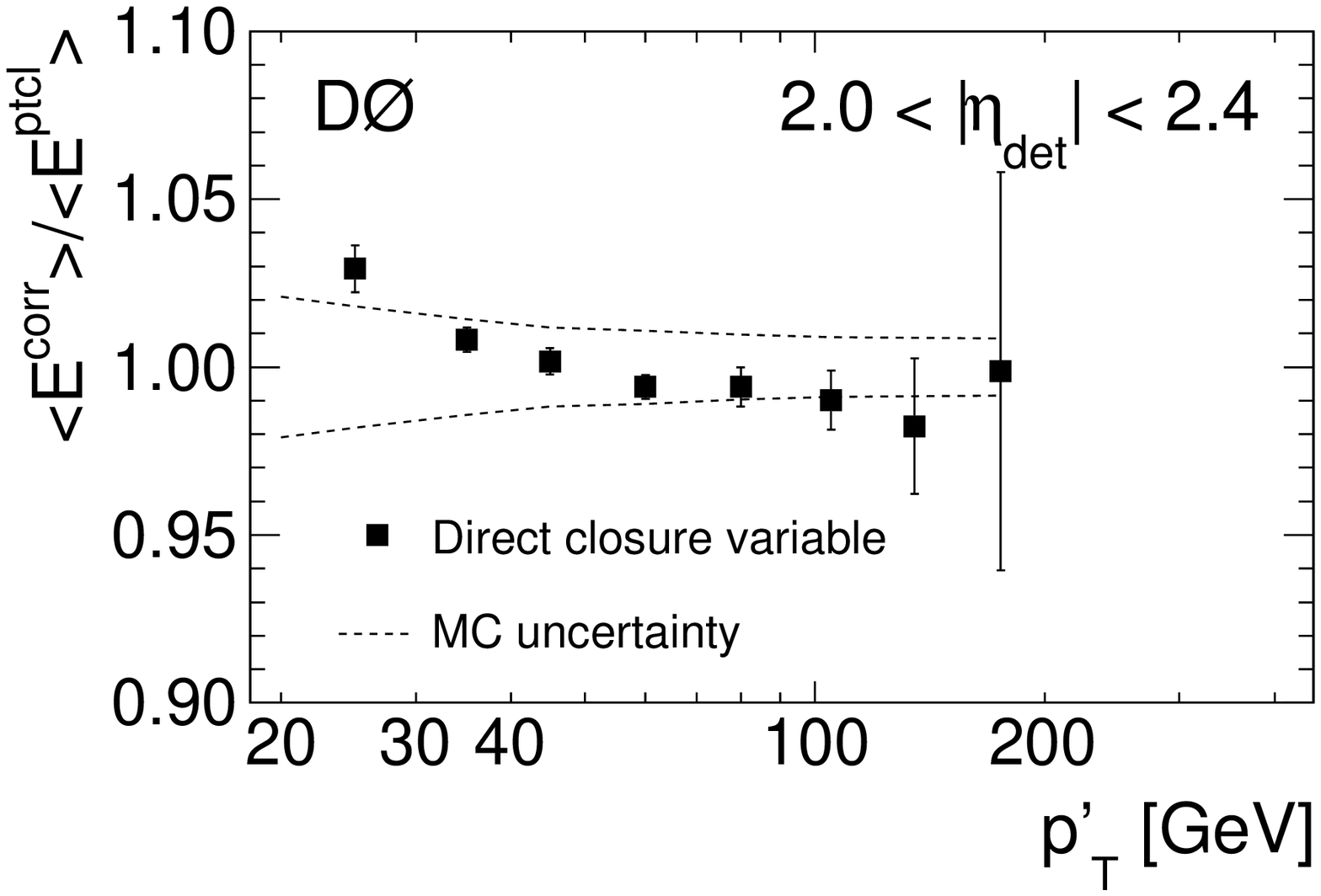}
	\caption{
		Direct closure variable for $\rcone=0.7$ jets
		as a function of \pTprime and in two \mdeta regions.
		The band represents the total uncertainty on the energy scale from 
                the corrections in data and MC added in quadrature.
	}
	\label{direct_mc_jcca}
\end{figure*}


\Cref{direct_mc_jcca} shows
results of direct closure tests for $\rcone=0.7$ in two \mdeta bins, taken as an example.
The jet energy scale corrections have been evaluated using the internally remapped \Eprime 
(see Eq.~\ref{eq_eprime_mapping}).
The small rise observed at low \pTprime (mostly at high $|\eta|$) is caused by the low-\pT bias, as
discussed in \cref{closure_selection}. Taking this into consideration, closure is in general
achieved within the quoted $1\%-2\%$ uncertainties. 

\subsection{Closure tests in data}
\label{closure_data}

In contrast with MC, for data there is no possibility 
to directly check the absolute energy calibration, as no information
regarding particle jets is available. Therefore, closure tests are
based on the comparison of the corrected jet energies between data and
MC. Provided that the jet energy calibration works properly in MC (see \cref{closure_direct}),
it is then possible to relate the data-to-MC intercalibration
to the absolute energy calibration in data
by reconstructing the particle jet energy, \Eptcl.

As for direct closure in MC, we use \photonjet events 
(\cref{closure_selection}). The closure observable
is defined as the ratio of the average corrected jet energies between data and MC,
$\hfrac{\average{\Ecorr[data]}}{\average{\Ecorr[MC]}}$,
computed as a function of \pTprime in different \mdeta regions.

Since the goal of the closure tests is to validate the jet energy calibration
in pure \photonjet events, it is important to properly account for any
differences between data and MC which could result in biases in the
closure observable.
The most relevant effect is related to the presence of \dijet background in data.
Since the closure observable does not directly involve the reconstructed
photon \pT, the main difference results from the flavor composition
of the jet. For example, as will be discussed in \cref{sec:qcd_specific}, 
low \pT jets from dijet background are dominated by jets from gluons, whereas
jets from \photonjet signal are dominated by jets from quarks, and there is up to $8\%$ difference
in response for low \pT jets. 
In order to account for this bias, 
data are not compared to pure \photonjet MC, but rather to a mixture of \photonjet 
and \dijet (\photon-like) MC, combined using the estimated sample purity. 
This allows a correction for the leading difference between data and MC. 
Also, the measured photon energy in MC is corrected to ensure its energy scale
is consistent with data. 

\begin{figure*}[t]
	\includegraphics[width=\columnwidth]{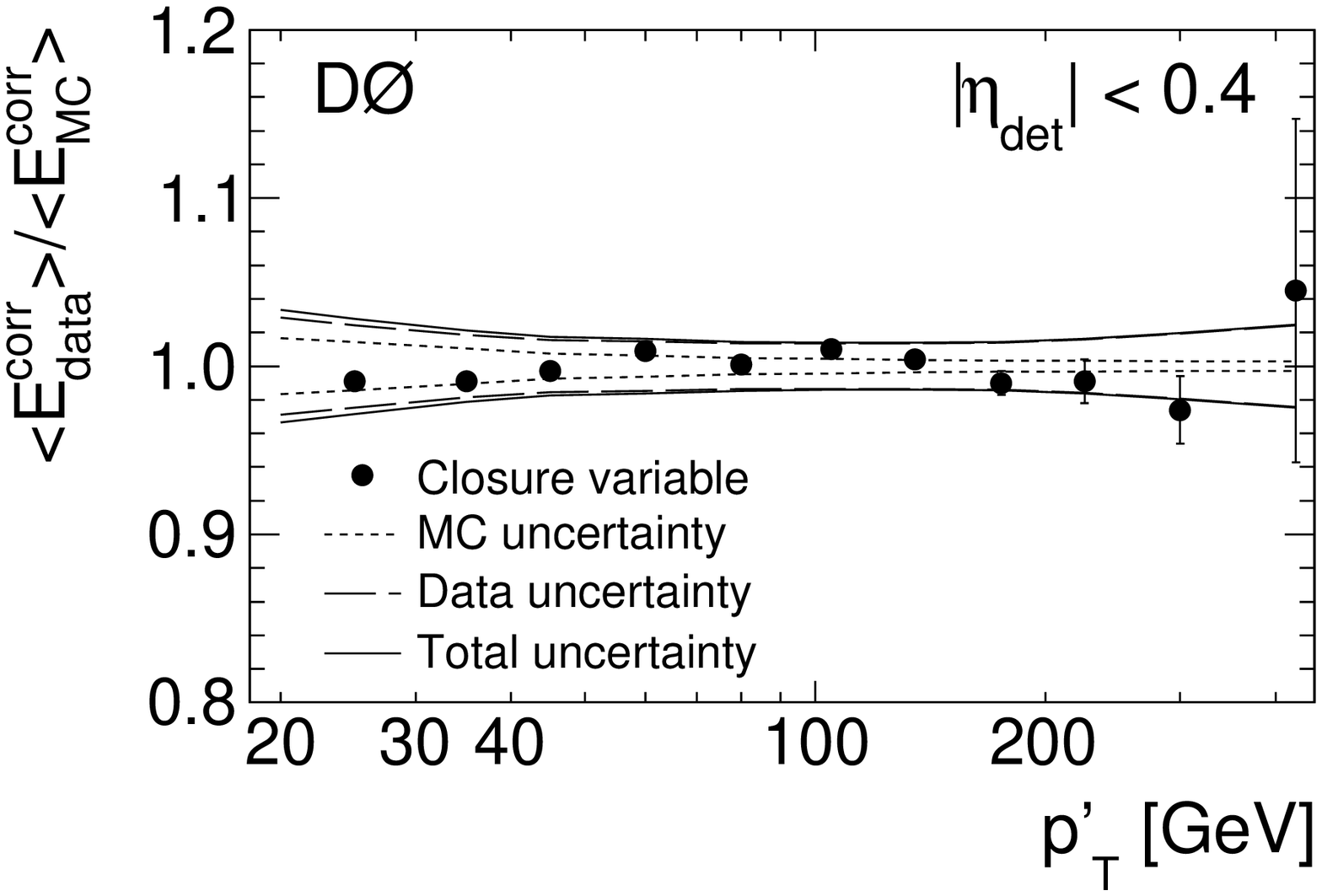}
	\includegraphics[width=\columnwidth]{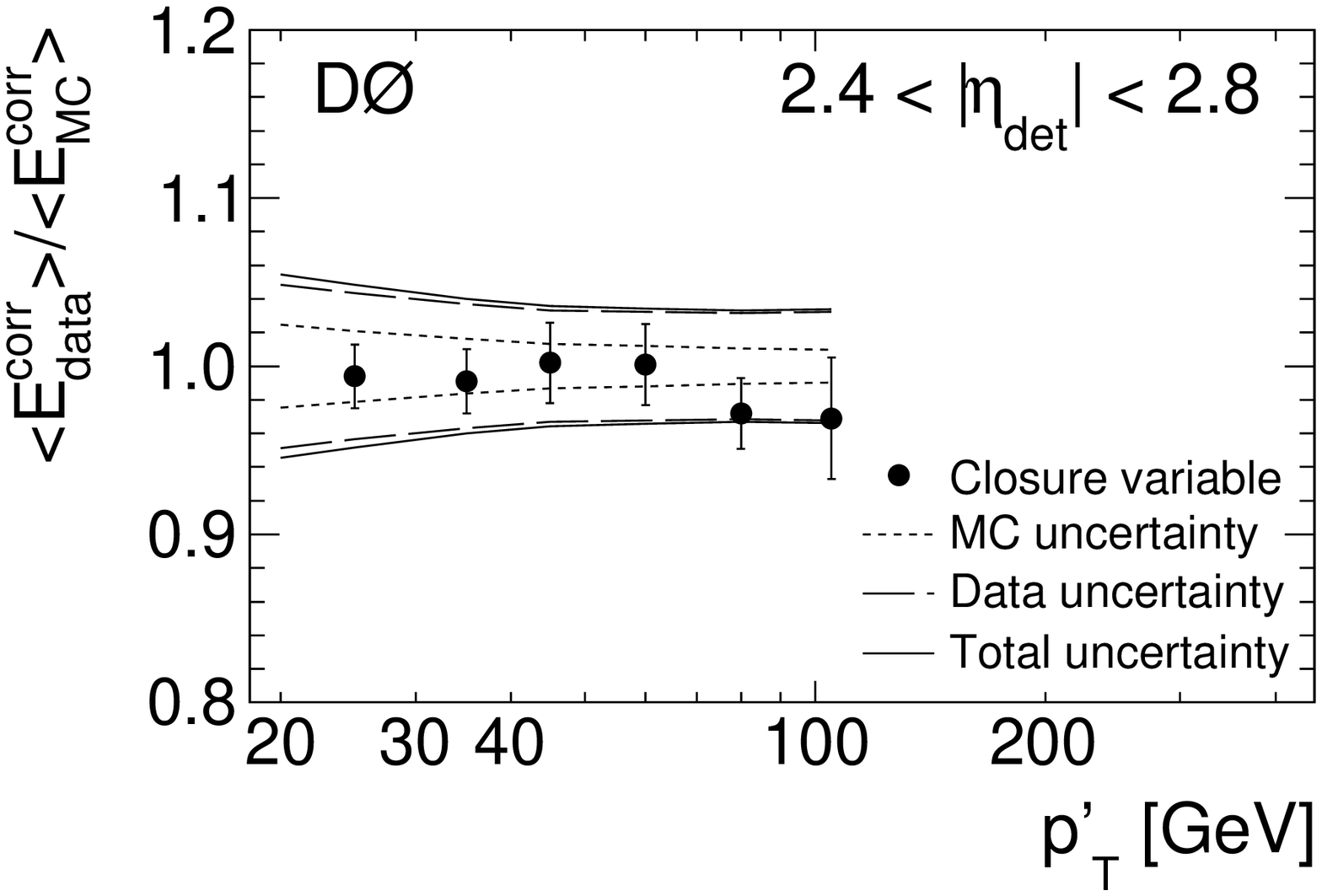}
	\caption{
		Relative data-to-MC closure variable for $\rcone=0.7$ jets
		as a function of \pTprime and in two \mdeta bins.
		The inner and intermediate bands represent the energy scale uncertainty
		for MC and data, respectively,
		while the outer band represents the total uncertainty.
	}
	\label{closure_data_vs_mcmixed_jcca}
\end{figure*}

\begin{figure*}[t]
	\includegraphics[width=\columnwidth]{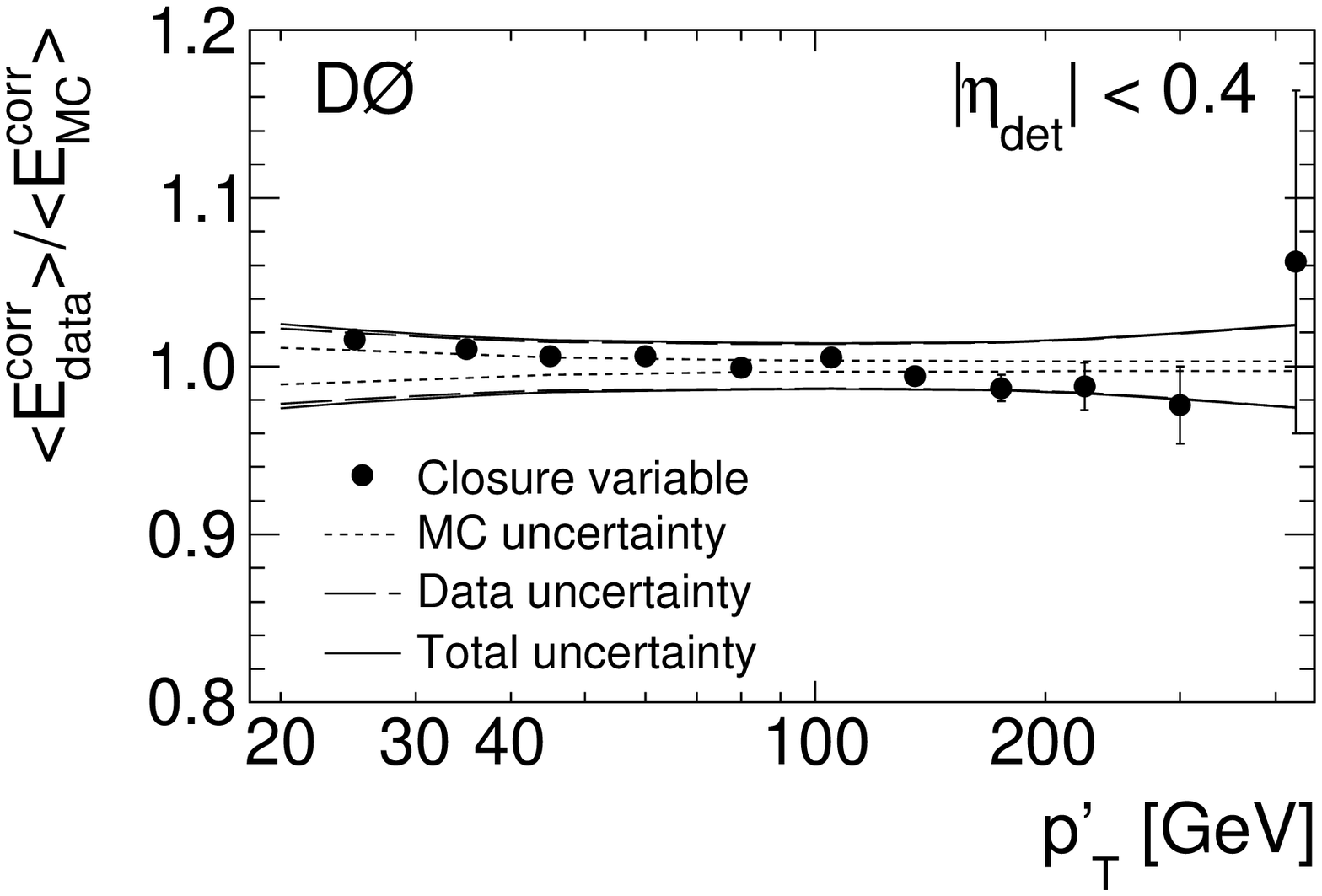}
	\includegraphics[width=\columnwidth]{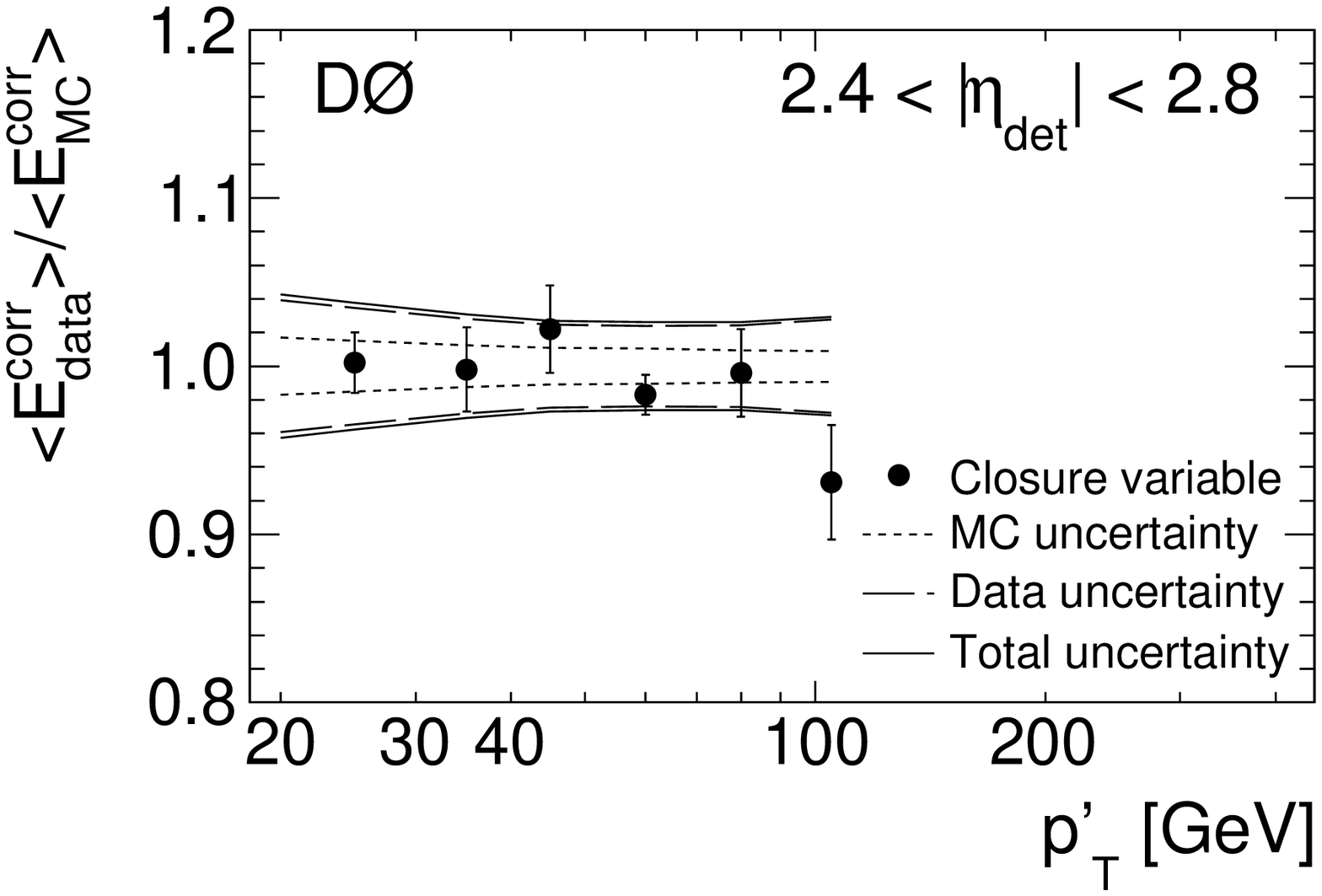}
	\caption{
		Same as on \cref{closure_data_vs_mcmixed_jcca} for jets with $\rcone=0.5$.
	}
	\label{closure_data_vs_mcmixed_jccb}
\end{figure*}

As already indicated, this correction enters the closure test indirectly, via the binning
of the closure variable in terms of \pTprime.
\Cref{closure_data_vs_mcmixed_jcca} presents the results of relative data-to-MC 
closure for $\rcone=0.7$ jets as a function of \pTprime in two
\mdeta bins. The corresponding plots for $\rcone=0.5$ jets are shown in 
\cref{closure_data_vs_mcmixed_jccb}. Since \emph{a~priori} jet energy calibration uncertainties 
for data and MC are largely uncorrelated, the uncertainty on the closure observable is 
defined as the sum in quadrature of data and MC uncertainties. 
Data and MC appear intercalibrated, \ie, reproduce \Eptcl within the estimated uncertainties. 

%% file: fdc_corrections.tex
\section{Flavor-dependent corrections in Monte Carlo}
\label{sec:fdc}

The default jet energy calibration, which is separately derived for data and MC simulations
as described in previous sections, corrects the detected jet energies to the particle level,
ignoring differences due to the specific flavor of the parton that initiated the jet
(known as ``jet parton flavor'').
As described in \cref{ssec:SampleDependencyIntro}, samples with different composition 
than \photonjet may require additional corrections to the jet energy scale. 
An estimation of single-particle responses should reduce the bias in jet energy that arises due to dependence 
on the jet parton flavor.

The MC simulation evaluates the energy deposited by a jet in the calorimeter
summing over the estimated response from all of the particles in the jet.
%
%
Estimation of single-particle responses should reduce the bias in jet energy that arises due to dependence 
on the physics process and (more generally) on the jet parton flavor.
It would allow measurements to rely on MC simulations to correct for this bias (see \eg,~\cite{Top_mass}).

This section describes a procedure to tune the MC simulated single-particle response 
and presents the additional energy correction 
required for various jet flavors. 

\subsection{Method}

The procedure described below uses single-particle MC samples (see \cref{sssec:SPRMC}) 
and the \photonjet and dijet MC and data samples described in \cref{sec:datasets}.
We define a correction factor $F$ 
for each reconstructed calorimeter jet in the simulation:
\begin{equation}
	F=\frac{\sum_{i}E_i\cdot R_i^\textnormal{data}}{\sum_{i}E_i\cdot R_i^\textnormal{MC}},
	\label{eqn:method1}
\end{equation}
where the subscript $i$ runs over the particles in the particle jet, 
$E_i$ is the particle energy, and $R_i^\textnormal{data}$ and $R_i^\textnormal{MC}$ represent the response to 
the particle in the calorimeter for data and MC simulation, respectively.
To calculate $F$, the particle and calorimeter jets are spatially matched using 
$\deltar < \rcone/2$.

To preserve the jet energy scale obtained with \photonjet{} events, 
we introduce the relative correction factor $F_\textnormal{corr}$:
\begin{equation}
	F_\textnormal{corr} = \frac{1}{\average{F}_{\photonjet}} \cdot \frac{\sum_{i}E_i\cdot R_i^\textnormal{data}}{\sum_{i}E_i\cdot R_i^\textnormal{MC}},
	\label{eqn:method12}
\end{equation}
where $\average{F}_{\photonjet}$ is the average of $F$ in the \photonjet{} sample, 
parameterized as a function of jet \pT and \deta. 
%
The correction factor $F_\textnormal{corr}$ is independent of corrections described in the previous sections and can be applied on the jet energy scale corrected jets.

\subsubsection{Single-particle response in MC}
\label{sssec:SPRMC}
Single-particle responses in MC are measured from MC samples of single
\photon, $e^\pm$, $\mu^\pm$, $\pi^\pm$, $K^\pm$, $K^0_\textnormal{S}$, $K^0_\textnormal{L}$, $p^\pm$, $n$, $\Lambda$, $\Sigma$, and $\Xi$ production
simulated with no zero suppression, calorimeter noise, and no ZB overlay. 
For each particle, the energy of the calorimeter cells contained
in the cone with radius $\rcone$ around the particle is summed to obtain the reconstructed energy.
The MC single-particle response $R_i^\textnormal{MC}$ is the ratio between the reconstructed energy and the particle energy.
The responses are parameterized versus particle energy and extracted independently for different pseudorapidities.
%
%
%
%
For example, the response to a given hadron $h$ (\eg, $\pi^{+}$) with energy $E$ is parametrized as:
\begin{equation}
	R_\textnormal{h}^\textnormal{MC} = p_{\textnormal{h}}^{(0)}\cdot\sqbrackets{1 - p_{\textnormal{h}}^{(1)}\cdot(E / 0.75)^{p_{\textnormal{h}}^{(2)} - 1}},
	\label{eq:SPRhadronsMC}
\end{equation}
where $p_\textnormal{h}^{(k)}$ are three free parameters for each hadron $h$ that have to be determined.
The functional form in Eq.~\ref{eq:SPRhadronsMC} has been varified using MC simulation.

\begin{figure*}
	\subfloat{
		\includegraphics[width=\columnwidth]{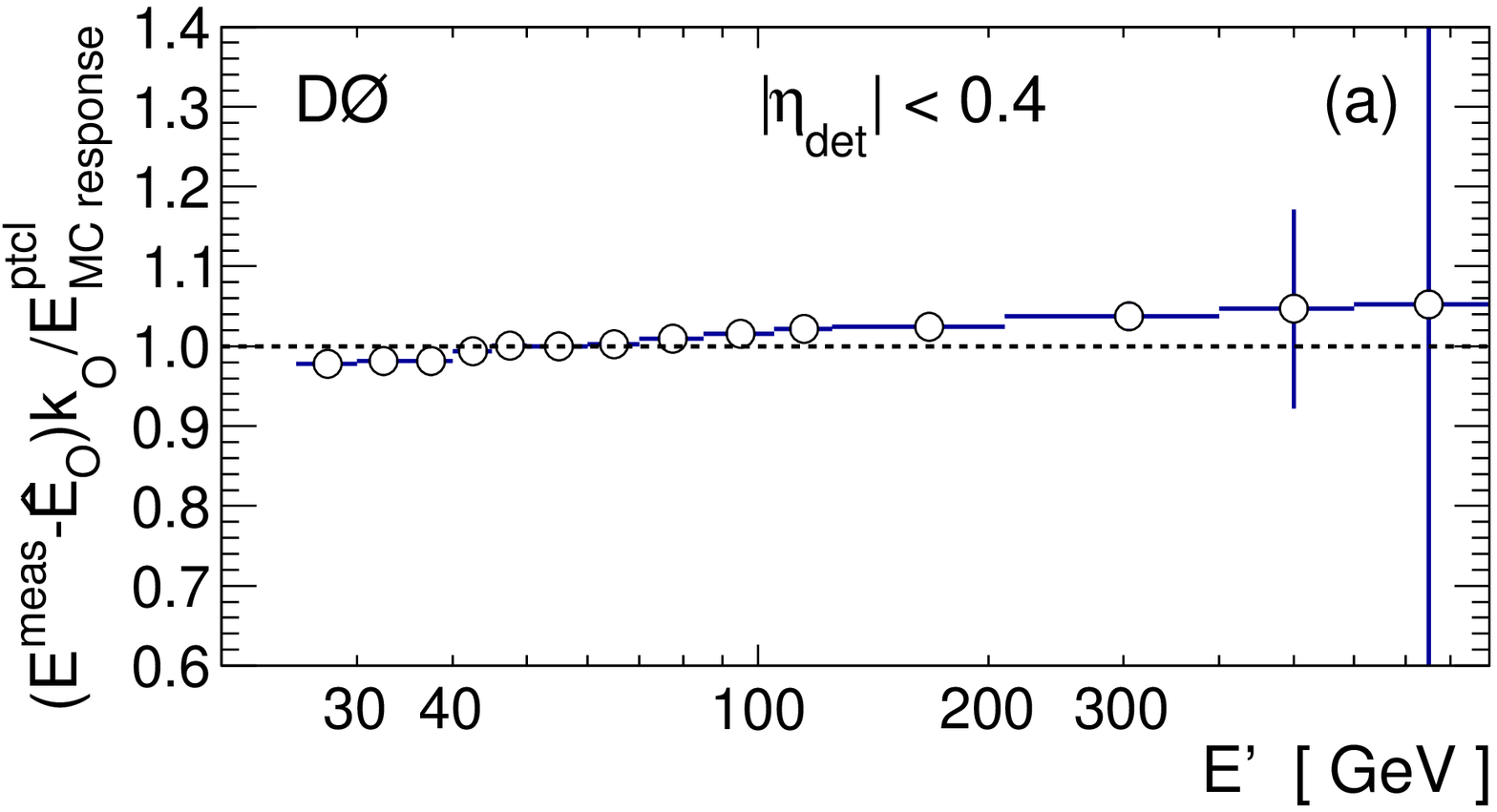}
		\label{fig:closure_no0sup-CC1}
	}
	\subfloat{
		\includegraphics[width=\columnwidth]{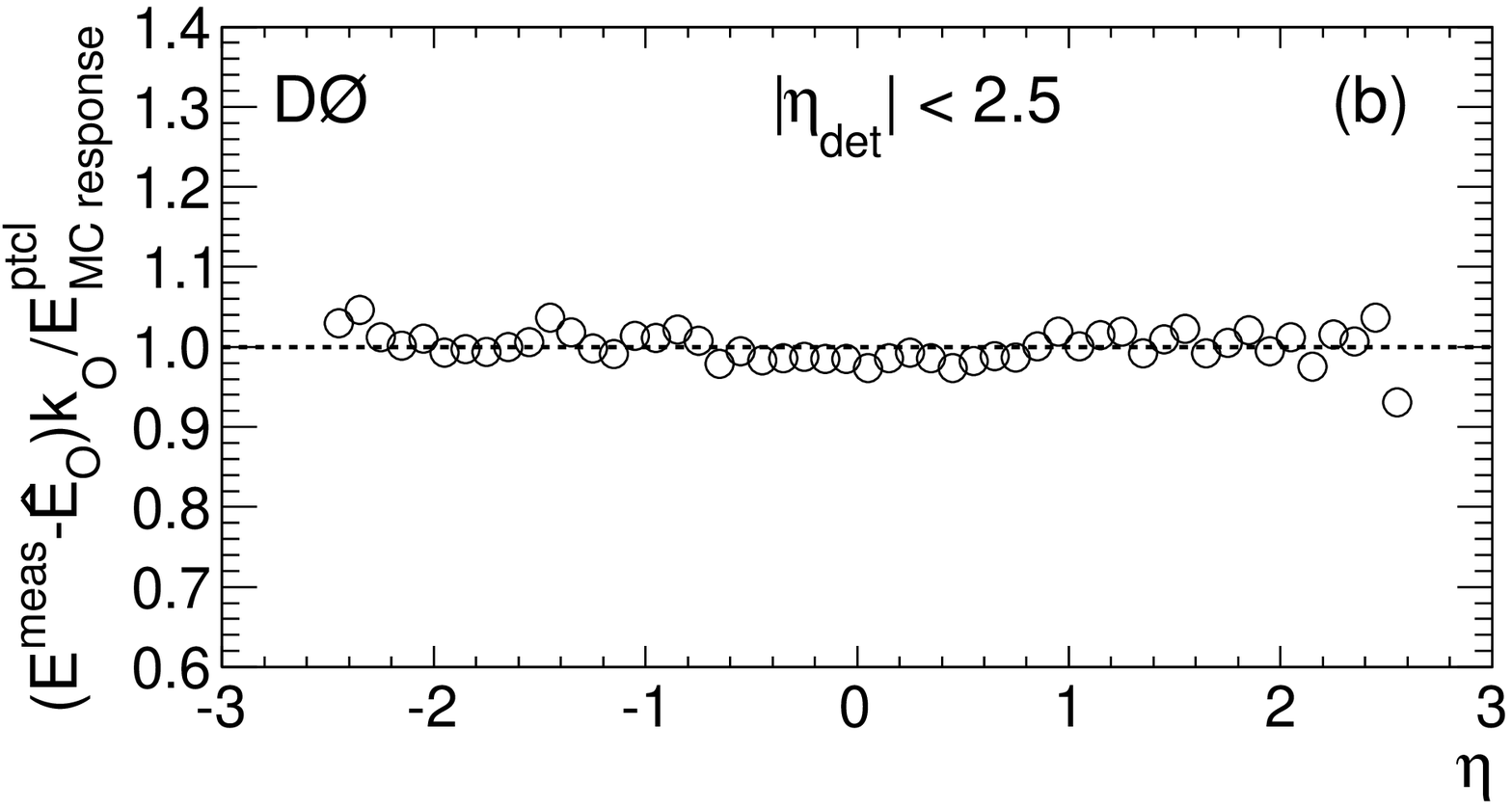}
		\label{fig:closure_no0sup-All}
	}
	\caption{
		Ratio between the jet energy as reconstructed, including offset corrections,
		and the jet energy as computed using the MC single-particle responses
		(see \cref{eq:SPRMCtest}) for
		\protect\subref{fig:closure_no0sup-CC1}~jets with $\mdeta < 0.4$
		in the \photonjet{} sample as a function of jet \Eprime (\cref{eq_eprime}), and
		\protect\subref{fig:closure_no0sup-All}~jets with $\mdeta < 2.5$
		in the \photonjet{} sample as a function of jet \deta.
                The shown uncertainties are statistical.
	}
	\label{fig:closure_no0sup}
\end{figure*}

Using MC simulation, we compare the sum of the single particle responses $\sum_{i}E_i\cdot R_i^\textnormal{MC}$ in \cref{eqn:method12}
with the reconstructed calorimeter jet energy corrected for offset contribution and the offset bias:
$(\Emeas-\hat{E}_\textnormal{O})\cdot \kO$.
\Cref{fig:closure_no0sup} shows that their ratio 
\begin{equation}
	\frac{\kO(\Emeas - \hat{E}_\textnormal{O})}{\sum_{i}E_i\cdot R_i^\textnormal{MC}}
	\label{eq:SPRMCtest}
\end{equation}
agrees with unity within 2\% for most of the jet energies and pseudorapidities. 
The residual disagreement is assigned as systematic uncertainty. 

\subsubsection{Single-particle response in data}
\label{sssec:SPRdata}

The single-particle responses in data cannot be determined directly.
Therefore the MC single-particle responses need to be tuned to reproduce the data.

\begin{figure*}
	\begin{minipage}{0.5\textwidth}
	\subfloat{
		\includegraphics[width=\textwidth]{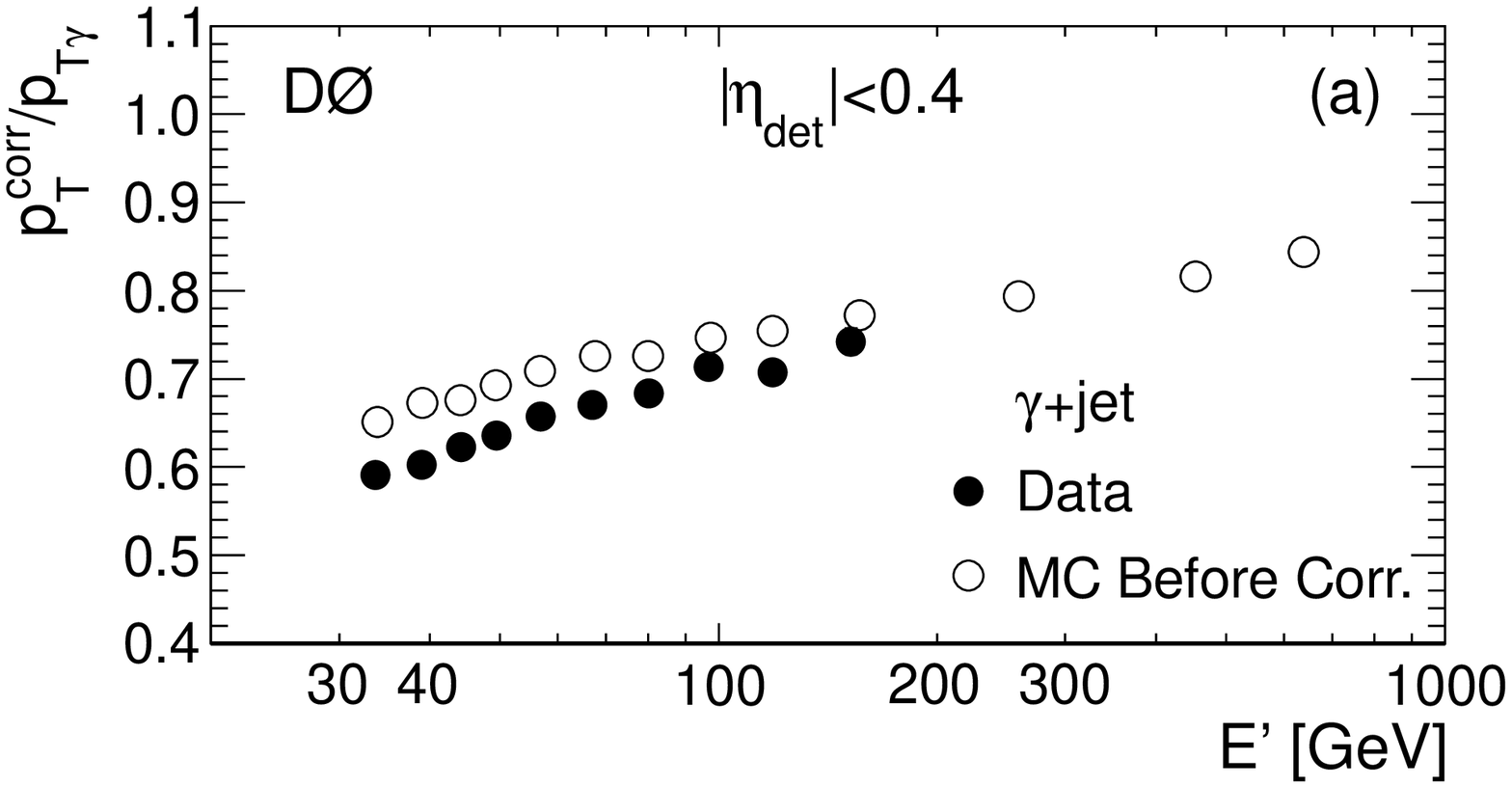}
		\label{fig:fdc:data_closure1_gamjet_default}
	}
	\\
	\subfloat{
		\includegraphics[width=\textwidth]{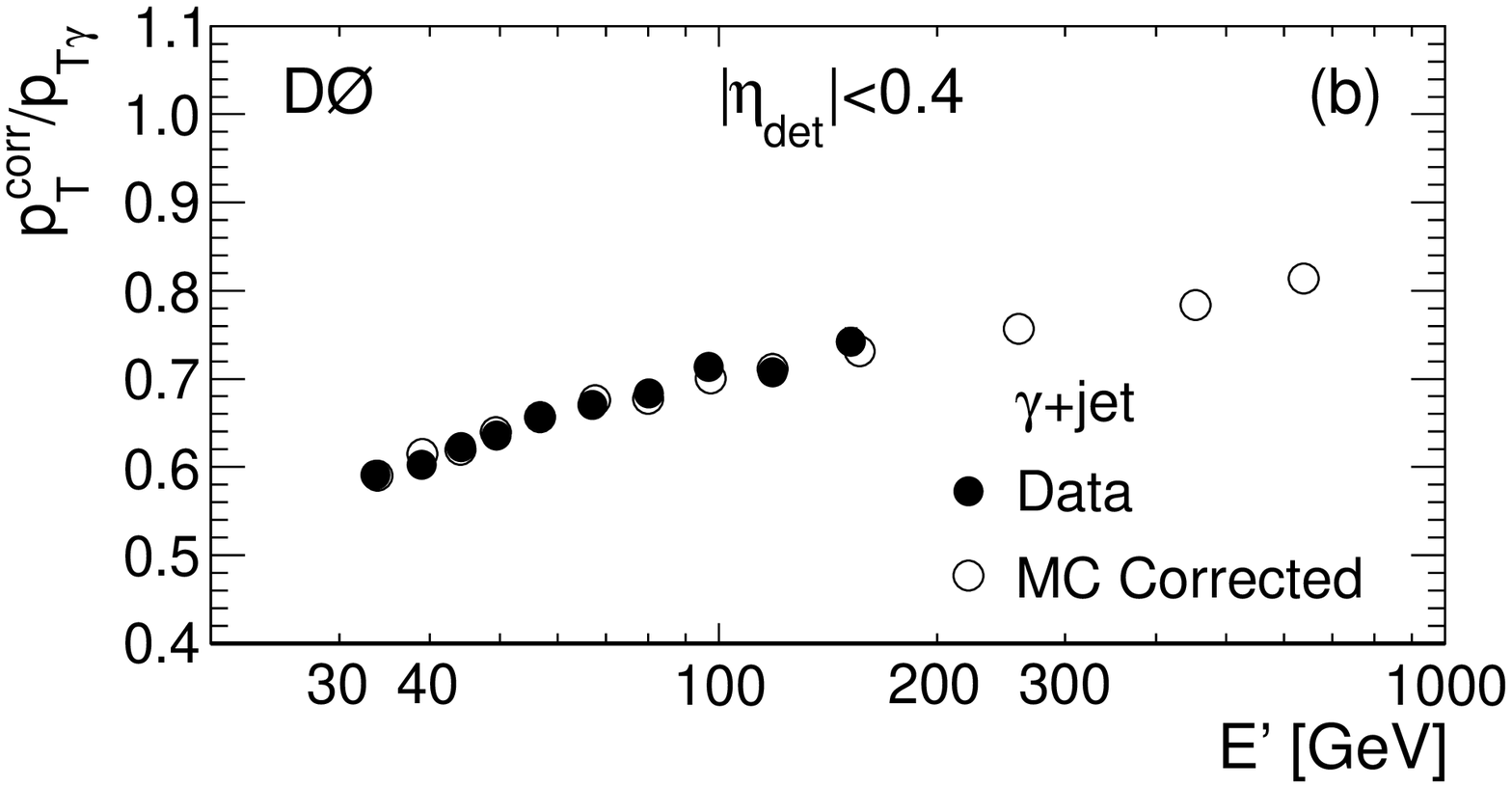}
		\label{fig:fdc:data_closure1_gamjet_newTune}
	}
	\end{minipage}
	\begin{minipage}{0.5\textwidth}
	\subfloat{
		\includegraphics[width=\textwidth]{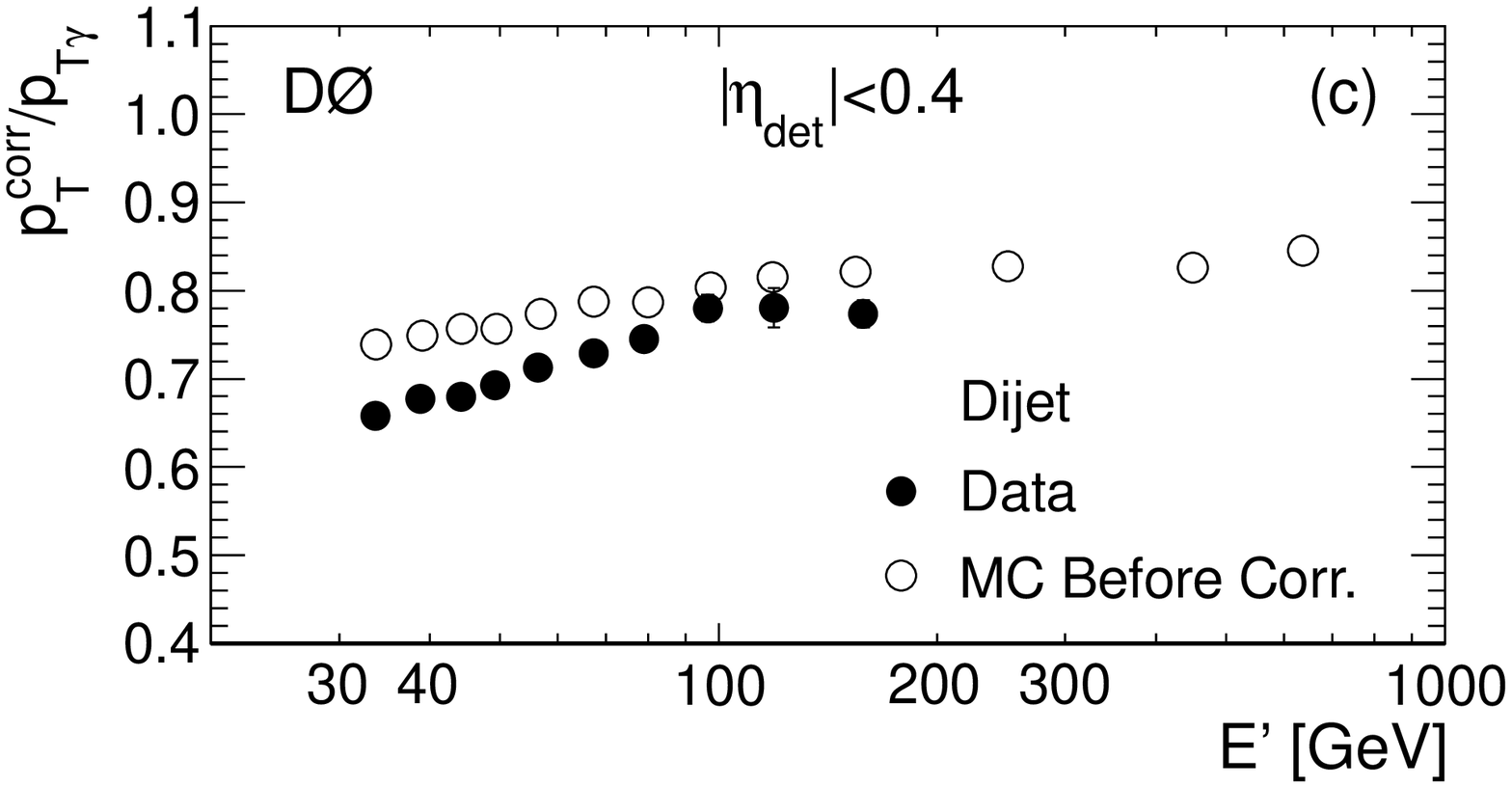}
		\label{fig:fdc:data_closure1_antgamjet_default}
	}
	\\
	\subfloat{
		\includegraphics[width=\textwidth]{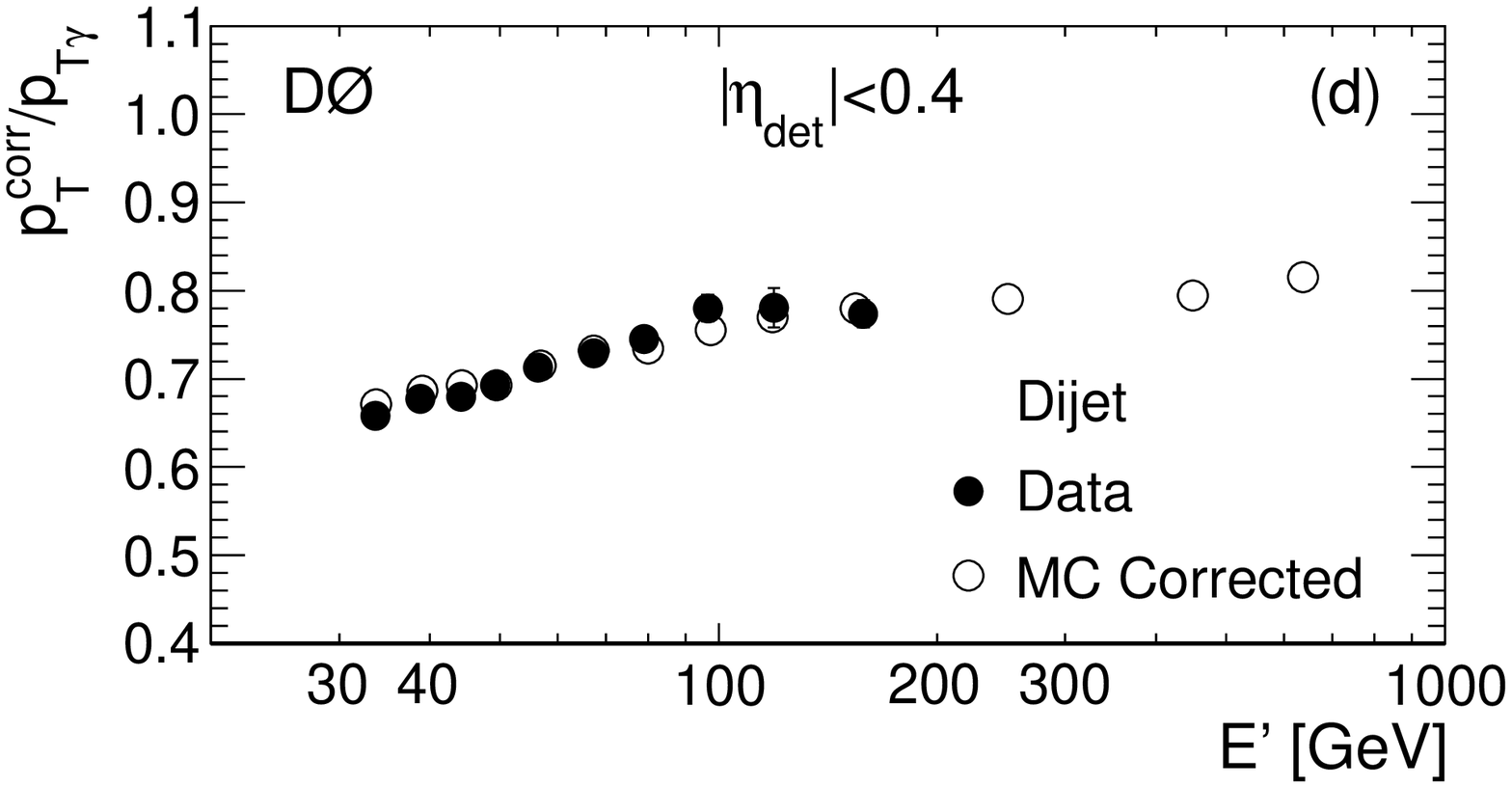}
		\label{fig:fdc:data_closure1_antgamjet_newTune}
	}
	\end{minipage}
	\caption{
		Tuning of single-particle responses (\cref{sssec:SPRdata}) using the ratio
		$\pT[corr]/\pTgamma$ as a function of \Eprime (\cref{eq_eprime}) for jets with $\mdeta < 0.4 $
		in a high-purity \photonjet{} sample, 
		\protect\subref{fig:fdc:data_closure1_gamjet_default}~before the MC response correction and
		\protect\subref{fig:fdc:data_closure1_gamjet_newTune}~after the correction. 
		Plots \protect\subref{fig:fdc:data_closure1_antgamjet_default}
		and~\protect\subref{fig:fdc:data_closure1_antgamjet_newTune}
		show similar results for the \dijet events.
	}
	\label{fig:fdc:data_closure1}
\end{figure*}

The distribution of the ratio $\pT[corr]/\pTgamma$
is first extracted in the \photonjet{} and \dijet data samples (see \cref{fig:fdc:data_closure1}).
Here \pT[corr] is the reconstructed jet \pT with the offset correction,
and \pTgamma is the \pT of the EM cluster that passed tight photon selection criteria.
To select \dijet events, the track isolation requirement has been reversed.
The purity of the \photonjet{} sample is not 100\% in spite of the tight photon selection criteria applied.
Adding the MC \photon-like \dijet events to the MC \photonjet{} sample 
according to the measured purity provides an accurate representation of the selected data sample.
%
The value of \pT[corr] is computed from the known particle composition
of the jet using single-particle responses described in \cref{sssec:SPRMC}. 
The responses of \photon, $e^{\pm}$ and $\mu^\pm$ are assumed to be the same in data as in MC. 
The hadron responses introduce three additional parameters $A$, $B$ and $C$ as compared to 
the MC response parameterization (\cref{eq:SPRhadronsMC}):
\begin{equation}
	R_\textnormal{h}^\textnormal{data} = C\,p_{\textnormal{h}}^{(0)}\cdot\sqbrackets{1 - A\,p_{\textnormal{h}}^{(1)}\cdot(E / 0.75)^{p_{\textnormal{h}}^{(2)} + B - 1}}.
	\label{eq:SPRhadronsData}
\end{equation}
These parameters are varied to reproduce the data distribution 
of the $\pT[corr]/\pTgamma$ ratio in MC.

\subsection{Results}
\label{sssec:SPresults}

The tuning of parameters $A$, $B$, and $C$ is performed using a fit, which is performed simultaneously for the \photonjet{} and \dijet samples.
The procedure is applied for four different \mdeta regions of the detector.
\Cref{fig:fdc:data_closure1} shows the result of the tuning for jets with $\mdeta < 0.4$ 
in samples dominated by \photonjet{} and \dijet{} events. 
The $\pT[corr]/\pTgamma$ ratios are shown before and after MC tuning.
Good agreement between MC and data is obtained. 
\begin{figure*}
	\subfloat{
		\includegraphics[width=0.33\textwidth]{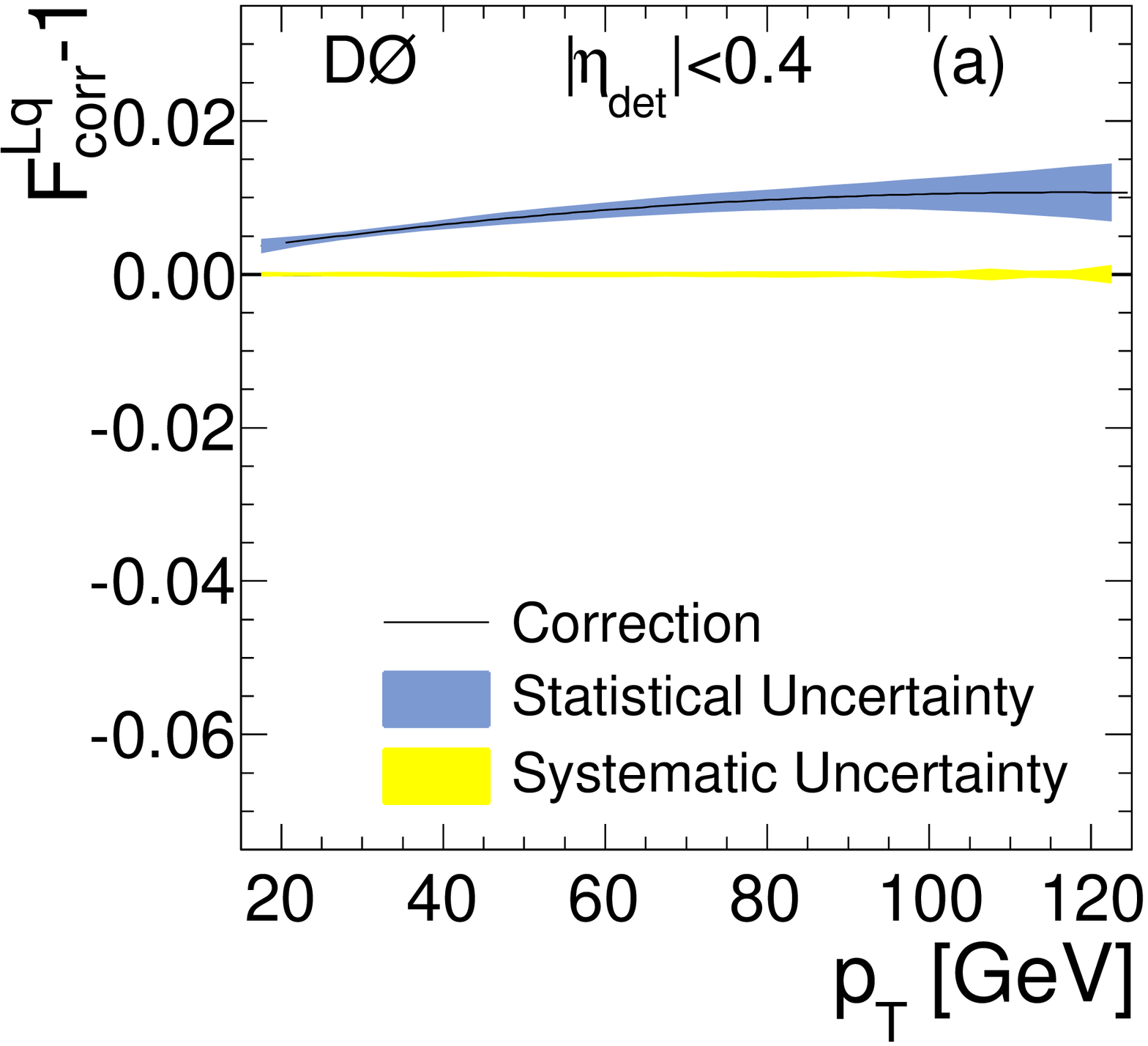}
		\label{fig:fdc:final1_light}
	}
	\subfloat{
		\includegraphics[width=0.33\textwidth]{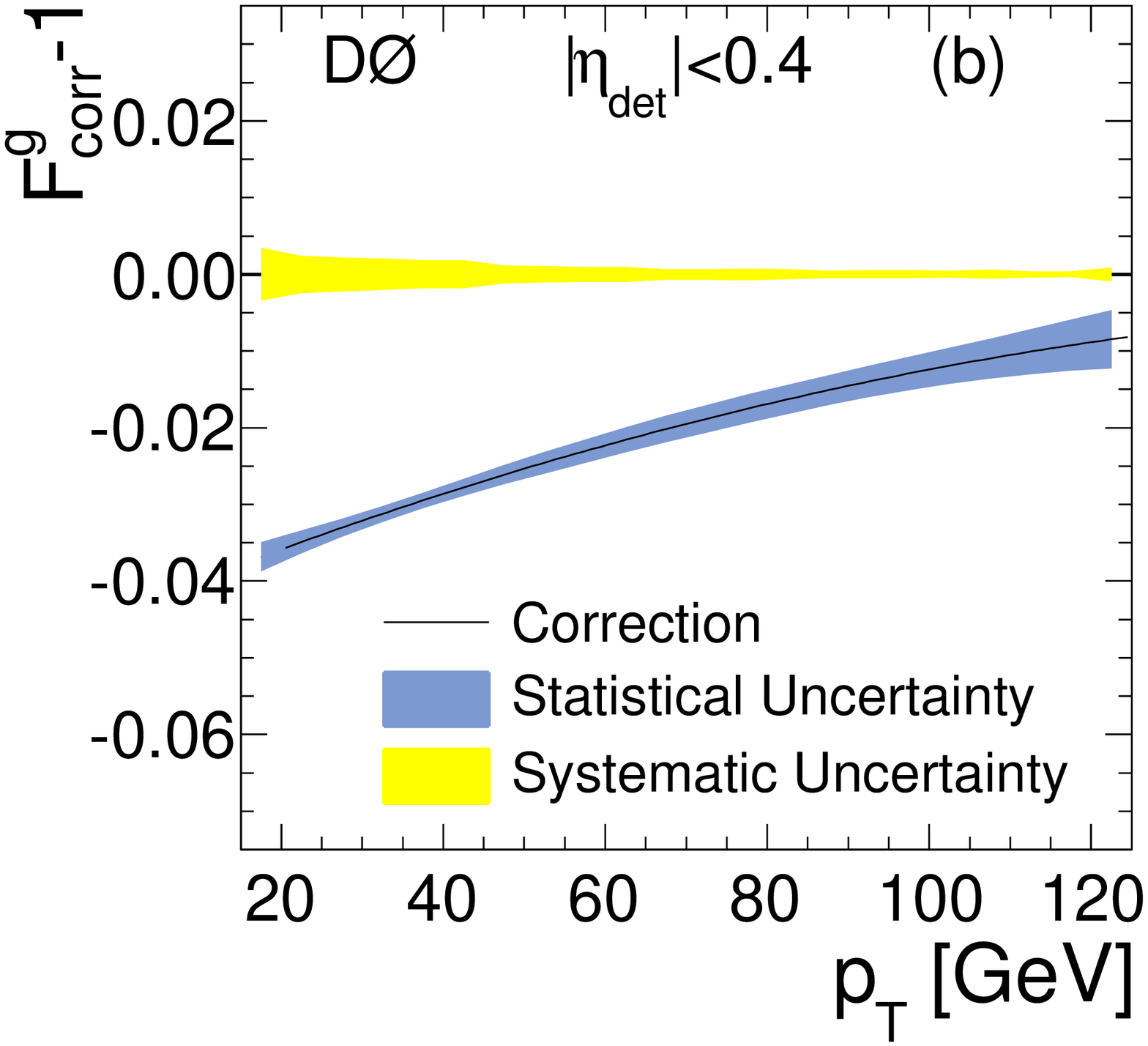}
		\label{fig:fdc:final1_gluon}
	}
	\subfloat{
		\includegraphics[width=0.33\textwidth]{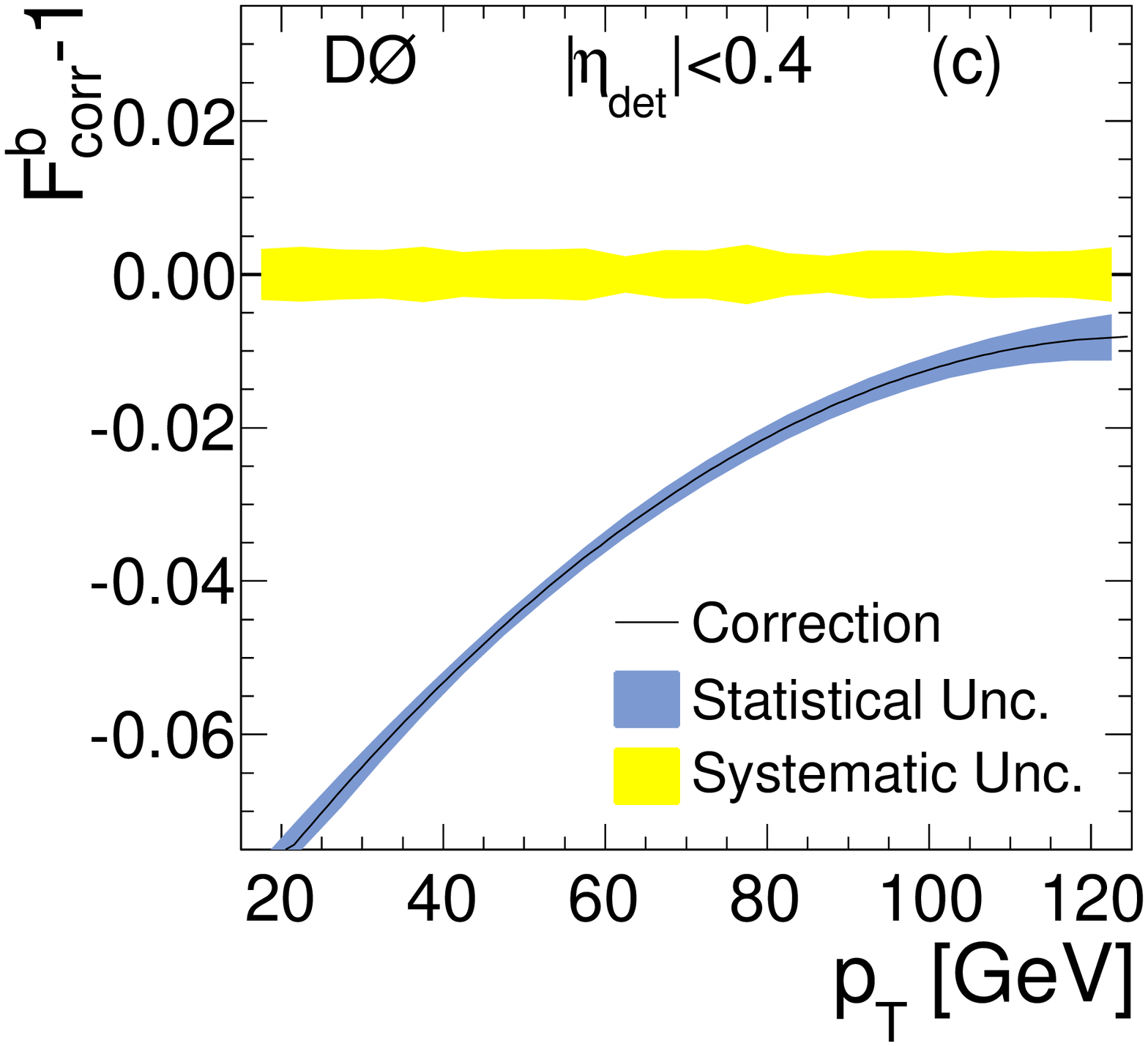}
		\label{fig:fdc:final1_b}
	}
	\caption{
		(color online) Correction factor $F_\textnormal{corr}$ derived using tuned MC single-particle responses 
		for central jets ($\mdeta < 0.4$) and different jet flavors, shown separately for jets from
		\protect\subref{fig:fdc:final1_light} light quarks ($u$, $d$, $s$, $c$),
		\protect\subref{fig:fdc:final1_gluon} gluons, and
		\protect\subref{fig:fdc:final1_b} $b$ quarks.
		The bands represent the statistical and systematic uncertainties.
	}
	\label{fig:fdc:final1}
\end{figure*}

The resulting relative jet energy correction factors $F_\textnormal{corr}$ for different jet flavors (light quark, gluon, and  bottom quark) 
are shown in \cref{fig:fdc:final1}. 
The relative correction required for the light quarks is 1\%, but it is
significantly larger for gluon and $b$-quark jets, where energies are undercorrected by a few percent,
especially at low \pT.

The dominant contribution to the uncertainty is obtained by propagating the fit errors
on the three parameters $A$, $B$, and $C$ as estimated by the covariance matrix and is typically smaller than 0.5\%.
The assumption that the \photon and $e^{\pm}$ 
responses are well simulated in MC needs to be verified.
The \photon{} and $e^{\pm}$ responses are varied according to the accuracy 
of their energy scale calibration (0.6\% and 0.3\% respectively), and
the effect is found to be negligible. 

%% file: jssr.tex
\section{Jet-\pT shifting and smearing in Monte Carlo}
\label{sec:jssr}

\newcommand{\pTZ}[1][]{\JESmathSym{\MakeTransverse{p}{Z\textnormal{#1}}}}
\newcommand{\dS}[1][]{\JESmathSym{\Delta S_{\textnormal{#1}}}}

\subsection{Introduction}
\label{ssec:SSR:Introduction}

In many physics analyses results are derived from a comparison between data and MC simulations.
Given the limitations of MC simulation
(\eg, in the modeling of the parton showering and approximations in the modeling of the \DZero{} detector),
it is necessary to modify the standard simulation in order to match the performance observed in data.
This is especially true for jets. A method has been developed to correct the simulated jets
by the residual data/MC difference in energy scale, resolution, and reconstruction efficiency,
known as Jet Shifting Smearing and Removal (JSSR) method.

To derive these JSSR corrections, we select \Zjet{} events in data and MC (see \cref{sec:datasets}).
The observable used in this study is the transverse momentum imbalance in a two-body process, given by:
\begin{equation}
	\dS = \frac{\pT - \pTZ}{\pTZ},
	\label{eq:deltas}
\end{equation}	
where \pTZ and \pT are the $Z$-boson and jet transverse momenta.

\begin{figure}
	\includegraphics[width=\columnwidth]{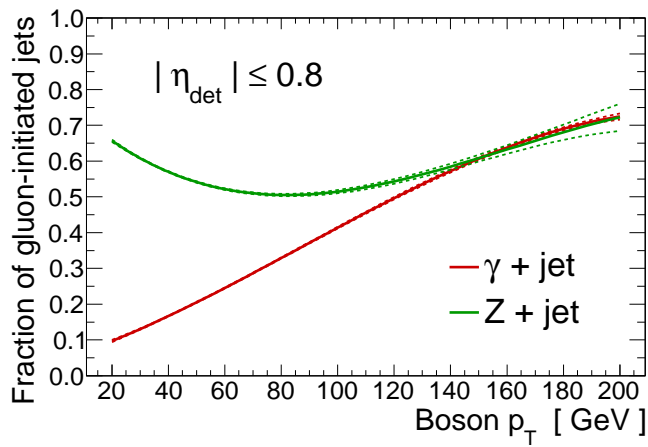}
	\caption{ (color online)
		Fraction of gluon-initiated jets in \Zjet and \photonjet events with the jet in the central calorimeter according to \PYTHIA{} simulation.
	}
	\label{fig:qg_frac_zj}
\end{figure}

In our data/MC consistency checks (\cref{sec:closure}) we confirmed that jet energies in data and MC are
intercalibrated, \ie, \Eptcl is reproduced within the estimated uncertainties in \photonjet events. Therefore,
the value of the \dS variable is sensitive to differences in the flavor compositions of jets balancing 
the $Z$ boson versus those balancing the photon. Since at small photon \pT the jets produced in 
\photonjet events
are mostly ``quark'' jets, $qg\to q\photon$ (see \cref{fig:qg_frac} in \cref{sec:qcd_specific}),
we calibrate mostly quark jets when applying the MPF procedure,
and the \dS for calibrated quark jets should be consistent between data and MC, yielding
$\dS_{\textnormal{data}}^{q} - \dS_{\textnormal{MC}}^{q} = 0$.
This is also confirmed by the small residual shift, $(+1.3 \pm 0.8 )\%$, observed
by constraining the $W$ boson mass reconstructed from decays into two quark jets in \RefCite{ttbar}.
However, as shown in \cref{fig:qg_frac_zj}, 
the flavor composition predicted by the {\sc pythia} simulation in \Zjet{} events is different than in \photonjet,
and a relative data/MC shift in \dS is observed 
in \Zjet{} events which is related to the larger fraction of gluon jets in the final state. 
At high energies the jet response for both quark and gluon jets increases, and 
jet energy scale corrections become smaller (see \cref{jcca_rtrue_gamjet_vs_dijet}, Sec.~16). Also, the
jet flavor compositions in \photonjet and \Zjet{} events converge.
Thus, the calibrated \dS shift is expected to vanish at large jet energies in \Zjet{} events.

\subsection{Event selection}
\label{ssec:SSR:Selection}

This study is performed using the \Zjet{} sample with $Z$ bosons decaying into a $\mu^+\mu^-$ pair.
The muon momentum is calibrated with an accuracy of about 0.3\% \cite{bib:RunII_muon}.
The selected events are required to contain only one jet 
and two muons.
The two muons are required to have $\pT[$\mu$] \geq \GeV{15}$, $\abs{\eta^{\mu}}<2.0$, 
and the dimuon invariant mass is required to be in the range $80 < M_{\mu\mu} < \GeVcc{110}$ .
The jet should be within $\mdeta < 2.4$.
Finally, the $Z$ boson must be back to back with respect to the jet in the transverse plane: $\Delta\azim(Z,\textnormal{jet}) \geq \rad{2.8}$.
The data and MC are first calibrated with the energy scale correction as described in the previous sections,
including the single-particle response correction (\cref{sec:fdc}) for MC jets.
In both data and MC, the muon momentum is calibrated using $Z\to \mu\mu$ events.
The muon energy resolution in MC is also tuned to data using the reconstructed width
of the $Z\to \mu\mu$ mass peak.

\subsection {Procedure}
\label{ssec:SSR:Procedure}

\begin{figure}[t]
	\subfloat{
		\centerline{\includegraphics[width=\columnwidth]{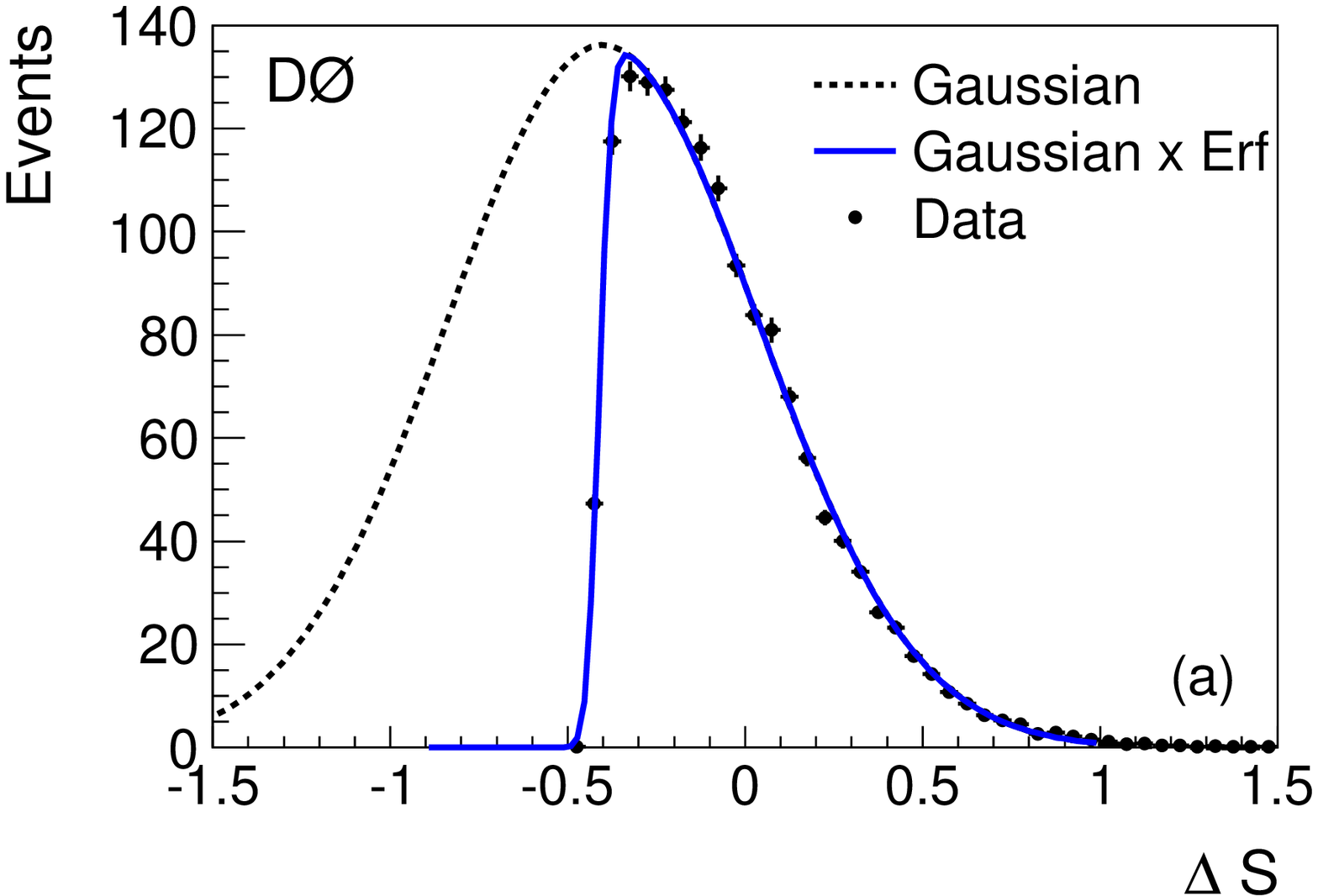}}
		\label{fig:dS_19_22}
	}
	\\
	\subfloat{
		\centerline{\includegraphics[width=\columnwidth]{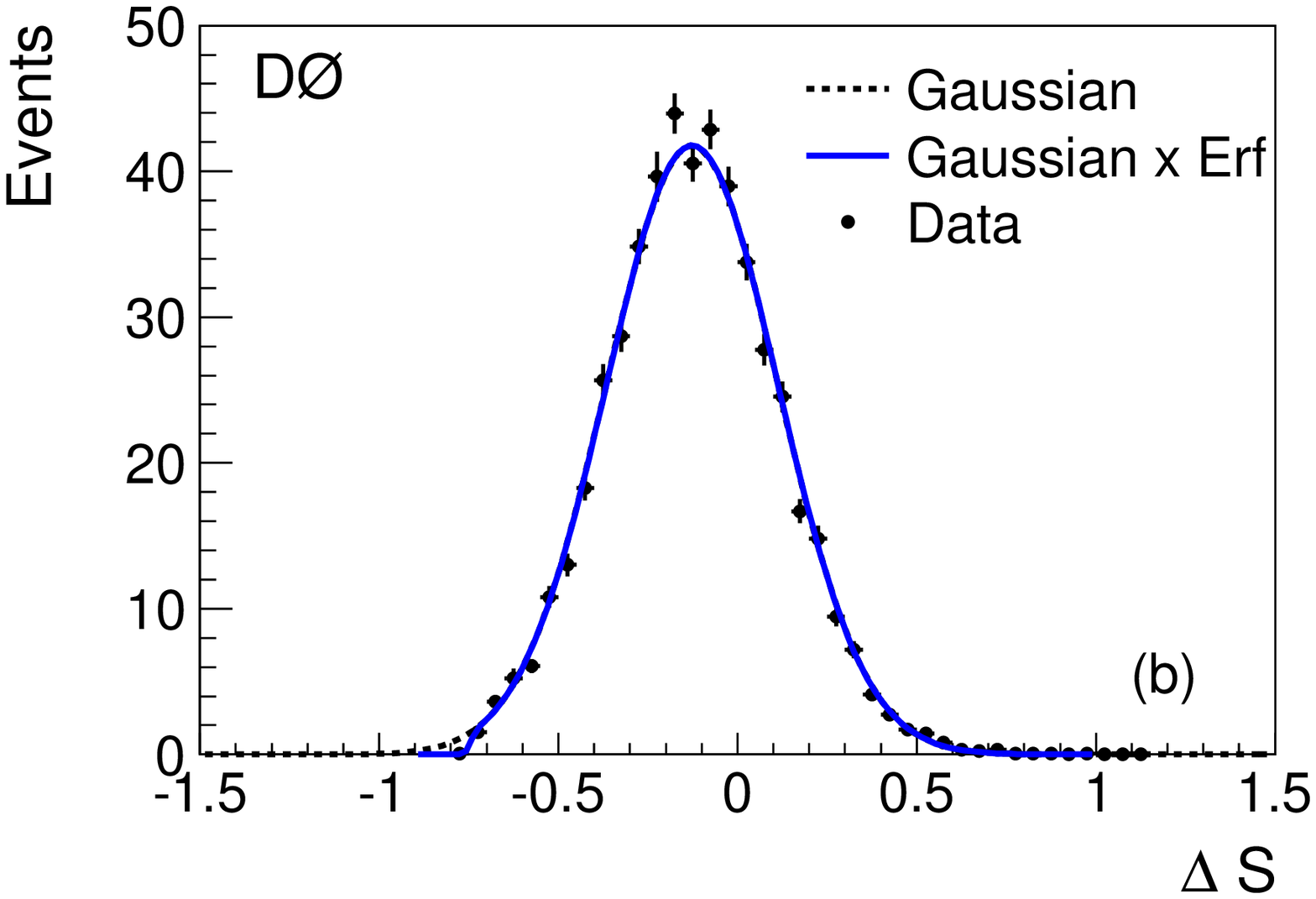}}
		\label{fig:dS_45_50}
	}
	\caption{
		The \dS distribution in the MC sample,
		for \pTZ{} in the ranges \protect\subref{fig:dS_19_22} \GeVc{19-22}
		and \protect\subref{fig:dS_45_50} \GeVc{45-50}.
		The solid line shows the best fitting function according to \cref{eq:deltas},
		while the dashed line shows only its Gaussian component.
		The threshold effect is negligible for \pTZ{} larger than \GeV{40}.
	}
	\label{fig:dS}
\end{figure}

The \Zjet{} events are split into different samples with similar \pTZ,
covering the range $10 \leq \pTZ \leq \GeV{250}$.
For each subsample, a \dS (\cref{eq:deltas}) distribution is obtained.
\Cref{fig:dS} shows that $\Delta S$ for $\pTZ > \GeVc{40}$ is well described by a Gaussian distribution,
as expected due to  the \pT imbalance and reconstruction resolution of both jet and $Z$ boson.
However, as the \pTZ{} decreases, this simple description fails
because of the jet \pT reconstruction threshold used in the jet finding algorithm ($\pT[meas] > \GeVc{6}$, see \cref{jetid}).
Therefore, in the more general case, \dS should be described by a function composed of a Gaussian distribution, 
combined with an Error Function ($\erf$) used to represent the jet \pT threshold effect:
\begin{multline}
	A\,\exp\sqbrackets{-\frac{\sqr{\dS - \average{\dS}}}{2\,\sigma_{\dS}^{2}}} 
\sqbrackets{1 + \func[\erf]{\frac{\dS - T}{\sqrt{2}\,\sigma_{T}}}},
	\label{eq:dSshape}
\end{multline}
with five free parameters: the relative average shift of jet \pT{} with respect to \pTZ, \average{\dS},
the resolution smearing of the jet \pT, $\sigma_{\dS}$,
the parameters of the Error Function, $T$ and $\sigma_{T}$ (used to model the removal of low \pT jets),
and a normalization factor, $A$.
The muon resolution is negligible compared to jet resolution. 

The shifting, smearing, and removal parameters are derived independently 
for data and MC samples in two steps, described below, and are combined into the final correction.
The corrections, for application in MC events, are extracted independently 
for jets falling into three distinct detector regions:
$\mdeta < 0.8$, $0.8 \leq \mdeta < 1.6$ and $\mdeta \geq 1.6$.

\subsubsection {Jet reconstruction threshold}
\label{sssec:SSR:Removal}

\begin{figure}
	\centerline{\includegraphics[width=0.9\columnwidth]{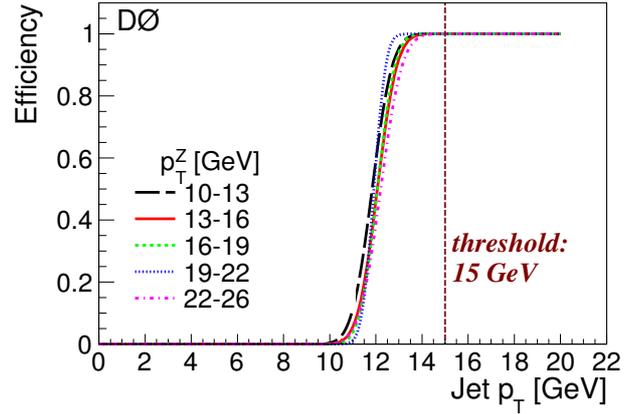}}
	\caption{
		(color online) The jet reconstruction efficiency turn-on curve
		as function of the jet \pT{} for different \pTZ{} ranges of the MC sample.
		The efficiency curve for data is very similar.
		The physics analysis selection threshold of \GeVc{15} is also shown.
	}
	\label{fig:SSR:turnon}
\end{figure}

The five parameters of \cref{eq:dSshape} are first obtained
from the fit to the \dS distribution in data and MC,
for each \mdeta{} and \pTZ{} range.
The threshold parameters $T$ and $\sigma_{T}$ are found to be almost independent of \pTZ{}
(\cref{fig:SSR:turnon}) and $\eta$ region, with a mean value $T$ of about \GeVc{12} and 
standard deviation $\sigma_{T}$ of about \GeVc{1}.
Based on this, the value of \GeVc{15} ($\approx\!3\,\sigma_{T}$ above $T$)
is recommended as the \pT threshold for jet selection in physics analyses.
This ensures close to 100\% jet reconstruction efficiency for both data and MC.

\subsubsection {Relative energy scale and resolution}
\label{sssec:SSR:ShiftingAndSmearing}

Due to the universality of the threshold $T$,
the second step of the procedure can be simplified by fixing the threshold to the value of \GeVc{12}.
In this step, a fit is performed to find the values of the other four parameters
that best describe the \dS distribution in each \pTZ{} range.
In regions with $\pTZ > \GeVc{40}$, where the threshold has no effect,
a simplified distribution made of a single Gaussian is employed in place of \cref{eq:dSshape}.

\begin{figure}[t]
	\includegraphics[width=0.9\columnwidth]{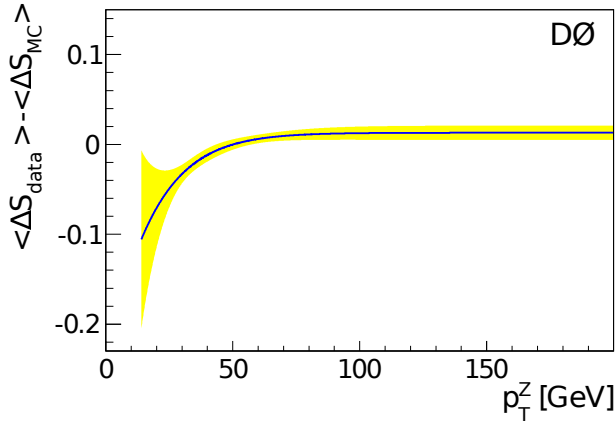}
	\caption{
		(color online) Residual jet energy shifting as a function of \average{\pTZ} for central jets ($\mdeta < 0.8$).
		The yellow band shows its statistical uncertainty.
	}
	\label{fig:SSR:shift}
\end{figure}

The difference between data and MC scale shifts 
$\average{\dS[data]} - \average{\dS[MC]}$
is used for the relative correction of MC jet \pT.
This is the ``shifting'' part of the correction.
\Cref{fig:SSR:shift} shows this correction
for jets in the central region of the calorimeter. 
In general, the residual shifting is small ($\approx 1\%$) for high energies,
while it becomes relevant for low energies. 

The procedure of optimization of the parameters in \cref{eq:dSshape} is unstable
for small \pTZ, where the threshold effect is so dominant that the parameters of
the shape must be inferred from just the tail of the Gaussian distribution of
\dS, as shown in \cref{fig:dS_19_22}.
Therefore to cover the region $\pTZ < \GeVc{20}$, an extrapolation is preferred to the direct measurement.

\begin{figure}[t]
	\includegraphics[width=0.9\columnwidth]{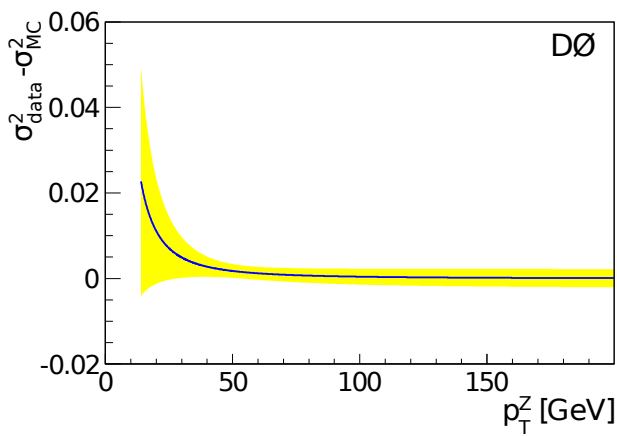}
	\caption{
		(color online) Jet resolution correction as a function of \average{\pTZ} for central jets ($\mdeta < 0.8$).
		The yellow band shows the statistical uncertainty.
	}
	\label{fig:SSR:smear}
\end{figure}



When the residual energy shifts are extracted,
the same procedure also provides the resolution parameter $\sigma_{\dS}$.
The quadratic difference in resolution between data and MC,
$\sigma^{2}_{\textnormal{data}} - \sigma^{2}_{\textnormal{MC}}$, is used to ``smear'' the jet \pT in MC simulation.
\Cref{fig:SSR:smear} shows that the resolutions for central jets are
similar in data and MC. 

\subsubsection {Application of the correction}
\label{sssec:ApplySSR}


The JSSR corrections are applied to all MC jets.
Since the corrections were derived as a function of \pTZ, but need to be applied to jets as a function of jet \pT,
a mapping is used 
between the average jet \pT and \pTZ.
This mapping is mostly linear, with a residual non-linearity 
in the low \pTZ regions
due to the jet reconstruction threshold discussed in \cref{sssec:SSR:Removal}.

The JSSR corrections are applied as follows.
First, the energy of the jets in MC is smeared according to $\sigma^{2}_\textnormal{corr} = \sigma^{2}_{\textnormal{data}} - \sigma^{2}_{\textnormal{MC}}$.
If the fitted $\sigma^{2}_\textnormal{corr}$ value is negative, the correction is not applied.
Secondly, the jet \pT{} is shifted by the amount corresponding to $\average{\dS_{\textnormal{data}}} - \average{\dS_{\textnormal{MC}}}$.
Finally, if the resulting jet \pT{} is below \GeVc{15}, the jet is removed from the list of jets in the event.

The JSSR corrections are applied in addition to all the jet energy scale corrections described in \cref{sec:overview},
which are derived with \photonjet events. 
The single-particle response correction (\cref{sec:fdc}) applied jet-by-jet 
separately for different jet flavors (see \cref{fig:fdc:final1}), significantly reduces
variation in jet \pT scale and resolution between data and MC,
with a negligible residual difference for \pT resolution and 
reduced relative \pT shifting. Instrumental and other detector effects in the ICD region
affect the agreement of the MC description with data resulting in larger systematic uncertainties in this region.

\subsection {Uncertainties}
\label{sssec:Syst}
As described in \cref{sssec:SSR:ShiftingAndSmearing},
the basic procedure to derive shifting and smearing corrections is the same.
When the fit is performed for data and MC,
it delivers the best-fit parameter values and the covariance matrix.
The uncertainty on the final correction is obtained by error propagation,
combining the two uncertainties from data and MC as independent.
The bands in \cref{fig:SSR:shift,fig:SSR:smear}
represent the uncertainty on the correction.
Both for shifting and smearing in the high \pTZ regions,
the uncertainty is of the order of 0.5\%,
while for $\pTZ < \GeVc{20}$ it increases up to 4\%--7\% for the jet \pT shifting and 2\%--3\% for the \pT smearing.



%% file: qcd_specific.tex
\section{Dijet-specific corrections}
\label{sec:qcd_specific}

\begin{figure}
\centerline{\includegraphics[width=0.9\columnwidth]{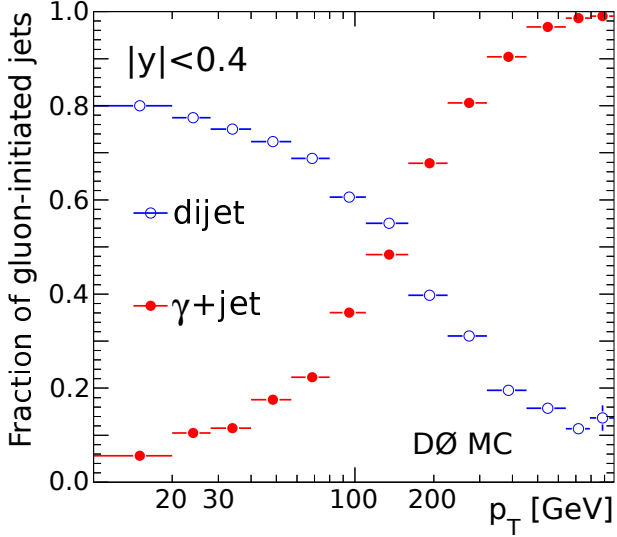}}
\caption{ (color online)
Fraction of gluon-initiated jets in \photonjet and \dijet events in the central calorimeter.}
\label{fig:qg_frac}
\end{figure}

The jet energy calibration corrections derived so far are designed to
correct the energy of jets with a flavor composition similar to that 
in the \photonjet{} sample. 
Event samples chosen for particular analyses may differ substantially from this case.  
For example, \Cref{fig:qg_frac} shows the fraction of gluon-initiated jets in the CC region
for \photonjet{} and \dijet{} events simulated with \PYTHIA{}.
Furthermore, these corrections are not guaranteed to properly calibrate 
to the particle level all the jet four-momentum components. These caveats do not
represent significant limitations for physics measurements relying
on the comparison of observables between data and MC, 
which benefit from JSSR corrections. 
Rather than on the
absolute energy scale calibration, these measurements depend on the relative 
intercalibration of MC and data (\cref{sec:fdc,sec:jssr}). 

In contrast, most of the QCD physics program is based on the comparison 
of observables in data, corrected to the particle level, with theoretical predictions.
These measurements depend on the absolute energy calibration of jets
in data. As discussed previously, the different flavor composition of jets
in \dijet events as compared to \photonjet production requires a dedicated
jet energy calibration to be derived for this sample. 
Furthermore, many of the observables considered in \dijet QCD measurements (jet \pT, $\eta$, \dijet mass)
involve components of the four-momentum other than the energy that are also required to be properly
calibrated to the particle level (see, for example, Refs.~\cite{Abazov:2008ae,IncJets,DiJetM}). 

This section outlines the calibration strategy of the jet four-momentum
in the dijet sample. These corrections are determined only for $\rcone=0.7$ jets,
used for the QCD jet measurements at \DZero.
A more complete description can be found in Ref.~\cite{IncJets}. 

\subsection{Jet energy calibration}

To understand what modifications are required to properly calibrate jet
energies in the \dijet{} sample, it is useful to start from the
expression of the corrected energy in case of the \photonjet{} sample, making
explicit the different subcorrections. 
Given a jet with detector pseudorapidity \deta, the corrected jet 
energy, following \cref{master3,master_resp}, is given by:
\begin{equation}
\Ecorr = \frac{\Emeas - \hat{E}_\textnormal{O}}
{\MPFCC{\photonjet}\,\Feta[\photonjet]\,\kR[topo,\photonjet]\,S^{\textnormal{\photonjet}}}
\cdot\frac{\kOZS}{\kRZS}
\end{equation}
where all subcorrections which are expected to be sample-dependent 
are denoted with the superscript ``\photonjet''. 

To calibrate the jet energy in the QCD \dijet sample, it is necessary 
to redetermine every sample-dependent subcorrection above. 
The relative MPF response correction 
($\Feta[\dijet]$) was already determined in \cref{sec:etadep} and is therefore
available. It is possible to estimate the showering correction directly from \dijet
data following a similar approach to the one used in \cref{sec:showering}.
However, the absolute MPF response correction can not be estimated from \dijet data. 
By taking into account the nature
of each of the subcorrections, it can be demonstrated that the energy for jets from QCD \dijet events
can be properly calibrated using the following modified formula:
\begin{equation}
\Ecorr = \frac{\Emeas - \hat{E}_\textnormal{O}}
{\MPFCC{\photonjet} \,\Feta[\dijet]\,\kR[topo,\photonjet]\,\Delta_{R}\,S^{\textnormal{\dijet}}}
\cdot\frac{\kOZS}{\kRZS}
\label{ecorr_dijet}
\end{equation}
where the main difference is the replacement of $S^{\photonjet}$
with $S^{\dijet}$ 
and the addition of the correction factor $\Delta_{R} = \MPFCC{\photonjet} / \MPFCC{\dijet}$, defined
as the ratio of true jet responses in the $\mdeta < 0.4$ region between \photonjet and \dijet events.

The two additional corrections are evaluated in MC, for both data and MC jet 
energy calibration. In the case of $S^{\dijet}$, this is justified by the consistency
of the direct measurements of $S^{\photonjet}$ in MC and in data
(see \cref{jetprofile_mc_example}).
Also, the MC simulation shows that
the different jet flavor composition between \dijet and \photonjet events results in a very small difference
in the true showering correction, typically less than $1\%$.
Nevertheless, this correction is explicitly included. 

\begin{figure}
\includegraphics[width=0.95\columnwidth]{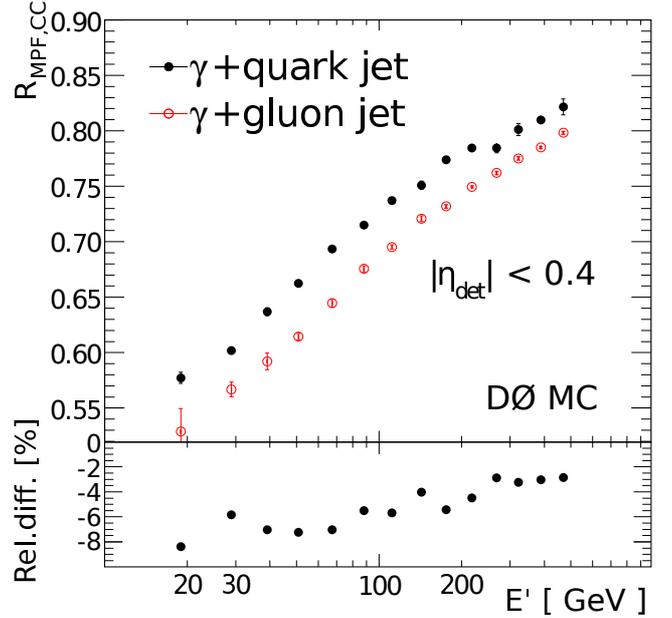}
\caption{
Quark- and gluon-initiated jet responses and their relative differences for jets with $\mdeta < 0.4$
jets as a function of \Eprime.}
\label{jcca_rtrue_gamjet_vs_dijet}
\end{figure}

In contrast, the MC simulation predicts a significant difference in true jet response for central jets between 
\photonjet and \dijet events, as shown in \cref{jcca_rtrue_gamjet_vs_dijet}.
Their ratio is estimated using MC with a tuned single-particle response (\cref{sec:fdc}). The same ratio is also used for data.

\subsection{Jet \pT{} calibration}
\label{ssec:qcd:pTcalibration}

\newcommand{\kRpT}[1][]{\JESmathSym{k_{\textnormal{R},\pT}^{\textnormal{#1}}}}
\newcommand{\kOpT}[1][]{\JESmathSym{k_{\textnormal{O},\pT}^{\textnormal{#1}}}}
\newcommand{\SpT}[1][]{\JESmathSym{S_{\pT}^{\textnormal{#1}}}}
\newcommand{\RpT}[1][]{\JESmathSym{R_{\pT}^{\textnormal{#1}}}}

\newcommand{\vecpTmeas}[1][]{\MakeTransverseExt{\vec{p}}{\,\textnormal{meas}}{#1}}

The procedure for the calibration of jet \pT{} relies
on the available subcorrections from energy calibration, and the corrected jet \pT{} 
is estimated from the measured one, \pT[meas], according to the following expression:
%
\begin{multline}
\pT[corr] = 
\frac{\pT[meas] - \MakeTransverseExt{p}{}{,O}}
{\MPFCC{\photonjet}\,\Feta[\dijet]\,\kRpT[topo,\photonjet]\,\Delta_{R}\,\SpT[\dijet]}
\cdot\frac{\kOpT[ZS]}{\kRZS}
\label{ptcorr_dijet}
\end{multline}
%
where \MakeTransverseExt{p}{}{,O} stands for the offset \pT{} correction,
\kOpT[ZS] is the zero-suppression bias correction to
the estimated offset \pT,
\kRpT[topo,\photonjet] is the \pT-based topology bias correction
to MPF response and \SpT[\dijet] is the true showering
correction to \pT.


Since the offset correction has been explicitly measured only for energy and not for \pT,
the following approximation is used:
\begin{equation}
\parentheses{\pT[meas] - \MakeTransverseExt{p}{}{,O}} \kOpT[ZS] \simeq 
\frac{\pT[meas]}{\Emeas} \parentheses{\Emeas - \hat{E}_{\textnormal{O}}} \kOZS,
\end{equation}
which is expected to be sufficiently accurate in the kinematic range of interest of
QCD measurements (typically $\pT[corr] > \GeVc{40}$).

Similarly to \cref{eq_mpfbias}, the \pT-based topology bias correction to MPF response is defined as:
\begin{equation}
\kRpT[topo,\photonjet] = \frac{\RpT[\photonjet]}{\MPF{\photonjet,noZB}},
\end{equation}
where \RpT[\photonjet] is the \pT-based true jet response (\cf \cref{rtrue}): 
\begin{equation}
	\RpT[\photonjet] = \frac{\norm{\sum_{i\in\textnormal{ptcljet}} \vecpTmeas[$i$]}}{\pT[ptcl]}.
\end{equation}
This correction is closer to unity than the one for energy, being typically 
1.0\%--2.5\% for $\mdeta < 2.5$.
This indicates that the MPF method
is more suitable for \pT{} calibration than for energy calibration.

Finally, the true showering correction to \pT{} is defined similarly to the (energy-based) true showering
correction (see \cref{strue}):
\begin{equation}
	\SpT = \frac{\norm{\sum_{i\in\textnormal{ptcljet}} \vecpTmeas[$i$] f_{i}
		+ \sum_{i\notin\textnormal{ptcljet}} \vecpTmeas[$i$] f_{i}}}
	{\norm{\sum_{i\in\textnormal{ptcljet}} \vecpTmeas[$i$]}}.
\end{equation}
As expected, the net \pT flow through the jet cone boundary is smaller than the net
energy flow, 
with \SpT values of $-3\%$ to $1\%$ for $\mdeta < 2.5$.

\subsection{Four-momentum correction}
\label{ssec:qcd:RapidityCalibration}

The MC simulation shows that the reconstructed jet rapidity
is generally biased towards the central calorimeter with respect to the true (particle level) jet rapidity,
with the largest deviations observed in the ICR region.
This bias is attributed to detector effects in the ICR in addition to
the jet cone algorithm itself. The absolute effect on the inclusive
jet measurement is small compared to the effect of jet \pT{} calibration,
being the largest (4\%) at $\mdeta \simeq 1.5$~\cite{IncJets}.
The correction $\Delta y$ for this bias is parametrized as a function of $\eta$ and applied to the reconstructed jets:
\begin{equation}
	y^\textnormal{corr} = y^\textnormal{meas} - \Delta y.
	\label{ycorr}
\end{equation}

The calibrated jet four-momentum $p^{\mu,\textnormal{corr}}$ is given by:
\begin{equation}
p^{\mu,\textnormal{corr}} = \parentheses{
	\Ecorr,\,
	\pT[corr]\cos\azim^{\textnormal{corr}},\,
	\pT[corr]\sin\azim^{\textnormal{corr}},\,
	p_{z}^{\textnormal{corr}}
	},
\end{equation}
where \Ecorr and \pT[corr] are defined, respectively, in \cref{ecorr_dijet,ptcorr_dijet}.
The measured jet azimuthal angle is assumed to be unbiased, and thus
$\azim^\textnormal{corr} = \azim^\textnormal{meas}$. Finally, given the calibrated jet energy
and rapidity (\cref{ecorr_dijet,ycorr}):
\begin{equation}
p_{z}^\textnormal{corr} = \Ecorr\ \frac{\func[\exp]{2\,y^\textnormal{corr}} - 1}{\func[\exp]{2\,y^\textnormal{corr}} + 1} \ .
\end{equation}


%% file: correlation.tex
\section{Correlations between jets}
\label{sec:correlation}

\newcommand{\Deltas}[1][]{\JESmathSym{\Delta_{\textnormal{s}}^{#1}}}

Knowledge of the correlation of the JES uncertainties between jets at different
energies and pseudorapidities is important for a proper comparison
of measured quantities with theoretical predictions. This information
is needed, for example, in the global fits of parton distribution functions.
Correlations also play an important role in various 
new physics searches and they have a direct impact on the expected sensitivity.

Determination of these correlations is a more complex task
than determining the individual uncertainties. At the same time, 
it facilitates the calculation
of the correlations in physics analyses.
The method of deriving the correlations is introduced
in \cref{sect_correl_method}, followed by the
presentation and discussion of the results.

\subsection{Method}
\label{sect_correl_method}

To evaluate correlations,
the total uncertainty is split into the
individual components of photon energy scale uncertainty,
statistical uncertainty of the central jet response fit, and other components.
These individual components will be called ``uncertainty sources''.

To simplify the description, each source
is assumed to be independent from the others. This is true in most cases. 
For example, the photon energy scale uncertainty has no connection to the showering
systematics. Each individual uncertainty source is assigned a function \Deltas{}
describing the relative change in jet energy correction that corresponds
to a ``$1\sigma$'' modification of source $s$ (where $\sigma$ is a standard deviation).
These functions depend on jet energy and pseudorapidity.
While in most cases this is straightforward, a few exceptions are discussed later, 
for example, the fit uncertainties,
which are driven by the covariance matrix of the fit parameters
and therefore can not be considered independent.

The functions $\func[\Deltas]{E,\eta}$ can be used to calculate correlations 
in jet energy uncertainties across different regions of jet energies, $E$, 
and pseudorapidities, $\eta$. They can also be used directly to evaluate 
jet energy scale correlations in an analysis. 

\subsubsection[Systematic uncertainties]{Treatment of systematic uncertainties}

In most cases, the sources of systematic 
uncertainties can be  treated as uncorrelated with each other.  
They are treated as fully correlated
in jet energy and $\eta$, which means that their effect on 
the jet energy scale can be described 
by the $\func[\Deltas]{E, \eta}$ functions.

Typically all the required information
is directly available from the studies of a given source of systematics.
For example, a $0.5\%$ shift in electron energy scale,
one of the dominant uncertainties for central jets,
directly introduces a $0.5\%$ shift in jet energy scale
for all jet energies and angles, \ie{},
$\func[\Deltas]{E, \eta} = 0.005$ for this particular source.

However, in some cases, such as for uncertainties in modeling of soft
underlying physics, the uncertainty is determined as an envelope covering 
various options of MC generators and their parameters. Different MC 
models give different correlations across jet energies and directions.
Since such sources do not give significant contributions to the final error,
an envelope function is assigned to $\func[\Deltas]{E, \eta}$ to simplify the model.


\subsubsection[Statistical uncertainties from the response fit]{Treatment of statistical uncertainties from the response fit}

Special treatment is needed for the error of the absolute response (\cref{sec:response})
fit where correlations across jet energies are introduced.
In general, these correlations are fully described by the
error matrix of the fit. To treat them properly in
the framework of independent sources, the error matrix
is diagonalized by appropriate linear transformation in the
space of fit parameters. The parametric response fit is thus
transformed into three independent uncertainty sources and 
error functions $\Deltas$ that correspond
to the modification in jet response due to one standard deviation change
in the respective transformed fit parameters. 


\subsubsection[Correlations across $\eta$]{Treatment of correlations across $\eta$}

The same treatment as for the jet absolute response fit
can be applied to the ``global fit'' of the relative $\eta$-dependent correction
(\cref{global_fit}).
However, this treatment would be unpractical, since the global
fit contains 57 parameters, requiring the introduction of as many sources of uncertainties. 

\begin{figure}
	\centerline{
	\includegraphics[width=0.8\columnwidth]{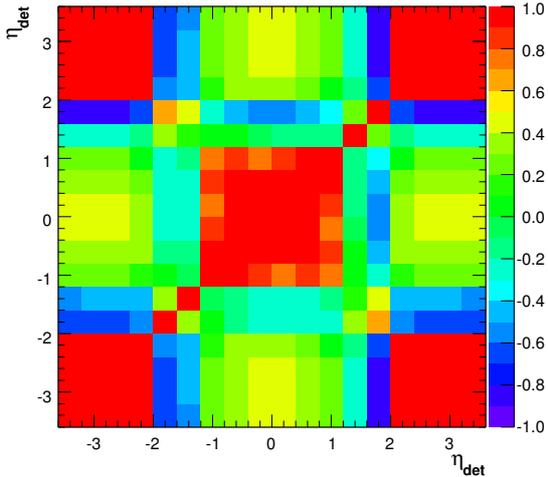}
	}
	\caption{
		(color online) Correlations between two jets at different
		pseudorapidities for $\pT = \GeV{50}$, from the global fit to the
		$\eta$-dependent corrections for $\rcone = 0.7$ jets in data.
	}
	\label{f_corr_etacorrel}
\end{figure}

Instead, the calorimeter
is divided into seven regions of detector pseudorapidity:
$\deta < -2.0$, $-2.0 \leq \deta < -1.6$, $-1.6 \leq \deta < -1.2$, 
$-1.2 \leq \deta \leq 1.2$, $1.2 < \deta \leq 1.6$, $1.6 < \deta \leq 2.0$,
and $\deta > 2.0$. Within these regions, jets are considered fully 
correlated in $\eta$. This partition is motivated by the analysis of the
correlations in pseudorapidity from the global fit of $\eta$-dependent corrections for 
a set of values of jet \pT{} (\cref{f_corr_etacorrel}). 

One uncertainty source is assigned to each of the seven calorimeter regions.
Each error function $\Deltas$ is defined to be equal to the error of the
global fit for jets within its particular calorimeter region, and equal to zero
outside.
To avoid abrupt changes in the correlations at the boundaries of the calorimeter
regions, smoothly falling functions are used around their edges.

Finally, all sources are decomposed into uncorrelated
ones by diagonalizing the correlation matrix.

\subsection{Results}

The jet energy scale uncertainty for \photonjet events is split into a total of 
48 independent sources
(54 for the \dijet{} events). This number is still  large, and further
simplification is possible by the identification of the most important sources
and the merging of the remaining ones into a single residual source. This approach 
is applied, for example, in the inclusive jet cross section 
measurement~\cite{Abazov:2008ae,IncJets}.

The method of calculating the correlations for an arbitrary
observable described in \cref{sect_correl_method} can be applied
to calculate the correlations between the jet energy scale corrections
for two jets at different energies and pseudorapidities,
$\parentheses{E_{1}, \eta_{1}}$ and $\parentheses{E_{2}, \eta_{2}}$.
In this case, for each jet $i$ the error $\sigma^{(i)}_\textnormal{s}$ from a given source $s$ is
the error function $\func[\Deltas]{E_{i}, \eta_{i}}$ multiplied by
the jet energy scale correction, while the correlation coefficient $\rho_{1 2}$ can be expressed as: 
\begin{equation}
\rho_{1 2} = \frac{
\sum_{\textnormal{s}} \func[\Deltas]{E_{1},\eta_{1}} \, \func[\Deltas]{E_{2}, \eta_{2}}
}{
\sqrt{\sum_{\textnormal{s}} \func[{\Deltas[2]}]{E_{1}, \eta_{1}}}\ 
\sqrt{\sum_{\textnormal{s}} \func[{\Deltas[2]}]{E_{2}, \eta_{2}}}
}\,.
\end{equation}

\begin{table}[t]
	\begin{center}
	\caption{
		Correlation coefficients $\rho_{12}$ 
		between two central ($\eta_1=\eta_2 = 0$) jets with $\rcone = 0.7$ 
                in data for different energies $E_1$ and $E_2$ (in GeV).
	}
	\label{t_correl_cone7_1}
		\begin{tabular}{c|ccccc}
			\hline\hline
			\backslashbox{$E_{1}$}{$E_{2}$} & 25 & 50 & 100 & 200 & 500 \\ 
			\hline
			 25 & 1     & 0.932 & 0.816 & 0.681 & 0.512 \\
			 50 & 0.932 & 1     & 0.957 & 0.843 & 0.651 \\
			100 & 0.816 & 0.957 & 1     & 0.951 & 0.79  \\
			200 & 0.681 & 0.843 & 0.951 & 1     & 0.932 \\
			500 & 0.512 & 0.651 & 0.79  & 0.932 & 1     \\
			\hline\hline
		\end{tabular}
	\end{center}
%
	\begin{center}
	\caption{
               Same as in \cref{t_correl_cone7_1} but for 
               central ($\eta_1 = 0$) and ICR ($\eta_2 = 1.4$) jets.
               Also shown is jet transverse momenta in parentheses (in GeV).
	}
	\label{t_correl_cone7_2}
		\begin{tabular}{c|ccccc}\hline\hline
			\multirow{2}{*}{\backslashbox{$E_{1}$}{$E_{2}$}}
			    &   25  &   50  &  100  &  200  &  500  \\
			    &  (12) &  (23) &  (47) &  (93) & (233) \\
			\hline
			 25 & 0.731 & 0.749 & 0.671 & 0.563 & 0.427 \cr
			 50 & 0.574 & 0.682 & 0.685 & 0.624 & 0.507 \cr
			100 & 0.429 & 0.568 & 0.628 & 0.639 & 0.586 \cr
			200 & 0.319 & 0.452 & 0.548 & 0.627 & 0.666 \cr
			500 & 0.221 & 0.326 & 0.433 & 0.559 & 0.687 \cr
			\hline\hline
		\end{tabular}
	\end{center}
%
	\begin{center}
	\caption{
               Same as in \cref{t_correl_cone7_1} but for 
               central ($\eta_1 = 0$) and forward ($\eta_2 = 2.4$) jets.
               Also shown is jet transverse momenta in parentheses (in GeV).
	}
	\label{t_correl_cone7_3}
		\begin{tabular}{c|cccc}\hline\hline
			\multirow{2}{*}{\backslashbox{$E_{1}$}{$E_{2}$}}
			    &   50  &  100  &  200  &  500  \\
			    &   (9) &  (18) &  (36) &  (90) \\
			\hline
			 25 & 0.663 & 0.714 & 0.701 & 0.582 \cr
			 50 & 0.483 & 0.603 & 0.666 & 0.612 \cr
			100 & 0.328 & 0.470 & 0.581 & 0.626 \cr
			200 & 0.228 & 0.371 & 0.514 & 0.655 \cr
			500 & 0.148 & 0.276 & 0.432 & 0.645 \cr
			\hline\hline
		\end{tabular}%
	\end{center}
\end{table}

The correlation coefficients between central jets at 
different energies are given in \cref{t_correl_cone7_1}. As expected, two central
jets are strongly correlated in energy. This reflects the fact that the uncertainty on
the central jet response  is dominated by the uncertainty on the electron and photon energy scales
in a wide range of energies. Central jets with energies of $E = \GeV{25}$
and $E = \GeV{500}$ still have a correlation of $50\%$.

The correlation drops with increasing distance in pseudorapidity between 
central and forward jets (see \cref{t_correl_cone7_2,t_correl_cone7_3}).
An interesting feature is that the strongest
correlation is for jets with the same \pT{} rather than the same energy.
This is a direct consequence of the calibration procedure, because the energy scale for forward
jets is derived from balancing the event in transverse momentum.


%% file: conclusions.tex
\section{Conclusions}
\label{sec:conclusion}

The correction of the energy of jets reconstructed with the \DZero detector
to the particle jet energy has been presented.
The study described in this paper is based on data taken by the \DZero{} experiment during the
2002--2011 \ppbar{} Fermilab Tevatron collider runs at \TeVenergy,
with a total luminosity of about \invfb{9.7} after imposing data quality requirements.
The corrections are derived for uncorrected jet $\pT \geq \GeVc{6}$ and $\mdeta \leq 3.6$
for data and MC samples for the two cone sizes, $\rcone = 0.7$ and $0.5$
(the first cone size is used mostly for QCD measurements while the latter is used for all other analyses).
Corrections are obtained for five run periods separately.
\Cref{fig:Summary_CorrVsEta_JCCB_DATA,fig:Summary_CorrVsEta_JCCA_DATA} 
show the magnitude of the total correction for jets in data,
and \cref{fig:Summary_ErrVsEta_JCCB_DATA} shows the size of the jet energy calibration 
uncertainty as a function of jet pseudorapidities for different jet energies. 
The overall correction factor to the jet energy in the central calorimeter varies
within $1.4-1.5$ and $1.25-1.3$ for jets with measured transverse momentum
equal to \GeVc{25\text{ and }100}, respectively,
with only a small
dependence on the jet cone size. The total uncertainties at the same energies are within $1.4\%-1.8\%$ in the central rapidity region,
while at larger rapidities ($\mdeta \approx 3.0$) the uncertainties increase to $3\%-3.5\%$.
The procedure is verified by a direct test in MC simulation and
comparisons of simulation with data, which demonstrate that the jet energy is
corrected to the particle level within the quoted uncertainties.
The jet energy scale correction also improves substantially the resolution of 
missing transverse energy, which important for various physics measurements and searches.

%% file: acknowledgement.tex
%
We thank the staffs at Fermilab and collaborating institutions,
and acknowledge support from the
DOE and NSF (USA);
CEA and CNRS/IN2P3 (France);
MON, NRC KI and RFBR (Russia);
CNPq, FAPERJ, FAPESP and FUNDUNESP (Brazil);
DAE and DST (India);
Colciencias (Colombia);
CONACyT (Mexico);
NRF (Korea);
FOM (The Netherlands);
STFC and the Royal Society (United Kingdom);
MSMT and GACR (Czech Republic);
BMBF and DFG (Germany);
SFI (Ireland);
The Swedish Research Council (Sweden);
and
CAS and CNSF (China).

%% file: biblio.tex
\bibliographystyle{unsrt}

%% file: jes_nim.bbl
\begin{thebibliography}{99.}
\label{bibliography}

\newcommand{\EtAl}{\textit{et al.}}
\newcommand{\VPY}[3]{\textbf{#1}, #2 (#3)} 
\newcommand{\CiteJournal}[4]{#1 \VPY{#2}{#3}{#4}}
\newcommand{\DZeroAuth}[1][V.~M.~Abazov]{#1 \EtAl{} (\DZero{} Collaboration)}
\newcommand{\NIMname}{Nucl. Instrum. Meth.~A}
\newcommand{\PRDname}{Phys. Rev.~D}
\newcommand{\PRLname}{Phys. Rev. Lett.}
\newcommand{\PLBname}{Phys. Lett.~B}
\newcommand{\JHEPname}{J.~High~Energy Phys.}
\newcommand{\NIM}[3]{\CiteJournal{\NIMname}{#1}{#2}{#3}}
\newcommand{\PRD}[3]{\CiteJournal{\PRDname}{#1}{#2}{#3}}
\newcommand{\PRL}[3]{\CiteJournal{\PRLname}{#1}{#2}{#3}}
\newcommand{\PLB}[3]{\CiteJournal{\PLBname}{#1}{#2}{#3}}
\newcommand{\JHEP}[3]{\CiteJournal{\JHEPname}{#1}{#2}{#3}}

\newcommand{\arXivHEP}[2]{arXiv:#1 [hep-ex] (#2)}


\bibitem{Abazov:2008ae}
  \DZeroAuth, \PRL{101}{062001}{2008}.

\bibitem{IncJets} 
  \DZeroAuth, \PRD{85}{052006}{2012}. 

\bibitem{DiJetM} 
  \DZeroAuth, \PLB{693}{531}{2010}. 


\bibitem{Top_mass}
  \DZeroAuth, \PRD{84}{012008}{2011}.

\bibitem{ttbar}
  \DZeroAuth, \PRD{84}{2011}{032004}.


\bibitem{Higgs} 
  \DZeroAuth, \PRD{88}{052008}{2013}; \PRL{109}{121804}{2012}.


\bibitem{JetAlgo}
  G.~C.~Blazey \EtAl, arXiv:hep-ex/0005012 (2000).

\bibitem{particle}
  C.~Buttar \EtAl, \arXivHEP{0803.0678}{2008}, section 9.


\bibitem{JES_run1} 
  \DZeroAuth[B.~Abbott], \NIM{424}{352}{1999}.


\bibitem{bib:RunI_nim} 
  S.~Abachi \EtAl, (\DZero{} Collaboration), \NIM{338}{185}{1994}.

\bibitem{bib:RunII_nim} 
  \DZeroAuth, \NIM{565}{463}{2006}.

\bibitem{l1cal}
  M.~Abolins \EtAl, \NIM{584}{75}{2008}.

\bibitem{l0} 
  R.~Angstadt \EtAl, \NIM{622}{298}{2010}.


\bibitem{SMT}
  S.~N.~Ahmed \EtAl, \NIM{634}{8}{2011}.

\bibitem{d0_coordinate}
The polar angle $\theta$ is defined
with respect to the positive $z$ axis, which is along the proton beam direction.
Pseudorapidity is defined as $\eta = -\ln[\tan(\theta/2)]$.
Also, \deta{} and \dphi{} are the pseudorapidity and the azimuthal angle
defined with respect to the geometric center of the \DZero{} detector.


\bibitem{bib:RunII_muon} 
  \DZeroAuth, \NIM{737}{281}{2014} 

\bibitem{LumiNIM} 
  B.~C.~K.~Casey
  \EtAl,
  \NIM{698}{208}{2013}. 

\bibitem{GS}
  A.~G\"{u}ntherschulze, 
  \CiteJournal{Z.~Phys.}{86}{778}{1933} (in German).

\bibitem{Malter}
  L.~Malter, \CiteJournal{Phys. Rev.}{50}{48}{1936}.


\bibitem{geant} R.~Brun and F.~Carminati, CERN Program Library, Long Writeup W5013 (1993).

\bibitem{Vlimant:2005ur} J.-R.~Vlimant,
  FERMILAB-THESIS-2005-52 (2005) (in French).

\bibitem{D0WMass2012}
  \DZeroAuth, 
  accepted by \PRDname, \arXivHEP{1310.8628}{2013}.

\bibitem{PDG}
  J.~Beringer \EtAl (Particle Data Group), \PRD{86}{010001}{2012}

\bibitem{TrackNIM}
  \DZeroAuth, 
  paper in preparation.

\bibitem{bID_nim}
  \DZeroAuth, \NIM{620}{490}{2010}.

\bibitem{Kalman} R.~E.~Kalman, \CiteJournal{J.~Bas. Eng.}{82 \textnormal{D}}{35}{1960};
  R.~E.~Kalman and R.~S.~Brucy, \CiteJournal{J.~Bas. Eng.}{83 \textnormal{D}}{95}{1961};
  P.~Billoir, \NIM{225}{352}{1984}.


\bibitem{Blazey:2000qt} We use the iterative, seed-based cone algorithm including midpoints, 
   as described on p.~47 in G.~C.~Blazey \EtAl, in \emph{Proceedings of the Workshop: 
   QCD and Weak Boson Physics in Run~II}, edited by U.~Baur, R.~K.~Ellis, and
   D.~Zeppenfeld, FERMILAB-PUB-00-297 (2000).

\bibitem{wigmans} R.~Wigmans, \textit{Calorimetry: Energy Measurement in Particle Physics}, 
  Clarendon, Oxford (2000).

\bibitem{mpf} F.~Abe \EtAl (CDF Collaboration), \PRL{69}{2896}{1992}.

\bibitem{pythia} 
  T.~Sjostrand, S.~Mrenna and P.~Z.~Skands, \JHEP{05}{26}{2006}.

\bibitem{cteq6l} J.~Pumplin \EtAl, \JHEP{07}{012}{2002}. 

\bibitem{IncGam} 
  \DZeroAuth,
  \PLB{639}{151}{2006}. 


\bibitem{alpgen} 
  M.~L.~Mangano \EtAl, \JHEP{07}{1}{2003}.

\bibitem{tuneA} R.~Field and R.~Craig Group, \arXivHEP{0510198}{2005}.

\bibitem{jetshapes}
  D.~Acosta \EtAl (CDF Collaboration), \PRD{71}{112002}{2005}. 


\bibitem{gj_PRD} 
  \DZeroAuth, \PRD{88}{072008}{2013}. 


\bibitem{HMCMLL}
  R.~J.~Barlow and C.~Beeston, \CiteJournal{Comp.~Phys.~Comm.}{77}{219}{1993}.


\bibitem{dphi}
  \DZeroAuth,
  \PRL{94}{221801}{2005}.


\bibitem{PRD88_072008_PhotonJet}
  \DZeroAuth, \PRD{88}{072008}{2013}

\bibitem{tuneDWT} D.~Acosta \EtAl, CMS Note 2006-067 (2006).




\bibitem{g3j_PRD} 
  \DZeroAuth, \PRD{81}{052012}{2010}; \VPY{83}{052008}{2011}. 



\end{thebibliography}
